\documentclass[11pt,a4paper]{article}
\usepackage{jheppub}
\usepackage[T1]{fontenc}
\usepackage[english]{babel}
\usepackage{lmodern}
\usepackage{float}
\usepackage{tikz-feynman}
\usepackage{bbm}
\usepackage{bm}
\usepackage{color}
\usepackage{slashed}
\usepackage{graphicx}
\usepackage{braket}
\usepackage{bbold}
\usepackage{dsfont}
\usepackage{todonotes}

\DeclareGraphicsExtensions{.pdf,.png,.jpg}

\allowdisplaybreaks[1]
\def\simg{{\ \lower-1.2pt\vbox{\hbox{\rlap{$>$}\lower6pt\vbox{\hbox{$\sim$}}}}\ }}
\def\siml{{\ \lower-1.2pt\vbox{\hbox{\rlap{$<$}\lower6pt\vbox{\hbox{$\sim$}}}}\ }}

\makeatletter \@addtoreset{equation}{section} \makeatother
\renewcommand{\theequation}{\arabic{section}.\arabic{equation}}

\renewcommand{\vec}[1]{{\bf #1}}
\newcommand{\rmi}[1]{{\mbox{\scriptsize #1}}}
\newcommand{\rmii}[1]{{\mbox{\tiny\rm{#1}}}}

\newcommand{\nB}{n_\rmii{B}}

\newcommand{\go}{g_1}
\newcommand{\gt}{g_2}

\newcommand{\vrel}{v_\rmi{rel}}

\newcommand{\msl}[1]{\,\slash\!\!\!{#1}\,}

\begin{document}

\flushbottom

\begin{titlepage}
\begin{centering}

\vfill

{\Large{\bf
Connecting $t$-channel Dark Matter Models 
\\
to the Standard Model Effective Field Theory
}} 

\vspace{0.8cm}

S.~Biondini$^a$, L.~Tiberi$^{b,c}$ and O.~Panella$^c$

\vspace{0.8cm}

$^{a}${\em Institute of Physics, University of Freiburg,
\\
Hermann-Herder-Straße 3, 79014 Freiburg, Germany} 
\\
\vspace{0.15 cm}
$^{b}${\em Dipartimento di Fisica e Geologia, Università degli studi di Perugia
Via A. Pascoli, I-06123, Perugia, Italy
}
\\
\vspace{0.15 cm}
$^{c}${\em INFN, Sezione di Perugia,
\\
Via A. Pascoli, I-06123, Perugia, Italy}

\vspace*{0.8cm}

\end{centering}

\vspace*{0.3cm}

\noindent
\textbf{Abstract}: 
We investigate the connection between simplified dark matter models featuring a $t$-channel scalar mediator and the Standard Model Effective Field Theory (SMEFT). We focus on scenarios with fermionic dark matter interacting with leptons, under the assumption of Minimal Flavor Violation. The dimension-six SMEFT Wilson coefficients are computed in the Warsaw basis at one loop, with the aid of \texttt{Matchete}. Assuming a compressed mass spectrum for the dark matter and the mediator, we incorporate coannihilations, Sommerfeld enhancement, and bound-state effects in the relic density calculation. We then analyze the interplay between the dark matter energy density, global SMEFT fits, and direct detection constraints. Our results show that SMEFT bounds, though loop-suppressed, can meaningfully constrain the parameter space for $m_\chi \gtrsim 0.5$~TeV and $\mathcal{O}(1)$ portal couplings.

\vfill
\end{titlepage}

\clearpage
\pagenumbering{gobble} 
\tableofcontents

\clearpage
\pagenumbering{arabic}
\setcounter{page}{1}

\section{Introduction}
\label{sec:intro}
The search for dark matter continues to foster strong synergy between the theoretical and experimental communities in particle physics. When we assume that dark matter exists in the form of a particle and interacts with the Standard Model (SM) degrees of freedom, we can rely on powerful detection strategies. These range from direct and indirect detection to collider searches. It is this complementary approach that makes minimal or next-to-minimal dark matter models quite constrained today \cite{Schumann:2019eaa, Arcadi:2017kky,Cirelli:2024ssz}. Moreover, if dark matter particles and other states of an extended dark sector are heavier than the electroweak scale, then possible contributions to generic low-energy observables may arise, which are captured by effective operators only made of SM fields. A systematic exploration of heavy and non-dynamical (new) physics can be performed with the aid of the Standard Model Effective Theory (SMEFT) \cite{Buchmuller:1985jz, Grzadkowski:2010es, Brivio:2017vri}. 

In the quest to uncover the nature of dark matter beyond its gravitational effects, one may adopt a model-specific approach, in which particle dark matter is embedded within a comprehensive theoretical framework built from first principles. While ultraviolet-complete models offer valuable theoretical structure, they are often intricate, particularly in terms of their free-parameter content, making it challenging to extract direct experimental implications. Simplified models are therefore commonly used as a bridge between experimental results and theoretical interpretations~\cite{Alwall:2008ag, LHCNewPhysicsWorkingGroup:2011mji, Morgante:2018tiq}. In this framework, a minimal set of fields and interactions is identified, enabling transparent predictions for theoretical observables while facilitating comparisons between the analyses in different experimental settings. 

In this work, we adopt the simplified model paradigm, generically assuming the existence of a DM particle that interacts minimally with the SM through a limited number of couplings and additional new physics states. More specifically, we consider a broad class of simplified models that is represented by $t$-channel dark matter models \cite{Arina:2020udz, Arina:2025zpi}. In contrast to $s$-channel models, both the actual dark matter particle and the mediator are odd under a stabilizing $\mathcal{Z}_2$ symmetry. Hence, the mediator couples to the dark matter particle and one SM particle.

These models were initially introduced for a fermion dark matter candidate, Majorana or Dirac, which couples either to SM quarks via a coloured scalar mediator \cite{Garny:2011ii, Garny:2013ama, Bai:2013iqa, DiFranzo:2013vra} or to leptons \cite{Dreiner:2012xm, Freitas:2014jla, Kopp:2014tsa} via electroweakly charged mediators. The systematization of different DM spin choices and couplings to chiral SM fermions, either quarks or leptons, was put forward in \cite{Baker:2015qna, Arina:2020udz}.\footnote{We stick to the common choice of coupling either to a quark or to a lepton, not to both fermion types at the same time. A relevant exception can be found in ref.~\cite{Arcadi:2021glq}, where $t$-channel DM models where exploited to accommodate the $B$-anomalies. In this case, the DM and mediator fields were ``integrated out'' and appeared in the matching coefficients of the dimension-six operators that induce new-physics contributions in semileptonic decays of $B$ mesons.}  In their minimal incarnation, the dark matter particle can be a fermion, a scalar, or a vector, while the mediator, either a scalar or a fermion. These models generally exhibit a rich phenomenology at colliders, as well as in direct and indirect detection \cite{Garny:2011ii,Garny:2013ama,Bai:2013iqa,DiFranzo:2013vra,Dreiner:2012xm,Freitas:2014jla,Garny:2014waa,Kopp:2014tsa,Baker:2015qna,Mohan:2019zrk,Arina:2020udz,Arina:2023msd,Arina:2025zpi}. They have also proven useful as a framework to explore various non-trivial aspects of dark matter production in the early universe. For thermal dark matter, the coannihilation channels of the mediator are also relevant in the case of small mass splittings with the dark matter particle \cite{Griest:1990kh,Edsjo:1997bg}. Here, non-relativistic annihilation of the mediator can be susceptible to sizable near-threshold effects, most notably Sommerfeld enhancement \cite{Hisano:2004ds,Iengo:2009ni,Cassel:2009wt} and bound-state effects \cite{Feng:2009mn,vonHarling:2014kha}. This is especially relevant whenever the mediator experiences strong interactions \cite{Ellis:2014ipa,Harz:2014gaa,Garny:2018ali,Biondini:2018ovz,Harz:2018csl,Biondini:2019int,Biondini:2018pwp,Garny:2021qsr}. In cases where feeble couplings between the dark matter and SM fields are considered, a systematic exploration of freeze-in production, including thermal masses and multiple soft scatterings, has recently been put forward \cite{Biondini:2020ric,Becker:2023vwd,Biondini:2023hek,Becker:2025lkc}. In the case of dark matter coupling to leptons through a weakly charged mediator, the effects on strengthening the character of the SM electroweak crossover were studied in refs.~\cite{Liu:2021mhn,Biondini:2022ggt}. 

In this work, we are especially interested in the implications of new fields, the dark matter and the mediator, being odd under the $\mathcal{Z}_2$ symmetry. The contributions of the heavy new-physics states to SMEFT operators are then one-loop induced \cite{Cepedello:2022pyx,Cepedello:2023yao,Gargalionis:2024jaw,Guedes:2024vuf} with the relative suppression. Hence, the constraints from a diverse and rich set of low-energy observables are weakened, i.e.~smaller new-physics scales are allowed. A further relaxation of the stringent limits on the Wilson coefficients of SMEFT is due to the additional assumption of Minimal Flavor Violation (MFV) \cite{Gerard:1982mm,Chivukula:1987py,Hall:1990ac,DAmbrosio:2002vsn}, which is usually adopted in phenomenological analyses of $t$-channel mediators.\footnote{Low-energy
probes and the LEP data provide quite stringent constraints on a
number of SMEFT operators \cite{Falkowski:2015krw,Falkowski:2017pss,Falkowski:2020pma}, which push the scale of new physics into the multi-TeV region if the Wilson coefficients are of order one, which is typical for contributions from tree-level processes. These bounds are pushed even to hundreds of TeV in case of lepton-flavour violating observables \cite{Crivellin:2017rmk} for tree-level generated Wilson coefficients.} Our main focus is to elucidate to what extent the constraints from SMEFT operators (namely, their matching coefficients) may exclude portions of the parameter space of $t$-channel models, and to examine how these compare with bounds from direct and collider searches. 
We consider \emph{leptophilic} $t$-channel models and aim to perform a matching of the renormalizable DM models onto SMEFT operators. In doing so, we can explore the bounds arising from global fits to \emph{all} matching coefficients. We consider fermionic Majorana and Dirac dark matter with a scalar mediator, which couple to either SM right-handed or left-handed leptons. 
 
The choice of leptophilic models is motivated by the following reasoning. First, we note that large DM masses further reduce the numerical values of the one-loop generated SMEFT matching coefficients, which scale parametrically as $C_i/(16 \pi^2 m_{\rmii{DM}}^2)$, thereby weakening their potential to probe the parameter space of these models. The mass of DM coupling to leptons is only mildly constrained by collider searches, $m_{\rmi{DM}} \lesssim 200$ GeV \cite{ATLAS:2019lff,ATLAS:2019lng,Arina:2025zpi,Barman:2021hhg}, due to the small cross section of electroweak-driven processes. Direct detection requires one-loop processes to induce a momentum transfer to nuclear target \cite{Kopp:2014tsa}. The corresponding limit for fermionic dark matter are quite different whether we consider the Dirac or Majorana option and strongly depend on the mass splitting between the dark matter and the mediator. In contrast, the corresponding exclusion limits for DM coupling to quarks (i.e., quark-philic models) are much more stringent, with the DM mass pushed to values of $\mathcal{O}(10)$ TeV  from collider and direct detection searches, see e.g.~\cite{Biondini:2018ovz,Arina:2020tuw,Becker:2022iso,Arina:2023msd,Arina:2025zpi}.
Second, the bounds on four-lepton dimension-six operators ($C_{\ell \ell}$, $C_{\ell e}$, and $C_{ee}$) are extracted from lepton-lepton scattering in the LEP data sets , which involve energies \emph{at most} $\sqrt{s}_{\rmi{LEP}} = 209$ GeV \cite{Electroweak:2003ram,ParticleDataGroup:2016lqr,VENUS:1997cjg,ALEPH:2013dgf}. 
Therefore, for dark matter masses comparatively larger than typical LEP energies, but still below $\mathcal{O}(10)$ TeV, one can resort to a SMEFT approach without suppressing too much the Wilson coefficients of the four-lepton operators.

The approach we undertake in this paper is similar to the earlier study presented in ref.~\cite{Freitas:2014jla}, where LEP data were used to constrain $s$-channel and $t$-channel dark matter models that produce effective four-lepton operators at tree and one-loop level, respectively. Here, the interplay with the cosmologically viable parameter space for the DM candidates was not considered, an aspect we aim to improve upon in our analysis. Our work is also complementary to more recent studies that focus on the identification of renormalizable new-physics models contributing to dimension-six four-fermion SMEFT operators and containing a viable dark matter candidate \cite{Cepedello:2022pyx,Cepedello:2023yao}. Also in this case~\cite{Cepedello:2023yao}, the bounds from low-energy observables, particularly the four-lepton operator $C_{\ell \ell}$, and the DM relic abundance were considered for an exemplary model realization: a SM extension with a fermion singlet and an inert doublet \cite{Ma:2006km,Herrero-Garcia:2018koq}.  In our work, we aim to provide a broader assessment of the potential of SMEFT operators to set bounds on $t$-channel models, without restricting exclusively on fully leptonic ones. Moreover, we include a consistent treatment of coannihilation for a compressed spectrum for the masses of the BSM fields, which appears to be missing in earlier works. We highlight that, whenever a single-scale matching is pursued, which implies that the DM and the mediator are close in mass, the effects of coannihilation must be included in the estimation of the dark matter energy density.

Finally, in a very recent analysis \cite{Kraml:2025fpv}, which extends the studies in refs~\cite{Huo:2015nka,Wells:2017vla}, the complete one-loop matching to SMEFT dimension-six operators has been carried out for the Minimal Supersymmetric Standard Model, also in this case by assuming a single mass scale for all the heavy super-partners. Other relevant examples of one-loop matching of BSM theories to dimension-six SMEFT operators, also detached from the dark matter problem, can be found in refs.~\cite{Fuentes-Martin:2016uol,Brivio:2021alv,Jiang:2018pbd,Gherardi:2020det,Ellis:2020ivx,Haisch:2020ahr}.  


The structure of the paper is as follows. In section~\ref{sec:model} we write the Lagrangians of the two classes of leptophilic $t$-channel models that we consider in our study. Section~\ref{sec:oneLOOP_match} is devoted to the one-loop matching of the models onto SMEFT operators, in particular in the Warsaw basis. The dark matter relic density is discussed in section \ref{sec:DM_energy_density}, whereas in section~\ref{sec:interplay} we present the interplay between the constraints from SMEFT global fits, direct searches and the cosmologically viable parameter space. Conclusions are summarized in section~\ref{sec:conclusions}. The appendices comprise the complete list of the Wilson coefficients  and further technical details of our work.

\section{$t$-channel model with fermionic dark matter}
\label{sec:model}
In this paper, we focus on a class of $t$-channel models where the dark matter interacts with SM leptons. This option is less studied than the case of interactions with quarks, even though a rich phenomenology arises at colliders \cite{Horigome:2021qof,Arina:2025zpi} and in direct detection experiments via the generation of anapole and magnetic-dipole moments \cite{Kopp:2014tsa}. Moreover, these models feature a quite interesting signature for indirect detection, where virtual internal bremsstrahlung \cite{Bringmann:2007nk,Bell:2010ei}  provides a sharp gamma ray pick (especially for a compressed mass spectrum of the DM and the mediator).

As required for a consistent application of the SMEFT framework, we assume the typical values of the dark matter particle mass to lie above the electroweak scale; more specifically, we take $m_{\rmii{DM}} \geq 500$ GeV.\footnote{In other works, see e.g.~\cite{Freitas:2014jla} and \cite{Cepedello:2023yao}, the dark matter/dark sector particle masses are taken to be as small as 200 GeV and 100 GeV, respectively. We avoid such low-mass limits and keep the new-physics masses above the electroweak scale, $m_{\rmii{DM}} \gtrsim 2 v_h$, where $v_h=246$ GeV is the SM Higgs vacuum expectation value.}
In the following, we explicitly write the models for the case where the dark matter is a Majorana fermion. However, we also provide the results for the relevant low-energy effective operators, as well the extraction of the relic density, for the Dirac DM case.

We consider $t$-channel leptophilic dark matter models that extend the SM with a gauge-singlet Majorana fermion~$(\chi)$ and a complex scalar field~$(\eta)$, which is in turn a singlet under ${\rm SU(3)}$. The scalar mediates the interaction between the dark fermion and the SM leptonic species,  the right-handed or left-handed leptons, which in turn define the corresponding quantum numbers of $\eta$. Since the complex scalar $\eta$ is at least charged under the ${\rm U(1)}\rmii{Y}$ gauge group and interacts with photons, it does not qualify as a DM candidate.
Assuming the Majorana fermion is lighter than the accompanying scalar state, $m\chi < m_\eta$, and imposing a stabilizing $\mathcal{Z}_2$ symmetry, the Majorana fermion is stable.\footnote{As usual, the SM particles are even under the symmetry, whereas the dark states, $\eta$ and $\chi$, are odd. Because of the $\mathcal{Z}_2$ symmetry, there is no Yukawa-like interaction of the DM fermion with the SM Higgs doublet and lepton doublet.}  For future use, we define the relative mass splitting between the DM and the mediator as $\delta m \equiv \Delta m /m_\chi$ with $\Delta m = m_\eta-m_\chi > 0$. 
Generically, the corresponding Lagrangian takes the form
\begin{align} 
\label{eq:lag_mod1}
\mathcal{L}_{\rmi{BSM}} & =
    \mathcal{L}_{\rmii{SM}}
  + \mathcal{L}_{\eta}
  + \mathcal{L}_{\chi}
  + \mathcal{L}_{\rmi{portal}}
  \;,
\end{align}
with the dark Majorana fermion and complex scalar terms given by
\begin{align} 
\mathcal{L}_{\chi} &=
    \frac{1}{2} \bar{\chi} (i \msl{\partial} - m_{\chi} ) \chi
    \;,\\
\mathcal{L}_{\eta} &=
    (D_\mu \eta)^\dagger (D_\mu \eta)
  - m^2_{\eta 0} \eta^\dagger\eta
  - \lambda_{2} (\eta^\dagger\eta)^2
     \, .
\end{align}
The covariant derivative can be specified only upon the assignment of the interacting lepton (cfr.~eqs.~\eqref{eq:L:covderivative:scalar_singlet} and \eqref{eq:L:covderivative:scalar_doublet}), and $\lambda_2$ is the $\eta$ scalar self-coupling ($\lambda_{1}$ is reserved for the SM Higgs doublet self-coupling). We also indicate the bare scalar mass $m_{\eta 0}$ in the Lagrangian and identify the physical mass, which is the one relevant in the experiments, as the one comprising the contribution from the electroweak symmetry breaking, cfr.~eq.~\eqref{eq:phys_eta_mass}. The portal Lagrangian $\mathcal{L}_{\rmi{portal}}$ comprises the interactions between the SM degrees of freedom $\chi$ and $\eta$, cfr.~eqs.~\eqref{eq:L:portal:yukawa_singlet}, \eqref{eq:L:portal:yukawa_doublet} and \eqref{eq:L:portal:scalar_doublet}.  

The two options we consider in this work are as follows, where we name the models in a similar fashion to ref.~\cite{Arina:2020udz}.\footnote{The naming indicates the type of mediator (here a scalar, S), the SU(2)$_\rmii{L}$ representation of the mediator (respectively singlet and doublet: 1 and 2), and the dark matter Majorana fermion (M). The corresponding Dirac fermion models would be labeled S1D and S2D.}
\subsubsection*{a) S1M model}
The first model comprises a Majorana dark matter particle that interacts with right-handed leptons, here denoted by $e^p$, where $p$ is the flavor index. The scalar field $\eta$ also carries a flavor index, while the DM field $\chi$ does not~\cite{Kopp:2014tsa,Garny:2015wea}.
In our analysis, we assume a MFV framework with a global symmetry $\mathcal{G}_{\text{flavor}} = U(3)^5$ \cite{Gerard:1982mm,Chivukula:1987py,Hall:1990ac,DAmbrosio:2002vsn}, which is broken only by the SM Yukawa interactions.
The BSM scalar particle sits then in the same flavor representation as the right-handed leptons and the scalar mediator comes in three identical copies $\eta^p$, one for each flavor
(more precisely $\eta^p$ is an anti-triplet of the flavor symmetry $U_e(3)$).

The portal Lagrangians and the covariant derivative read \cite{Kopp:2014tsa,Garny:2015wea}
\begin{eqnarray} 
\label{eq:L:portal:yukawa_singlet}
&& \mathcal{L}_{\rmi{Yukawa}} =
    -y\,\eta \, \bar{\chi} P_{\rmii{R}}\, e
  -y\, \bar{e}\, P_{\rmii{L}} \chi\, \eta^{\dagger}\, , \quad 
\mathcal{L}_{\rmi{scalar}} =
  -\lambda_{3} \, \eta^{\dagger} \eta H^\dagger H
  \;,
  \\ 
  && D_\mu \eta =
(\partial_\mu - i\go Y_\eta B_\mu)\eta \, ,
\label{eq:L:covderivative:scalar_singlet}
\end{eqnarray}
where $\go$ is the ${\rm U(1)}_\rmii{Y}$ gauge coupling.  A sum over the flavor index is understood when writing each term of the Lagrangian, e.g.~$y\,\eta^p \, \bar{\chi} P_{\rmii{R}}\, e^p$ for the Yukawa interaction. Here, the complex scalar has the following quantum number assignment $(\bm{1},\bm{1},-Y_e)$ of the SM gauge group SU(3)$\times$SU(2)$_\rmii{L} \times$U(1)$_\rmii{Y}$, where $Y_e=-1$.  The same model setting where all three lepton flavors couple to the dark matter fermion, in a flavor-diagonal way, was also considered in ref.~\cite{Kopp:2014tsa}. This is the model version that we implement in \texttt{Matchete} and for the matching onto SMEFT operators. The Yukawa couplings are taken as real. In so doing, no additional source of CP violation with respect to the SM is induced. This choice also reduces the number of electromagnetic moments at low-energy that are relevant for direct detection for Dirac dark matter~\cite{Kopp:2014tsa,Ibarra:2024mpq}. More specifically, no electric dipole moment is induced with real couplings (see ref.~\cite{Ibarra:2024mpq} for a more general treatment of complex couplings and interference among electromagnetic moments).

The option where dark matter particle only couples to one generation
of leptons at a time is considered for example in ref.~\cite{Garny:2015wea}, where the underlying symmetry would then be $U(1)_e \times U(1)_\mu \times U(1)_\tau$. In this case, one would need to use global fits that select a particular lepton flavor, see e.g.~\cite{Faroughy:2020ina,Greljo:2022cah,Greljo:2023adz}. 

\subsubsection*{b) S2M model}
The second option stems for Majorana dark matter that interacts with left-handed leptons, which we denote with $\ell^p=(\nu^p,l^p)^T$.  
Flavor considerations and coupling structures are the same as those of S1M, while in this case the scalar doublet is an anti-triplet of the $U_\ell(3)$ symmetry. The portal Lagrangian and the covariant derivative read~\cite{Garny:2015wea,Arina:2020tuw}
\begin{eqnarray} 
\label{eq:L:portal:yukawa_doublet}
&& \mathcal{L}_{\rmi{Yukawa}} =
    -y\, \bar{\chi} P_{\rmii{L}}\, (\ell \, i \sigma_2 \eta) 
  -  y  \, (\eta^\dagger \, i \sigma_2 \bar{\ell})\, P_{\rmii{R}} \chi\, , 
  \\
&& \mathcal{L}_{\rmi{scalar}} =
  -\lambda_{3} (\eta^\dagger\eta) (H^\dagger H) - \lambda_4 (\eta^\dagger H) (H^\dagger \eta)  - \frac{\lambda_5}{2} \left[ (H^\dagger \eta)^2 + (\eta^\dagger H)^2 \right] 
  \;,
  \label{eq:L:portal:scalar_doublet}
  \\ 
\label{eq:L:covderivative:scalar_doublet}
  && D_\mu \eta =
(\partial_\mu - i\go Y_\eta B_\mu -i \gt W_{\mu}^a \tau^a )\eta \, ,
\end{eqnarray}
where $\gt$ is the ${\rm{SU(2)}}_\rmii{L}$ gauge coupling, $W^{a}_{\mu}$ the corresponding gauge bosons, $\tau^a=\sigma^a/2$ are the generators and $\sigma^a$ the Pauli matrices. We use round brackets to signal the SU(2)$_{\rmii{L}}$ structures in the Lagrangian without explicitly using indices. Here, the complex scalar has the following quantum number assignment $(\bm{1},\bm{2},-Y_\ell)$, where $Y_\ell=-1/2$; $\eta$ has the same gauge charges as the SM Higgs boson. Similarly to the inert doublet dark matter model \cite{Deshpande:1977rw,Barbieri:2006dq,LopezHonorez:2006gr}, the SU(2)$_{\rmii{L}}$ doublet $\eta$ is odd under a $\mathcal{Z}_2$ symmetry. Hence, the dark-sector scalar does not participate to the electroweak symmetry breaking
directly, it does not acquire its own vacuum expectation values and does not mix with the Higgs boson. It appears that the last term in eq.~\eqref{eq:L:portal:scalar_doublet} was not included in the general set of scalar interactions in ref.~\cite{Garny:2015wea}.

The simplified models find a natural realization in supersymmetric extension of the SM. There, the mediator $\eta$ is the lightest slepton and the corresponding DM candidate is the lightest neutralino. In the case of no mixing between neutralinos, the yukawa coupling is fixed in terms of the SM gauge coupling $g_1$ and approximate values are $y = \sqrt{2} g_1 \simeq 0.48$ and $y = g_1/\sqrt{2} \simeq 0.24$ for the coupling to right-handed and left-handed leptons respectively \cite{Garny:2015wea} (for general neutralino mixing see discussions in refs.~\cite{Bringmann:2012vr,Garny:2015wea,Kopp:2014tsa}). In our work we will treat the coupling $y$ as a free parameter and include the effects of the scalar interactions, at least to some extent. More specifically we will consider $\lambda_3 \sim \mathcal{O}(1)$, while assuming $\lambda_4, \lambda_5 \ll 1$. In this way we can compare the two models with the same set of parameters $(m_\chi, m_\eta, y, \lambda_3)$ and carry out scans with a smaller number of unknowns. Despite scalar interactions for the scalar mediator $\eta$ are often discarded to simplify the treatment of the parameter space, they can induce important effects when dealing with a compressed mass spectrum. With our choice, the physical mass of the scalar $\eta$, which is accessible experimentally, is for both models
\begin{eqnarray}
    m^2_\eta \equiv   m_{\eta  0}^2 + \frac{\lambda_3}{2} v_{h}^2 \, ,
    \label{eq:phys_eta_mass}
\end{eqnarray}
where $v_h=246$ GeV is the SM Higgs vacuum expectation value and $m_{\eta0}$ is the mass parameter that enters the Lagrangian.\footnote{As shown in Loryons’ work (e.g.\ \cite{Banta_2022,Cohen:2020xca}), if the mediator’s mass is predominantly generated by electroweak symmetry breaking, one should use HEFT rather than SMEFT. For the masses and couplings that we use in our work, we fulfill the SMEFT validity criterion.}

\section{One-loop matching to SMEFT}
\label{sec:oneLOOP_match}
In this section, we address the matching of the leptophilic simplified models described in section~\ref{sec:model} to the operator content of SMEFT, focusing specifically on the dimension-six operators. 
The DM and mediator field excitations are assumed to be heavy compared to the electroweak scale and are therefore treated as non-dynamical. In EFT terminology, they are said to be integrated out. Under our working hypothesis, we treat the two masses as defining a single high-energy scale, $\Lambda \equiv m_{\chi} \simeq m_{\eta}$, meaning that the dark matter and mediator fields are integrated out simultaneously. We implement this requirement by imposing a relative mass splitting smaller than unity, taking $\delta m \leq 0.5$ (the value of the smallest splitting, $\delta m = 0.005$, was formerly used in the phenomenological analyses of these simplified models, see e.g.~\cite{Kopp:2014tsa, Garny:2015wea}).
This procedure allows us to exploit the powerful framework of SMEFT in combination with global fits and the associated bounds on operators.\footnote{An alternative and more general EFT at the electroweak scale is provided by the Higgs Effective Field Theory (HEFT)~\cite{Brivio_2016, FERUGLIO_1993}. However, global fits are typically performed within the context of SMEFT~\cite{deblas2024globalsmeftfitsfuture}.}

Consistent with the assumption of MFV for the models under consideration, which correspond to a $U(3)^5$ flavor symmetry in the low-energy theory, we use the global fit results for dimension-six operators as presented in ref.~\cite{Bartocci:2024fmm}. The result of the global fit is given in the Warsaw basis~\cite{Grzadkowski_2010} and, under the MFV assumption, the number of SMEFT operators reduces to 47 (41 CP-even and 7 CP-odd).\footnote{Typically, the cross-talk between CP-even and CP-odd operators is quite small~\cite{Ethier:2021ydt}, since dedicated observables are required to constrain them.} The global fit is further restricted to the CP-even operators and, most importantly, renormalisation group evolution (RGE) of the Wilson coefficients is included, together with prediction of relevant observables at NLO~\cite{Bartocci:2024fmm}. Our choice for the global fit with MFV is grounded in a first systematic exploration of $t$-channel models and their SMEFT operators, without delving in (many) possibilities of flavor structures in this work.  We leave this aspect to be the focus for future research on the subject. 

We perform the matching from the $t$-channel DM models to SMEFT using the automated package \texttt{Matchete}~\cite{Fuentes-Martin:2022jrf}, which is based on functional methods~\cite{Fuentes-Martin:2016uol, Fuentes-Martin:2020udw, Cohen:2020fcu}. The models under consideration induce one-loop matching coefficients due to the $\mathcal{Z}_2$ symmetry, ensuring the stability of the DM candidates. To organize the results in the Warsaw basis, the direct matching of diagrams from a given UV model may need to be supplemented by the use of the SM equations of motion (EOM)~\cite{Jenkins:2013zja}. We summarize the main additional steps required to map to the Warsaw basis for the S1M model in the following section.
Even though we rely on the automation of the matching to SMEFT, we have checked intermediate steps analytically and confirmed agreement with the output of \texttt{Matchete}. Lepton- and baryon-number-violating operators, which are generically allowed in the Warsaw basis, are not generated in our models. CP-odd operators are also not induced. The resulting SMEFT operators generated at dimension six for the leptophilic models considered in this work are: 24 operators for S1M and S1D and 31 operators for S2M and S2D respectively, see appendix~\ref{Appendix_A_S1M} and \ref{Appendix_A_S2M}.   
In truncating the EFT expansion at dimension six and one-loop order, the theoretical uncertainties from neglected contributions are parametrically suppressed by $g^2_{\rmii{SM},\rmii{BSM}}/(4 \pi)^2$ for two-loop operators at dimension six, and by $v_h^2/\Lambda^2$ for one-loop generated dimension-eight operators.
\subsection{Steps to the Warsaw basis and four-lepton operators}
\label{sec:details_Warsaw}
The Lagrangian obtained by integrating out the heavy fields $\chi$ and $\eta$ with \texttt{Matchete} contains terms that are not directly written in the Warsaw basis.  In the following, we express the dimension-six operators by taking the mass of the scalar mediator as the expansion parameter $v_h/m_\eta$, because it is the (slightly) larger mass scale of the BSM models (though in the compressed mass spectrum, both $m_\chi$ and $m_\chi$ can be taken as indicative for the high-energy scale). For the leptophilic $t$-channel models with Majorana fermions, additional structures appear in the intermediate steps due to a richer set of diagrams involving more fermionic field contractions. An excerpt of the Lagrangian for the S1M model reads:
\begin{eqnarray}
    \label{eq:eff_L}
&&\mathcal{L}_{\rmi{SMEFT,S1M}} \supset - \frac{g_{1}^2}{60(4\pi)^2m_{\eta}^2} D^{\mu} B_{\rho \mu}D_{\sigma} B^{\rho \sigma} + \frac{g_{1} y^2} {60(4\pi)^2m_{\eta}^2} \mathcal{F}_2 \left(\frac{m_{\chi}}{m_{\eta}} \right)D^{\mu}B_{\nu \mu} (\bar{e}^p \, \gamma^{\nu}P_{R} \,  e^p) 
    \nonumber
    \\
   &&~~~~~~~~~~~~~~~~+  \frac{y^4}{2 (4\pi)^2 m_{\eta}^2}   \mathcal{F}_{3} \left( \frac{m_\chi}{m_\eta} \right)  \left( \bar{e}^{p} C P_{\rmii{L}} \bar{e}^{s T} \right) \left( e^{s T} C P_{\rmii{R}} e^{p}\right) \, ,
\end{eqnarray}
where the loop functions $\mathcal{F}_2(x)$ and $\mathcal{F}_3(x)$ are defined in eqs.~\eqref{f2_loop} and \eqref{f3_loop}, and $p$ and $s$ are flavor indices.  
The first term is typical of the electroweak Euler-Heisenberg Lagrangian~\cite{Manohar:1997qy}, while the second is sometimes referred to as a gauge current operator \cite{Grzadkowski_2010}.

The first two terms are not in the Warsaw basis, and the use of the SM EOMs for the gauge field $B_\mu$ (see e.g.~\cite{Jenkins:2013zja}) 
\begin{align}
\label{eq:SM_EOM}
    & D^{\mu} B_{\mu \nu}=g_1 \sum_{\psi=u,d,q,e,\ell} \bar{\psi} y_{i} \gamma_{\nu} \psi+ \frac{g_1}{2} H^\dagger \overset{\leftrightarrow}{D}_{\nu} H \,, 
\end{align}
in eq.~\eqref{eq:eff_L} induces many operators in the Warsaw basis, more specifically for our case
\begin{equation}
    \mathcal{O}_{\ell e}, \mathcal{O}_{ue}, \mathcal{O}_{de}, \mathcal{O}_{qe}, \mathcal{O}_{\ell \ell}, \mathcal{O}_{\ell q}, \mathcal{O}_{qq}, \mathcal{O}_{uu}, \mathcal{O}_{dd} \, .
\end{equation}
Moreover, in the second line of eq.~\eqref{eq:eff_L}, an additional contribution appears that is also not in the Warsaw basis. One can verify that this spinor structure arises from Majorana-like contractions of the dark fermion $\chi$ in one-loop box diagrams with external right-handed lepton fields (see figure~\ref{fig:box_Maj_Dir} in Appendix~\ref{appendiix_A}). Upon using the Fierz identities outlined in refs.~\cite{Nishi:2004st, Isidori:2023pyp} for Dirac matrices in $D=4$, one can show\footnote{The derivation of this identity in $D=4$ is safe because we do not find any $1/\varepsilon$ poles in the one-loop diagrams generating $\mathcal{O}_{ee}$. Hence, we do not need to worry about evanescent contributions, which in general could combine with $1/\varepsilon$ and give a finite term~\cite{Fuentes-Martin:2022vvu}. The only divergences generated at one loop in all models appear in the kinetic terms of the gauge field $B_{\mu}$ and the right- and left-handed leptons, and are removed in the $\overline{\textrm{MS}}$ scheme.}
\begin{eqnarray}
   (\Bar{e}^p  \, C P_{\rmii{L}} \, \bar{e}^{s T} ) \,  ( e^{s T} \, C P_{\rmii{R}} \, e^p )  = \frac{1}{2}  ( \bar{e}^p  \, \gamma^\mu P_{\rmii{R}} e^p) \, (\bar{e}^s \, \gamma_\mu P_{\rmii{R}} \, e^s)  \, ,
\end{eqnarray}
for generic flavor indices. To the best of our knowledge, this identity was not explicitly derived in previous studies, and we provide a detailed derivation in appendix~\ref{sec::Fierz_4f}.  
For the S2M model, similar procedures are followed, although the SU(2)$_{\rmii{L}}$-charged scalar $\eta$ introduces additional gauge structures, requiring Fierz identities in the internal SU(2)$_{\rmii{L}}$ space as well.

\begin{figure}
    \centering
    \includegraphics[width=0.5\linewidth]{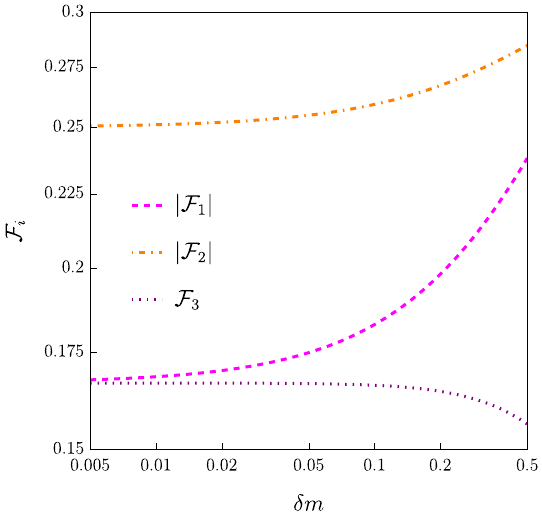}
    \caption{Loop function in eqs.~\eqref{f1_loop}-\eqref{f3_loop} expressed in term of the relative splitting according to the relation $m_\chi/m_\eta=(1+\delta m)^{-1}$. For negative loop functions the absolute value is displayed.}
    \label{fig:loop_func_deltam}
\end{figure}

In our study, a relevant class of operators are fully leptonic four-fermion operators, which exhibit a rich structure in terms of loop functions and SM/BSM couplings.  For the S1M model, they read\footnote{In ref.~\cite{Freitas:2014jla} the one-loop matching for the model S1M and S1D was performed, however without considering a complete map to the Warsaw basis. Contributions proportional to $g_1^2$ were not included. For the terms proportional to $y^4$, that were obtained in ref.~\cite{Freitas:2014jla}, we find a factor of 2 difference between eqs.~\eqref{f1_loop} and \eqref{f3_loop} with their loop functions.}
\begin{eqnarray}
&&  C_{ee}  \mathcal{O}_{e \, e} = \frac{1}{(4\pi)^2 m_{\eta}^2}\left[ - \frac{g_{1}^4}{20} + \frac{y^4}{4} \left( \mathcal{F}_{1} \left( \frac{m_\chi}{m_\eta} \right) + \mathcal{F}_{3} \left( \frac{m_\chi}{m_\eta} \right) \right)
\right. 
\nonumber \\ 
&&\left. 
~~~~~~~~~~~~~~~~~~~~~~~~~~~~~~~~~~- \frac{g_{1}^2 y^2}{6}  \mathcal{F}_{2} \left( \frac{m_\chi}{m_\eta} \right) \right]  \left( \bar{e}^{p}\gamma_{\mu} P_{\rmii{R}} e^{p} \right) \left( \bar{e}^{s} \gamma^{\mu} P_{\rmii{R}} e^{s} \right) \, ,
\label{O_ee_Majorana}
    \\
&& C_{\ell \, e}    \mathcal{O}_{\ell \, e} = - \frac{g_1^2}{60(4\pi)^2 \, m_\eta^2} \left[ g_1^2   + 5 y^2 \mathcal{F}_{2}\left( \frac{m_\chi}{m_\eta} \right) \right]   (\bar{e}^p \gamma_\mu  P_{\rmii{R}}  e^p )\, ( \bar{\ell}^s_i \gamma^\mu  P_{\rmii{L}} \ell_i^s) \, ,
\label{O_le}
    \\
&& C_{\ell \ell}    \, \mathcal{O}_{\ell \, \ell} = -\frac{g_1^4}{80(4\pi)^2 \, m_\eta^2} \,   (\bar{\ell}_i^p \gamma_\mu  P_{\rmii{L}}  \ell_i^p )\, ( \bar{\ell}_j^s \gamma^\mu  P_{\rmii{L}} \ell_j^s) \, ,
\label{O_ll}
\end{eqnarray}
where $p,s$ are flavor indices and $i, j$ are SU(2)$_{\rmii{L}}$ indices. $C'_{\ell \ell}$ is not generated. 
The loop functions $\mathcal{F}_{1}(x)$, $\mathcal{F}_{2}(x)$, and $\mathcal{F}_{3}(x)$ depend on the mass ratio of the BSM particles and are plotted in figure~\ref{fig:loop_func_deltam}.
These are extracted as combinations of internal loop functions from \texttt{Matchete}, which we have independently verified. They read 
\begin{eqnarray}
   && \mathcal{F}_1(x) = -\frac{1 - x^4 + 2 x^2 \ln x^2}{2(1 - x^2)^3} \, ,
     \label{f1_loop}
   \\
  && \mathcal{F}_2(x)  = - \frac{2 - 9x^2 + 18 x^4 - 11 x^6  + 6 x^6 \ln x^2 }{6(1 - x^2)^4} \, ,
   \label{f2_loop}
   \\ 
   && \mathcal{F}_3(x) = - x^2 \, \frac{2 - 2x^2 + (1 + x^2) \ln x^2}{(1 - x^2)^3} \, ,
    \label{f3_loop}
\end{eqnarray}
with $x = m_\chi / m_\eta$. For the S1D model, the only difference lies in $C_{ee}\mathcal{O}_{ee}$ in eq.~\eqref{O_ee_Majorana}, where the contribution from $\mathcal{F}_3(x)$ is absent; this is due to fewer diagrams compared to the Majorana case. The full set of Wilson coefficients generated at one loop is listed in appendix~\ref{Appendix_A_S1M} for the SU(2)$_{\rmii{L}}$ singlet scalar mediator models, and in appendix~\ref{Appendix_A_S2M} for the SU(2)$_{\rmii{L}}$ scalar doublet case.

We note in passing that the mass ratio of the dark matter and mediator can be expressed in terms of the relative mass splitting, so that the argument of the loop functions $\mathcal{F}_i(x)$ becomes $x = 1 / (1 + \delta m)$. For small relative splittings, the leading-order expansions of the loop functions are:
\begin{eqnarray}
    \mathcal{F}_1(\delta m) \simeq -\frac{1 + \delta m}{6} \, , \quad  \mathcal{F}_2(\delta m) \simeq -\frac{1}{4} - \frac{\delta m}{10} \, , \quad \mathcal{F}_3(\delta m) \simeq  \frac{1}{6} \, .
    \label{loop_fun_expanded}
\end{eqnarray}
These expansions are useful for analyzing the interplay with the dark matter relic density and when comparing the Majorana and Dirac dark matter options in section~\ref{sec:interplay}.

\subsection{Constraints on the WCs from experiments}
\label{subsec:constraints_on_WC}.
Let us now discuss how to use the constraints on the Wilson coefficients that arise from global fits. More specifically, we adopt the analysis of SMEFT operators as performed in \cite{Bartocci:2023nvp, Bartocci:2024fmm}. The limits on the Wilson coefficients are given as usual for a fixed UV scale, and with a 2$\sigma$ confidence. We comment on the general validity of applying global fit results to the model parameter space, with particular attention to the mass range considered, $m_\eta > m_\chi \geq 0.5$~TeV. 

In general, different classes of Wilson coefficients are constrained by various sets of experimental data, each associated with one or more characteristic energy ranges. In the case of purely leptonic four-fermion operators, constraints arise from \emph{lepton scattering} processes, with energies ranging from a fraction of a GeV up to the LEP collision energies (the maximal energy of collected data being 209 GeV \cite{Electroweak:2003ram}). The global fit that we use relies on muon neutrino-electron scattering, the weak mixing angle measured in parity-violating electron scattering~\cite{ParticleDataGroup:2016lqr}, $\tau$ polarisation measured in $e^- e^+ \to \tau^- \tau^+$~\cite{VENUS:1997cjg}, as well as differential cross sections and asymmetries in $e^- e^+ \to l^- l^+$~\cite{ALEPH:2013dgf, Electroweak:2003ram}. In other words, for a new physics scale larger than $0.5$~TeV, one can reasonably apply the bounds from the global fit to constrain the leptonic operators in the $t$-channel models.

One has to be more cautious when considering other operators generated in the models under study. We have scanned over all operators involving the dark matter and scalar mediator masses, $y$ and $\lambda_3$, and we find that $C_{ed}$ and $C_H$ impose constraints on the model parameter spaces, such as $(m_\chi, y)$ and $(m_\chi, \lambda_3)$ respectively, when naively applying the bounds from the global fit in the region $m_\chi \geq 0.5$TeV. However, the data sets used to extract the bounds on these coefficients involve experiments at energies larger than 1 TeV. The Wilson coefficients of the Higgs sector
are constrained by LEP measurements and electroweak precision observables (EWPO), but the most stringent bound on $C_H$ arises from di-Higgs production, which probes the invariant mass range $ 0.5 \, \text{TeV} \lesssim m_{hh} \lesssim 1$ TeV \cite{ATLAS:2018dpp, ATLAS:2018rnh, ATLAS:2018uni, CMS:2017hea, CMS:2021ctt, CMS:2022cpr}. A similar limitation arises for $C_{ed}$, where both Parity Violation experiments (PVE) and Drell-Yan processes are used to obtain the bound of the global fit. Constraints
from Drell-Yan dominate over those from PVE throughout the whole parameter space, and are taken from LHC data up to 3 TeV \cite{CMS:2021ctt,ATLAS:2020zms}.  This, of course, invalidates the direct use of the corresponding global fit results for masses smaller than such energies. If one insists on using the bounds on $C_{ed}$ and $C_H$, the energy scale of the new physics, $m_\eta \approx m_\chi$ for our case, must be increased accordingly. The constraints on the models parameter space are given in section~\ref{sec:interplay}. We anticipate here that no significant exclusion is provided by the coefficient $C_H$, whereas $C_{ed}$ plays a role for dark matter masses larger than 3 TeV for the S1M and S1D models.

In the following, we list the coefficients that are relevant for the analysis and are included in the final results in section~\ref{sec:interplay}. These are $C_{ee}$ and $C_{ed}$ for the S1M and S1D models, and $C_{\ell \ell}$ for the S2M and S2D models, which are found in eqs.~\eqref{App_A_Cee_S1M} and \eqref{App_A_Cee_S1D},  \eqref{C_ED_S1M},  and \eqref{App_A_Cll_S2M} and \eqref{App_A_Cll_S2D} respectively.  We adopt the improved values that account for RGE and NLO effects in the extraction of the global fit~\cite{Bartocci:2023nvp, Bartocci:2024fmm}. For the leptonic coefficients $C_{ee}$ and $C_{\ell \ell}$, and the four-fermion operator involving right-handed leptons and $d$-quarks, the bounds read:
\begin{eqnarray}
\frac{C_{ee}(\Lambda)}{\Lambda^2} = \frac{[-0.14, 0.03]}{\textrm{TeV}^2}
\, ,
\quad \frac{C_{\ell \ell}(\Lambda)}{\Lambda^2} = \frac{[0.0, 0.21]}{\textrm{TeV}^2} \, , \quad \frac{C_{ed}(\Lambda)}{\Lambda^2} = \frac{[0.0, 1.2]}{\textrm{TeV}^2}
\, .
\label{bounds_relevant_WCs}
\end{eqnarray}
The coefficients from the global fit are given at a fixed matching scale, here $\Lambda = 4$ TeV \cite{Bartocci:2024fmm}. To apply these bounds at a different high-energy scale hypothesis, we evolve the matching coefficients using the RGE in SMEFT. This approach is more rigorous than simply rescaling the bounds in eq.~\eqref{bounds_relevant_WCs} as $C_i(m_\eta) = C_i(\Lambda) \times (m_\eta/\Lambda)^2$, which neglects potential operator mixing effects. 

For the new-physics scale $m_\eta$, we consider the range $m_\eta = m_{\chi}(1 + \delta m)$, with $m_ {\chi}\in [0.5,10]$ TeV. The evolution leads to the following expression:
\begin{equation}
\label{eq::RGE_WC}
C_{ee}(m_\eta) = \left( \frac{m_\eta}{\Lambda} \right)^2 U_{ee,j} C_j(\Lambda) \, ,
\end{equation}
for the coefficient $C_{ee}$, where $U$ denotes the evolution matrix. The sum over $j$ includes all relevant matching coefficients, as extracted from~\cite{Bartocci:2024fmm}.
Analogous expressions are used to evolve $C_{\ell \ell}$ and $C_{ed}$ from $\Lambda = 4$~TeV to the scale $m_\eta$.

The quantitative RGE evolution of the matching coefficients has been performed using \texttt{DsixTools}~\cite{Fuentes_Mart_n_2021} and cross-checked with \textit{RGESolver}~\cite{Di_Noi_2023}. We find that, for the fully leptonic operators, the running induces effects at the level of a few percent across the full range $m_\eta \in [0.5 , 10] (1+ \delta m)$~TeV. The largest correction is a $3\%$ shift for $C_{ee}$, while a slightly larger effect of about $5\%$ is observed for $C_{ed}$. Although small, these effects are included in the summary plots in section~\ref{sec:interplay}, and \emph{a posteriori}, one may safely adopt the simplified rescaling procedure for the matching coefficients.
In order to be consistent with eq.~\eqref{eq::RGE_WC}, we also consider the running of SM gauge couplings entering the analytic expressions of the Wilson coefficients. The corresponding beta function in SMEFT can be found in \cite{Jenkins:2013zja}. In doing so, the couplings $g_1$ and $g_2$ depend on the new-physics scale $\Lambda = m_\eta$. Also in this case, the running of the SM couplings is at per-cent level. 

\section{Dark matter energy density}
\label{sec:DM_energy_density}
In the matching of the simplified models to SMEFT, we have considered a single high-energy scale.  In terms of the relative mass splitting, this corresponds to the regime where $\delta m$ is taken smaller than unity.
From the point of view of dark matter freeze-out, this implies the need to account for the so-called coannihilation regime \cite{Griest:1990kh}, especially when the relative mass splitting becomes $\delta m \lesssim 0.2$ \cite{Griest:1990kh, Baker:2015qna}. In this situation, the scalar particle remains abundantly present in the thermal bath while the DM fermions are annihilating. For sufficiently large $y$, a thermal contact is established between the populations of dark fermions and scalars. As a result, one must include coannihilation processes involving one scalar and one dark fermion in the initial state, as well as scalar pair annihilations, in order to correctly estimate the freeze-out of the dark matter.

The effects of the coannihilating particle can still be captured by a single Boltzmann equation \cite{Griest:1990kh,Edsjo:1997bg}
\begin{equation}
    \dot{n} = -\langle \sigma_{\textrm{eff}} \, \vrel \rangle (n^2 -n_{\textrm{eq}}^2) \, ,
    \label{BE_start}
\end{equation}
where $n$ stands for the total number density of the dark sector, $n=n_\chi+ 6 n_\eta$. Here we used that the dark fermion is Majorana and that $n_\eta=n_{\eta^\dagger}$, with the multiplicity given by the three flavors with the same mass $m_\eta$. In the following equations we refer to the Majorana fermion option and we comment on explicit differences with the Dirac case when relevant throughout the paper.
The effective cross section is a combination of the various annihilation processes that involve the dark sector states, namely 
\begin{eqnarray}
   \sigma_{\textrm{eff}} \, \vrel   = \frac{1}{(\sum_k n_k^{\textrm{eq}})^2}\sum_{i,j} n_i^{\textrm{eq}}n_j^{\textrm{eq}}\sigma_{ij}  \vrel \, ,
\end{eqnarray}
where $i,j$ run over the dark sector particles. By employing a Maxwell-Boltzmann distribution for the non-relativistic particles, and retaining only the leading order term in the momentum over the mass, the number density of the Majorana fermion and complex scalar read
\begin{eqnarray}
    && n_\chi^{\textrm{eq}} \simeq g_\chi  \left( \frac{m_\chi T}{2 \pi} \right)^{\frac{3}{2}} e^{-m_\chi/T} \, , 
    \quad n_\eta^{\textrm{eq}} \simeq g_\eta \left( \frac{m_\chi T}{2 \pi} \right)^{\frac{3}{2}} e^{-m_\eta/T} \, .
    \label{eq_number_density_app}
\end{eqnarray}
The degrees of freedom are $g_\chi=2$ (spin polarizations) and $g_\eta=3$ (flavor).  
Accounting for all coannihialtions processes, we find the following expression for the effective cross section:
\begin{eqnarray}
    &&\sigma_{\textrm{eff}} \vrel = \left[ \sigma_{\chi \chi} \vrel +  \frac{4 g_\eta}{g_\chi} e^{-\delta m \, \frac{m_\chi}{T}}  \left( 1+\delta m \right)^{\frac{3}{2}} \sigma_{\chi \eta^\dagger} \vrel  \right.
    \nonumber
    \\
    &&\left. +  \frac{ 2 g_\eta^2}{g_\chi^2} e^{-2 \delta m \, \frac{m_\chi}{T}} \left( 1+\delta m \right)^{3} \left(  \sigma_{\eta \eta} \vrel + \sigma_{\eta \eta^\dagger} \vrel \right) \right] \frac{1}{\left[ 1+ 2 \frac{g_\eta}{g_\chi} e^{-\delta m \, \frac{m_\chi}{T}} \left( 1+\delta m \right)^{\frac{3}{2}} \right]^2} \, ,
    \label{eff_Xsection_coann}
\end{eqnarray}
where, in the second line of eq.~\eqref{eff_Xsection_coann}, we explicitly split the particle-antiparticle annihilation channel, $\eta \eta^{\dagger}$, from the particle-particle process, $\eta \eta$ and its conjugate (collectively denoted with $\sigma_{\eta \eta}$ for simplicity). More specifically they are $\eta \eta \to e e$ and its conjugate. 

The structure of the effective cross sections agrees with, e.g., ref.~\cite{Baker:2015qna}.\footnote{In the limit of large mass splittings, only the first term in eq.~\eqref{eff_Xsection_coann} survives, $n_\eta$ becomes exponentially suppressed and hence one recovers the single-specie Boltzmann equation for a Majorana fermion (self-conjugate particle), $\dot{n}_\chi = - \langle \sigma_{\chi \chi} \vrel \rangle (n_\chi^2-n_{\chi,\rmii{eq}}^2)$ \cite{Gondolo:1990dk}. The same applies to the Dirac case, where one finds $\dot{n}_\chi = - (1/2)\langle \sigma_{\chi \bar{\chi}} \vrel \rangle (n_\chi^2-n_{\chi,\rmii{eq}}^2)$ \cite{Gondolo:1990dk}} The exponential factors in the numerator of the cross section in eq.~\eqref{eff_Xsection_coann} clearly show how smaller relative splittings make the coannihialtion processes of the scalar increasingly more important around the freeze-out. The approximation for the equilibrium number densities in eq.~\eqref{eq_number_density_app} is particularly useful to elucidate this last point; however, in the numerical implementation of the Boltzmann equation, we employ number densities in terms of the Bessel function of the second kind.\footnote{The equilibrium number density, which better describes the non-relativistic situation for $1<m_\chi/T \lesssim 25$ reads $n_{\textrm{eq}} \simeq g_\chi m_\chi^2 T /(2 \pi^2) \mathcal{K}_2(m_\chi/T)$. The corresponding expression holds for the scalar $\eta$. The two approximations differ by a factor of about 10\% at $m_\chi/T =20$.}  

For the model S1M, the cross sections for the various annihilation channel has been calculated for all $2 \to 2$ processes in refs.~\cite{Garny:2015wea, Kopp:2014tsa, Biondini:2022ggt}. We take over such results with minor modifications due to the coupling of the dark matter to all lepton flavors. In order to give a minimal context, we recall that the dark matter fermion pair annihilation $\chi \chi \to e \bar{e}$ proceeds via a leading $p$-wave, hence it is velocity suppressed \cite{Kopp:2014tsa, Garny:2015wea}. The coannihilation processes are induced by SM U(1)$_Y$ gauge interactions, as well as by the BSM Yukawa portal interaction and the SM Higgs-$\eta$ portal coupling $\lambda_3$. We checked the corresponding cross section for the Dirac option, which brings in two main differences: the dark matter pair annihilation $\chi \bar{\chi} \to e \bar{e}$ features a leading $s$-wave contribution and the processes $\eta \eta \to e e $ and its conjugate are absent.
For future reference, we list the annihilation cross sections for the Majorana and Dirac fermion, S1M and S1D respectively, that read 
\begin{equation}
    \sigma \vrel (\chi \chi \to e \bar{e}) =
  \frac{|y|^4}{16 \pi}
  \frac{m_\chi^2 (m_\chi^4+m_\eta^4)}{(m_\chi^2+m_\eta^2)^4} \vrel^2 \, , \quad   \sigma \vrel (\chi \bar{\chi} \to e \bar{e}) =
  \frac{3|y|^4}{128 \pi}
 \frac{m_\chi^2}{(m_\chi^2 + m_\eta^2)^2} \, .
 \label{DM_ann_M_and_D}
\end{equation}

As far as the corresponding cross sections for the scalar-doublet option, S2M and S2D, we find that they are a factor of 2 larger than the corresponding ones in eq.~\eqref{DM_ann_M_and_D}, as due to the SU(2)$_{\rmii{L}}$ multiplicity (see also \cite{Garny:2015wea} for the Majorana option). We note that, since the scalar sector of the models S2M and S2D is equivalent to that of the inert Higgs doublet, $\eta^T=(\eta^+,\eta^0)$, we can use the known cross sections from the annihilation processes $\eta \eta^\dagger$ to SM gauge bosons, e.g.~refs.~\cite{Hambye:2009pw, Biondini:2017ufr}. We then have to compute only the coannihilation processes, such as $\chi \eta^- \to W^- \nu$, $\chi \eta^0 \to W^- l^+ $, $\eta \eta \to \ell \ell$, and their conjugates, which are absent in the inert doublet model (basically the contributions arising from the Yukawa portal $y$).

The Boltzmann equation~\eqref{BE_start} is as usual recast in terms of
the yield parameter $Y = n/s$, where
$s=2 \pi^2 h_{\rmi{eff}} \, T^3/45$ is the entropy density, and
the time evolution is traded for the variable $z = m_\chi/T$.
As for the temperature-dependent relativistic degrees of freedom
$h_{\rmi{eff}}$ entering the entropy density, we use the SM values from
ref.~\cite{Laine:2015kra}.\footnote{
  In the freeze-out case this is well justified since
  the states $\chi$ and $\eta$ are non-relativistic particles for
  the relevant temperature window.}
The relativistic degrees of freedom for the energy density $g_{\rmi{eff}}$, that  enter
the Hubble rate $H=\sqrt{8\pi e/3}\,m_{\rmii{Pl}}$, where
$e=\pi^2 T^4  g_{\rmi{eff}}/30$, are also taken from~\cite{Laine:2015kra}. 
\begin{figure}[t!]
    \centering
    \includegraphics[width=0.47\linewidth]{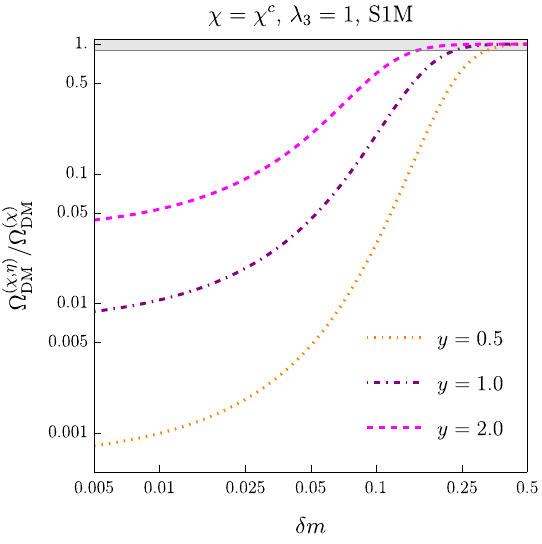}
     \includegraphics[width=0.47\linewidth]{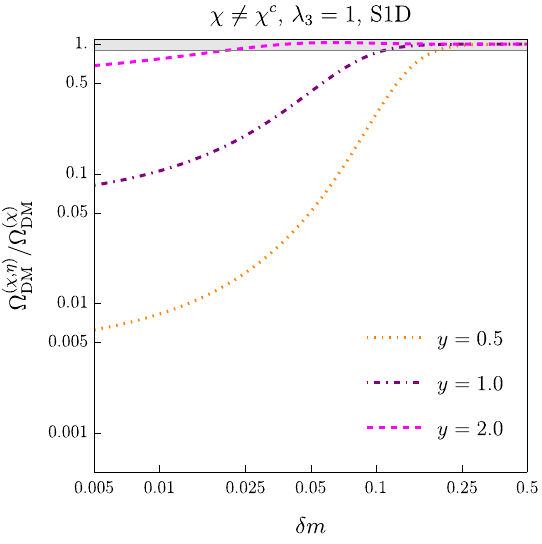}
    \caption{Ratio of the predicted energy density with and without coannihailtion processes. In the left (right) panel we show the numerical output for the Majorana (Dirac) dark matter fermion for different choices of the Yukawa coupling and $\lambda_3=1.0$.}
    \label{fig:co_annihilation_example}
\end{figure}

In figure~\ref{fig:co_annihilation_example}, we summarize the effect of coannihilations on the predicted energy density for different values of the portal coupling $y$ and for one illustrative choice of the scalar coupling,  $\lambda_3=1$. The left panel corresponds to the Majorana fermion case, whereas the Dirac case is shown in the right panel. The gray shaded area indicates where coannihilations produce a correction factor within 10\% to the predicted DM energy density. We display the effect of coannihilation by taking the ratio of the energy densities with and without coannihilation processes ($\Omega_{{\rmii{DM}}}^{(\chi,\eta)}$ and $\Omega_{{\rmii{DM}}}^{(\chi)}$ respectively). The curves do not depend on the dark matter mass, but they strongly depend on the choice of the couplings and $\delta m$. The larger the value of $y$ for a fixed $\lambda_3$, the smaller the effect of the coannihilations. This is due to the dependence of the cross section for the processes $\chi \chi \to e \bar{e}$ and $\chi \bar{\chi} \to e \bar{e}$, both of which scale solely with the Yukawa coupling as $y^4$. Coannihialtion processes involve instead all couplings, also $g_1$ and $\lambda_3$. For example, for the Majorana case, coannihilating scalars reduce the predicted DM energy density by a factor of about 20 for the smallest splitting considered in our work and for $y=2.0$. Larger values of $\lambda_3$ imply even stronger effects, since many scalar-pair annihilation cross sections scale as $\lambda_3^2$ (see e.g.~\cite{Biondini:2022ggt} for the list of cross sections). One may notice that the effects of the coannihilation for the same values of the couplings is less important for the Dirac option (right panel of figure~\ref{fig:co_annihilation_example}). This is due to larger DM pair annihilation cross section, which is not velocity suppressed as for the Majorana case, cfr.~eq.~\eqref{DM_ann_M_and_D}. For large relative mass splittings, where coannihilations get suppressed, the prediction of the relic density is the same and the ratio indeed goes to unity. 

\subsection{Thermal masses and near-threshold effects}
\label{sec:DM_energy_density_SOMBSF}
 There are two aspects that we consider in this work when dealing with coannihilation. First, we observe that a thermal contribution to the mass splitting is induced in this model, because the scalar particle interacts with SM degrees of freedom.\footnote{We do not consider the contribution from the Yukawa interaction involving the DM fermion and the leptons. In this case, the thermal corrections are exponentially suppressed in the non-relativistic limit, $T \ll m_\chi$, due to the DM particle running in the loop. Accordingly, we do not compute the thermal correction to the DM mass, which could arise only from the Yukawa-type interaction with the scalar $\eta$. In-vacuum loop contributions are understood to be included in the physical masses $m_\chi$ and $m_\eta$.} Since the effective cross section is exponentially sensitive to the mass splitting, it is worthwhile to include thermal corrections to the mass difference at freeze-out. These were included in ref.~\cite{Biondini:2022ggt} for the model with the SU(2)$_{\rmii{L}}$ singlet scalar, building on earlier studies of thermal dark matter coannihilation~\cite{Kim:2016kxt, Biondini:2017ufr}.
Second, the $\eta$ particles interact with lighter SM gauge and Higgs bosons ($m_\eta > m_\chi \geq 500$ GeV in this work), which may induce repeated soft exchanges between the heavy, slowly-moving particles. As a result, pair annihilations of the type $\eta \eta^\dagger$ are affected by Sommerfeld enhancement~\cite{Hisano:2004ds,Iengo:2009ni,Feng:2010zp} and the possible formation of bound states~\cite{Detmold:2014qqa,vonHarling:2014kha,Petraki:2015hla}. The latter provides an additional channel for depleting the heavy scalars, which in turn affects the DM abundance for small mass splittings. For Majorana dark matter, scalar pair annihilations of the type $\eta \eta$ are suppressed due to a repulsive potential in the case of vector boson exchange. Bound states cannot form in this case, and the same holds for the charge-conjugate configuration $\eta^\dagger \eta^\dagger$.

The thermally corrected scalar mass can be inferred from earlier works. Here, we explicitly list the expression for the models S1M and S1D, which reads, in the non-relativistic regime relevant for freeze-out~\cite{Kim:2016kxt,Biondini:2017ufr,Biondini:2022ggt}:
\begin{eqnarray}
    &&M_{\eta} \simeq m_\eta + \frac{\lambda_3}{m_\eta} \int_{\vec{p}}\frac{ \nB(E_{p,h})}{E_{p,h}} + \frac{\go^2 Y_\eta^2}{ m_\eta} \int_{\vec{p}} \left(
      \frac{\tilde{c}^2 \, \nB(E_{p,\gamma})}{E_{p,\gamma}} 
    + \frac{\tilde{s}^2 \, \nB(E_{p,\rmii{$Z$}})}{E_{p,\rmii{$Z$}}}
    \right)
  \nonumber
  \\
  &&~~~~~~~~~~~~~-\frac{\go^2 Y_\eta^2}{8 \pi} \left(
        \tilde{c}^2 M_{\gamma}
      + \tilde{s}^2 M_{\rmii{$Z$}}
      - s^2 m_{\rmii{Z}}
    \right) 
    \, . 
    \label{thermal_mass_eta_singlet}
\end{eqnarray}
where
$E_{p,i} = \sqrt{\vec{p}^{2} + M_{i}^2}$. We denote thermal masses with $M_i$, while in-vacuum masses are denoted with $m_i$.
The thermal contributions are computed in the SM electroweak broken phase and smoothly approach those of the symmetric phase, where the Higgs mechanism ``melts away,'' i.e., $v_h \to 0$.
The term in the second line of eq.~\eqref{thermal_mass_eta_singlet} arises from the contribution of screened soft gauge bosons at the scale $gT$ (also known as the Salpeter correction, see e.g.~\cite{Brown:1996qs}). Thermal corrections to the mass splittings have a quite moderate impact on the relic density, within a few percent for large values of $\lambda_3 \gtrsim 2$. 

The relevant thermal masses $M_i$, the in-vacuum weak mixing angles $c$ and $s$, and their thermal counterparts $\tilde{c}$ and $\tilde{s}$ can be found in refs.~\cite{Kim:2016kxt,Biondini:2017ufr}.
From eq.~\eqref{thermal_mass_eta_singlet}, we can define a temperature-dependent mass splitting to be used in the numerical extraction of the dark matter energy density.
The corresponding thermal mass for the SU(2)$_{\rmii{L}}$ scalar, namely in the S2M and S2D models, includes an additional contribution from the charged $W$ boson.
In this case as well, the relevant thermal masses are given in refs.~\cite{Kim:2016kxt,Biondini:2017ufr}.

Let us elaborate on the second aspect, which has been previously scrutinized for the model S1M in ref.~\cite{Biondini:2022ggt}. The heavy scalar particle $\eta$ interacts with the U(1)${\rmii{Y}}$ gauge boson $B_\mu$ and the Higgs boson. The former interaction is more conveniently described in the SM broken phase, in terms of the mass-diagonal fields, namely the $Z$ boson and the photon. Since pair annihilations occur in the non-relativistic regime, the BSM scalars move with small velocities of order $v \sim \sqrt{T/m_\eta}$. In this setting, repeated soft exchanges of the force carriers, SM gauge bosons and the Higgs boson, can affect the annihilation cross section of various heavy-scalar pair combinations. The Sommerfeld enhancement (for pairs in an attractive channel) and bound-state effects have the main phenomenological consequence of boosting pair annihilations at small velocities, allowing for larger viable dark matter masses~\cite{vonHarling:2014kha, Petraki:2015hla}.

The relevant potentials are generally Yukawa-like, with SM gauge and Higgs boson masses that depend on temperature, scaling parametrically as $M^2_i \sim g^2 v_h^2 + g^2 T^2$ (for the photon one does not get the Higgs contribution). Electroweak symmetry is restored around $T_c \simeq 160$ GeV \cite{DOnofrio:2015gop}, and we follow the implementation of a temperature-dependent Higgs vacuum expectation values as outlined in refs.~\cite{Kim:2016kxt, Biondini:2017ufr,Biondini:2025ihi}. For temperatures above $T_c$, the tri-linear vertex responsible for Higgs-mediated soft exchange vanishes, as $v_h \vert_{T \geq T_c} = 0$, and the gauge bosons only acquire thermal masses of order $g T$, which are parametrically much smaller than $m_\chi$ in the non-relativistic regime. A dark matter/scalar mass $m_\chi \simeq m_\eta \gtrsim 4$ TeV is an indicative benchmark distinguishing between these two situations.

Previous analyses~\cite{Mitridate:2017izz, Biondini:2017ufr, Bollig:2021psb, Biondini:2022ggt} have shown that mild Sommerfeld enhancements are induced by photon and $Z$-boson exchange, while the Higgs-induced potential yields rather small effects for this class of models (typically at the percent level). The impact of Sommerfeld enhancement on the predicted dark matter abundance ranges from 15\% to 5\%, depending on the relative splitting, $0.005 \leq \delta m \leq 0.2$ (we also consider larger splittings; however, the effects of scalar coannihilations are negligible due to the exponentially suppressed scalar population at freeze-out and later stages, cf.~eq.~\eqref{eff_Xsection_coann}).

As for bound-state effects, their impact is even milder, for reasons that we briefly summarize as follows. In the broken phase, the $Z$ and $W$ bosons become massive due to the Higgs mechanism, and the condition for the existence of a ground state in a Yukawa-like potential requires $1/m_{\rmi{boson}} \gtrsim 0.84 \, a_0$~\cite{Rogers:1970xx} (here $m_{\rmi{boson}}$ is one of the SM gauge boson or Higgs boson). The Bohr radius $a_0$ scales inversely with the coupling responsible for the binding and the constituent particle mass, here $m_\eta$, namely $a_0 \simeq 2/(\alpha m_\eta)$, with $\alpha = g^2/(4\pi)$. Here $g$ stands for a generic electroweak coupling and is $g_1 c_w$ for the $Z$ boson and $g_2$ for the $W$ boson respectively. For the massive gauge bosons, one finds that bound states only form if $m_\eta \gtrsim 15$~TeV and  $m_\eta \gtrsim 4$~TeV respectively for $Z$ and $W$ force mediator. For the photon such a constraint does not hold. 
As discussed earlier, for $m_\eta \gtrsim 4$TeV, we transition to the symmetric phase, where the potentials become essentially Coulomb-like. Even in this regime, bound-state effects remain quite moderate due to the typical strength of the electroweak couplings $\mathcal{O}(\alpha) \approx 10^{-2}$ for all gauge bosons. We find effects that range from few-percent to about 10\% on the dark matter yield, in agreement with the analysis in refs.~\cite{vonHarling:2014kha} and confirmed by subsequent studies~\cite{Petraki:2015hla, Kim:2016kxt, Biondini:2017ufr, Bollig:2021psb}.\footnote{The larger effects of order 10\% are mainly due to large values of $\lambda_3$, which increases the hard annihilation multiplying the bound-state formation enhancement factor, see e.g.~\cite{Biondini:2017ufr, Biondini:2019int}.}

\section{Interplay of experimental constraints and observed DM energy density}
\label{sec:interplay}
In this section, we summarize the constraints on the model parameter space imposed by global fits on the SMEFT Wilson coefficients, the observed dark matter relic abundance $\Omega_{\rmii{DM}}h^2 = 0.1200 \pm 0.0012$ \cite{Planck:2018vyg}, and direct detection limits. The Wilson coefficients are taken from the global fit in ref.~\cite{Bartocci:2024fmm}. Regarding indirect detection, the sensitivity to dark matter masses larger than $0.5$ TeV is quite limited~\cite{Kopp:2014tsa, Garny:2015wea}, and we do not include it in the following analysis. The same applies to collider searches, where the latest ATLAS analyses constrain dark matter masses only below 200 GeV \cite{ATLAS:2019lff, ATLAS:2019lng, Arina:2025zpi}, whereas we are interested in larger masses. 

Interestingly, as pointed out in ref.~\cite{Kopp:2014tsa}, this class of leptophilic models can be probed by direct detection searches of dark matter scattering off nuclear targets. Although the dark matter fermion does not couple directly to quarks, one-loop processes induce an effective coupling between the dark matter and the photon. This, in turn, generates effective low-energy interactions with nuclear constituents in the form of electromagnetic moments~\cite{Fukushima:2013efa,Ibarra:2022nzm}. A detailed study of the emergence of electromagnetic moments, with an application to $t$-channel models, was recently presented in ref.~\cite{Ibarra:2024mpq}.
The situation is rather different for Dirac and Majorana dark matter. In the first case,  electric and magnetic dipole moments, an anapole moment and charge radius may in general appear. In the case of Majorana dark matter, only the anapole moment is induced~\cite{Radescu:1985wf, Kayser:1983wm}.
For the models of this study, we directly adopt the expressions for electromagnetic moments as computed in ref.~\cite{Kopp:2014tsa,Ibarra:2024mpq}.\footnote{It is worth noting that the scalar coupling $\lambda_3$ may induce an effective coupling to the Higgs boson. This effect was not included in the calculations of refs.~\cite{Kopp:2014tsa,Ibarra:2024mpq}. Moreover, if the dark matter electromagnetic moments are generated via loops involving electrons, the full momentum dependence should be taken into account. This has only been computed for the Majorana anapole moment, see ref.~\cite{Kopp:2014tsa}. Consequently, in our figures, the direct detection limits for the Dirac case include only the muon and tau contributions. Overall, the direct detection bounds presented in this study are therefore conservative. A complete treatment including the electron contribution lies beyond the scope of this work.
} We use the bounds on the electromagnetic moments as extracted in ref.~\cite{Ibarra:2024mpq} and we only present the most constraining experimental limit in the mass range of interest in the following section. More specifically, we consider the LUX-ZEPLIN (LZ)
experiment \cite{LZ:2022lsv}.  

The leptophilic models that we study in this work contribute to the anomalous magnetic moment of the SM leptons. Relevant constraints may arise from the electron \cite{Parker_2018} and muon $g-2$ \cite{Muong-2:2025xyk}. We have checked that they do not induce bounds on the parameter space of the models for $m_\chi > 500$ GeV.\footnote{As for the $g-2$ of the muon, the models induce a negative shift, at variance with the SM prediction that is (under debate) slightly larger than the experimental value. Hence, the negative contribution of the $t$-channel models can then be treated as a fluctuation  within 5$\sigma$ of the measurement, see also ref.~\cite{Bringmann:2012vr}. From the latest measurement \cite{Muong-2:2025xyk}, the combined uncertainty is $1 \sigma \simeq 1.48 \times 10^{-10}$.}


\subsection{S1M and S1D parameter space}
\label{sec:interplay_plots_S1}
Since we do not obtain meaningful constraints from the SMEFT Wilson coefficients involving the scalar coupling $\lambda_3$, cfr.~eqs.~\eqref{C_HB_S1M} and \eqref{C_H_S1M} for $C_{HB}$, $C_{H}$, $C_{H \Box}$, we focus on the parameter space $(m_\chi, y)$. This also facilitates comparison with previous studies that present results in a similar fashion, e.g.~ref.~\cite{Cepedello:2023yao}. In order to maximize the impact of the Wilson coefficient $C_{ee}$ on the parameter space, we consider large values of $y$, specifically up to the perturbative limit $y_{\rmi{max}} = \sqrt{4 \pi}$. Such large portal couplings are commonly used in dark matter phenomenology studies; see, e.g.~\cite{Kopp:2014tsa, Arina:2020tuw, Arina:2023msd, Cepedello:2023yao, Arina:2025zpi}. In addition to $m_\chi$ and $y$, there are two more free parameters: the relative mass splitting $\delta m$ and the scalar coupling $\lambda_3$. We fix $\delta m$ and $\lambda_3$ to specific values while scanning over $m_\chi$ and $y$. For the self-scalar coupling, we take values in the range $\lambda_3 \in [0, \pi]$. Such large values are also commonly used for DM and/or accompanying scalar states and are in any case below the corresponding perturbative value of $4 \pi$. 
\begin{figure}[t!]
    \begin{minipage}{0.445\linewidth}
      \includegraphics[width=\linewidth]{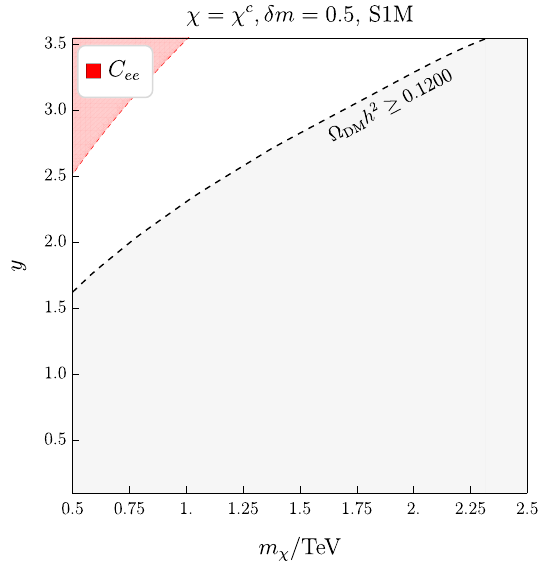}
       \label{fig::summary_plots_deltam_05_M}
    \end{minipage}
    \begin{minipage}{0.45\linewidth}
      \includegraphics[width=\linewidth]{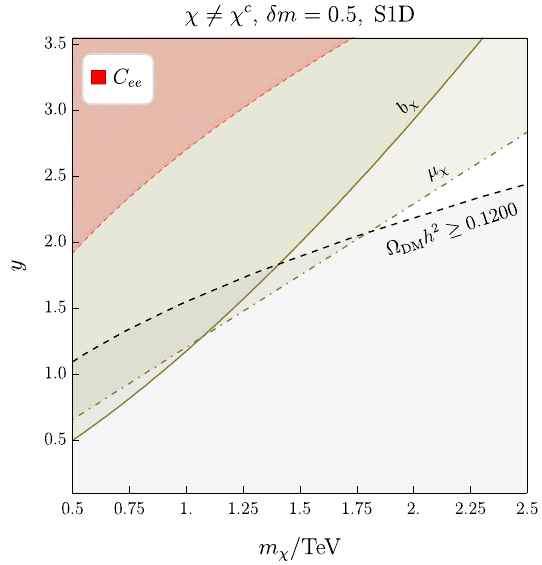}
       \label{fig::summary_plots_deltam_05_D}
    \end{minipage}
  
  \vspace{0.2pt}

      \begin{minipage}{0.44\linewidth}
      \includegraphics[width=\linewidth]{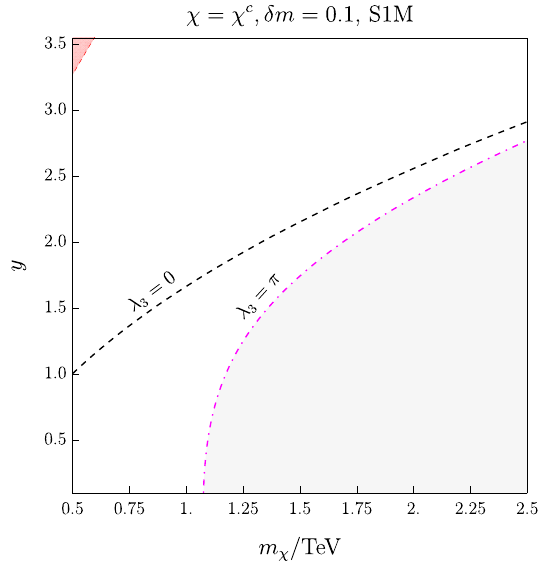}
       \label{fig::summary_plots_deltam_01_M}
    \end{minipage}
    \begin{minipage}{0.45\linewidth}
      \includegraphics[width=\linewidth]{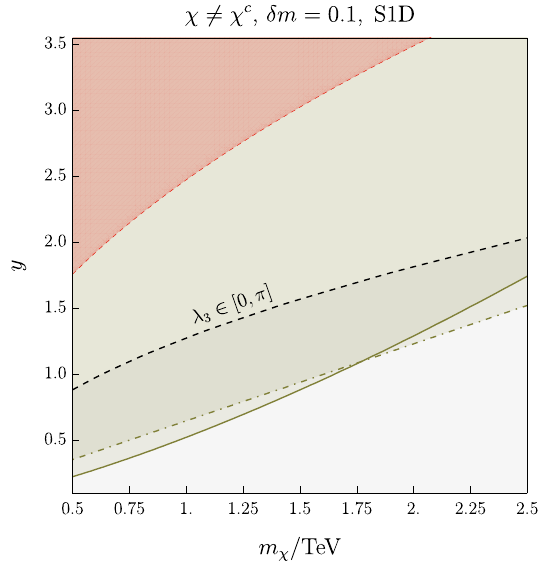}
       \label{fig::summary_plots_deltam_01_D}
    \end{minipage}
    \caption{ Model parameter space in the $(m_\chi, y)$ plane. Constraints from the global fit on $C_{ee}$ are shown as red-shaded regions. Areas excluded by overproduction of dark matter are indicated in gray shading, corresponding to the largest value of $\lambda_3$ used in each plot. The direct detection bound is displayed with olive-green shading: the dot-dashed line comes from the magnetic dipole moment ($\mu_{\chi}$) while continuous line stands for charge radius ($b_{\chi}$). Upper panel: Majorana (left) and Dirac (right) DM fermion for $\delta m = 0.5$. Lower panel: Majorana (left) and Dirac (right) DM fermion for $\delta m = 0.1$.
}
    \label{fig:summary_s1M_y_vs_mchi_largeDM}
\end{figure}

\begin{figure}[t!]
    \begin{minipage}{0.445\linewidth}
      \includegraphics[width=\linewidth]{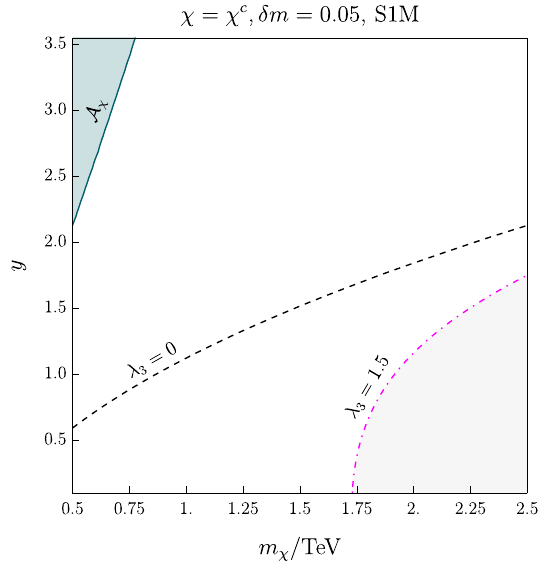}
       \label{fig::summary_plots_deltam_005_M}
    \end{minipage}
    \begin{minipage}{0.45\linewidth}
      \includegraphics[width=\linewidth]{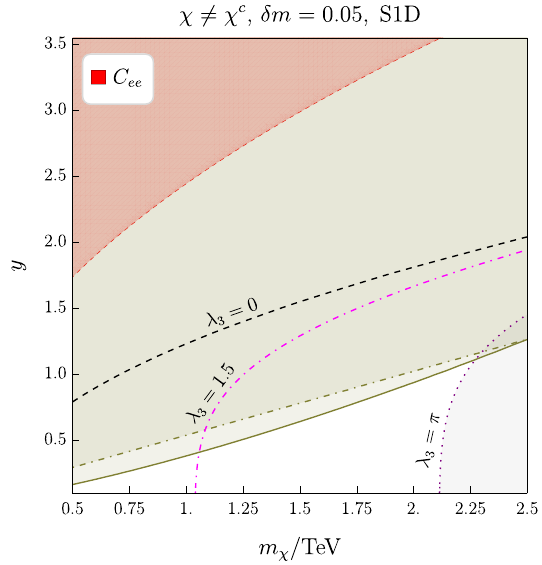}
       \label{fig::summary_plots_deltam_005_D}
    \end{minipage}
  
  \vspace{0.2pt}

      \begin{minipage}{0.44\linewidth}
      \includegraphics[width=\linewidth]{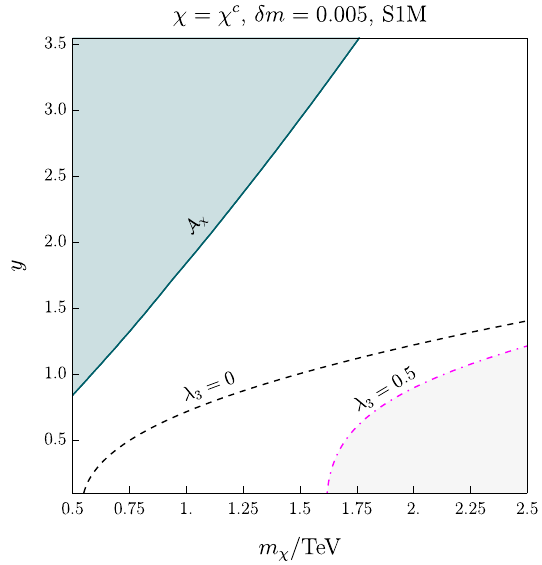}
       \label{fig::summary_plots_deltam_0005_M}
    \end{minipage}
    \begin{minipage}{0.45\linewidth}
      \includegraphics[width=\linewidth]{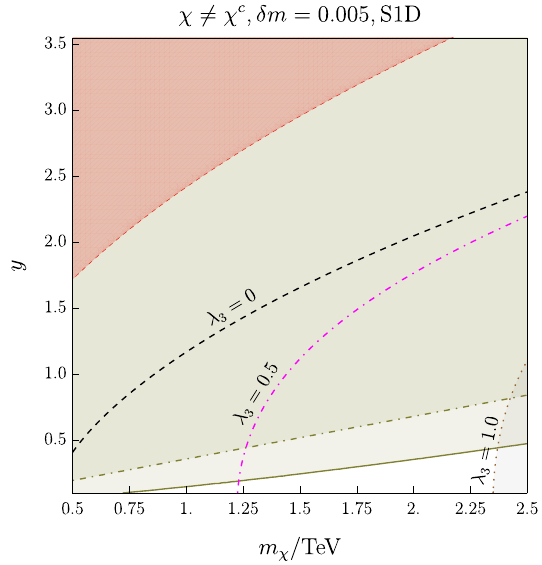}
       \label{fig::summary_plots_deltam_0005_D}
    \end{minipage}
    \caption{Model parameter space in the $(m_\chi, y)$ plane. Same color code and shading conventions as in the previous figure~\ref{fig:summary_s1M_y_vs_mchi_largeDM}. Exclusion limits from the anapole moment ($\mathcal{A}_\chi$) with petrol-blue shaded area.  Upper panel: Majorana (left) and Dirac (right) DM fermion for $\delta m = 0.05$. Lower panel: Majorana (left) and Dirac (right) DM fermion for $\delta m = 0.005$.
}
    \label{fig:summary_s1M_y_vs_mchi_smallDM}
\end{figure}
In the summary figures~\ref{fig:summary_s1M_y_vs_mchi_largeDM} and~\ref{fig:summary_s1M_y_vs_mchi_smallDM}, we collect the constraints imposed by the one-loop generated Wilson coefficient $C_{ee}$ (red-shaded regions), the observed relic density (gray-shaded regions, corresponding to $\Omega_{\rmii{DM}}h^2 > 0.1200$), and direct detection limits. Regarding the latter, the bounds on the magnetic dipole moment ($\mu_\chi$) and charge radius ($b_\chi$), relevant for the Dirac DM scenario, are shown as olive-green shaded regions, while the anapole moment ($\mathcal{A}_\chi$) constraint is indicated with petrol-blue shading.

Let us now discuss the bounds from the global fit. In general, the red-shaded regions cover the upper part of the $(m_\chi, y)$ plane, corresponding to large $y$ and relatively small dark matter masses. This can be readily understood from the scaling $C_{ee} \sim y^4/m_\eta^2$ (cf.~eq.~\eqref{O_ee_Majorana}); we recall that in the compressed spectrum regime, $m_\eta$ and $m_\chi$ are approximately interchangeable. The impact of $C_{ee}$ is found to differ significantly between the two DM cases. For the S1D model, the bounds extend down to $y \simeq 1.7$ and up to DM masses $m_\chi \simeq 2.1$~TeV for the smallest splitting considered in our study, $\delta m = 0.005$. Moreover, the exclusion region grows and saturates as $\delta m$ decreases. In contrast, much weaker constraints are found for the Majorana case: the bounds from $C_{ee}$ shrink rapidly with decreasing $\delta m$ and eventually disappear for $\delta m = 0.05$ and $\delta m = 0.005$ (see the small triangle-shaped region in the lower-left panel of figure~\ref{fig:summary_s1M_y_vs_mchi_largeDM}, which stands for $\delta m =0.1$). This behavior can be understood by inspecting more carefully the structure of $C_{ee}$ in eq.~\eqref{O_ee_Majorana}, where the relevant combination of loop functions is $\mathcal{F}_1(\delta m) + \mathcal{F}_3(\delta m)$. As shown in figure~\ref{fig:loop_func_deltam} and in the expansions in eq.~\eqref{loop_fun_expanded}, this combination vanishes for small splittings, which suppresses the $y^4$ contribution to $C_{ee}$. For the Dirac case, there is no $\mathcal{F}_3(\delta m)$ contribution, so the matching coefficient remains larger and tends to a constant as $\delta m \to 0$.

Despite the relatively good coverage of parameter space by $C_{ee}$ in the S1D model, we find that it is \emph{not} competitive with the direct detection limits, which always dominate (see olive-green shaded regions in figures~\ref{fig:summary_s1M_y_vs_mchi_largeDM} and~\ref{fig:summary_s1M_y_vs_mchi_smallDM}). A quite different situation is found for the model S1M, where the direct detection limit is absent for large splittings (see figure~\ref{fig:summary_s1M_y_vs_mchi_largeDM}), whereas it appears at lower splittings. The bounds from $C_{ee}$ are then complementary in the large-$y$, low-$m_\chi$ region to the direct detection constraints in the Majorana DM scenario.

When turning to the interplay with the DM energy density, we observe again marked differences between the two models. First, let us clarify that, whenever multiple relic density curves appear in a plot, corresponding to different $\lambda_3$ values, we shade in gray only the region below the rightmost curve (see the lower-left panel in figure~\ref{fig:summary_s1M_y_vs_mchi_largeDM} and figure~\ref{fig:summary_s1M_y_vs_mchi_smallDM}). This convention helps to identify the still-viable regions, i.e.~the white areas in the plots. At large splittings, dark matter pair annihilations dominate the freeze-out, and coannihilation effects are negligible. Pair annihilations are suppressed in the case of Majorana DM compared to the Dirac case (see eq.~\eqref{DM_ann_M_and_D}); therefore, larger values of $y$ are needed to achieve the observed relic abundance $\Omega_{\rmii{DM}} h^2 = 0.1200$. 
Moreover, coannihilation effects become prominent more quickly in the S1M model compared to S1D, as seen from the black dashed and magenta dot-dashed lines, which collapse into a single line in the lower-right plot in figure~\ref{fig:summary_s1M_y_vs_mchi_largeDM}. This indicates negligible net effects from coannihilation processes, which are sensitive to $\lambda_3$. The impact of coannihilations becomes stronger as the splitting decreases, and is reflected in the increasing separation between the relic density curves for different values of $\lambda_3$ (see figure~\ref{fig:summary_s1M_y_vs_mchi_smallDM}).

In summary, we find that the parameter space for the S1M model is generally less constrained by the SMEFT global fit (i.e.~$C_{ee}$) and by direct detection bounds. For Majorana dark matter, wide portions of the parameter space remain viable. In the Dirac scenario, we find a viable region for the large splitting $\delta m = 0.5$ at masses $m_\chi \gtrsim 1.8$~TeV, while this region closes for smaller splittings, as seen in the $\delta m = 0.1$ panel. As the dark matter and mediator masses become more compressed, the only viable regions lie at small values of $y$ and larger dark matter masses, which are necessary to evade direct detection limits. Accordingly, these corners of parameter space are compatible with the observed relic density only if coannihilation effects provide a significant contribution, which in turn requires $\lambda_3 \approx \mathcal{O}(1)$.

\begin{figure}[t!]
     \begin{minipage}{0.45\linewidth}
      \includegraphics[width=\linewidth]{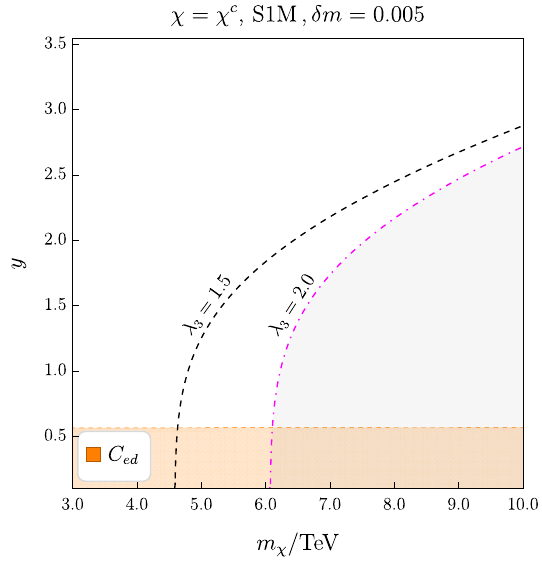}
       \label{fig::summary_plots_deltam_0005_maj_highmass}
    \end{minipage}
    \begin{minipage}{0.45\linewidth}
      \includegraphics[width=\linewidth]{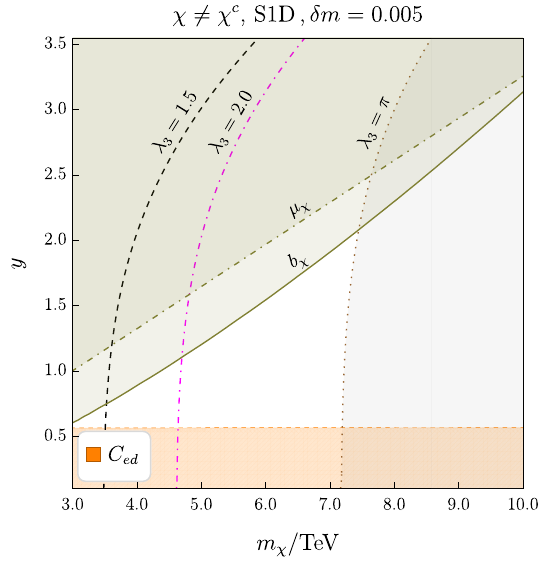}
       \label{fig::summary_plots_deltam_0005_Dirac_highmass}
    \end{minipage}    
    \caption{High-mass region for S1M and S1D models at $\delta m=0.005$, respectively left and right panel. In addition to  the SMEFT global fit exclusion from $C_{ed}$, we show the cosmological constraint for the relic density as well as direct detection bounds (for the Dirac DM scenario).}
    \label{fig:high_mass_S1M_S1D}
\end{figure}

In figure~\ref{fig:high_mass_S1M_S1D}, we show two plots where we include the bounds on the model parameter space $(m_\chi, y)$ arising from the lepton-quark four-fermion Wilson coefficient $C_{ed}$. 
Its expression reads
\begin{eqnarray}
    C_{ed} = -\frac{g_1^2}{90 (4 \pi)^2 m_\eta^2} \left[ 3 g_1^2 + 5 y^2 \mathcal{F}_2 \left( \frac{m_\chi}{m_\eta} \right) \right] \, .
\end{eqnarray}
As discussed in section~\ref{sec:oneLOOP_match}, larger DM masses are needed to apply this bound consistently (Drell-Yan data at $\mathcal{O}(1)$ TeV). We therefore consider the range $3\, \text{TeV} \leq m_\chi \leq 10\, \text{TeV}$, where $C_{ee}$ plays no significant role, and we fix $\delta m$ to the smallest value, consistent with the requirement of viable relic density at large masses. The bounds from $C_{ed}$ are shown as orange-shaded regions. Unlike $C_{ee}$, the constraint arises from requiring $C_{ed}$ to be positive (see eq.~\eqref{bounds_relevant_WCs}; values of $y \lesssim 0.56$ make the Wilson coefficient negative and induce the corresponding excluded region in the plots, i.e.~$0.1 \leq y \leq 0.56$. The dependence of this constraint on the DM mass is very mild (about $1\%$), while the dependence on $\delta m$ is more pronounced, varying between $3\%$ and $10\%$. For the Dirac case, the direct detection bound is still present at such large masses and significantly constrains the parameter space. For the same values of the scalar portal coupling $\lambda_3$, the DM energy density imposes stronger bounds for the Dirac case than for the Majorana option. 

\subsection{S2M and S2D parameter space}
\label{sec:interplay_plots_S2}
For models with an SU(2)$_{\rmii{L}}$-charged scalar mediator, we find meaningful constraints from the fully leptonic operator $C_{\ell \ell}$. They read, for the Majorana and Dirac models, as follows 
\begin{align}
   &   C^{\rmii{S2M}}_{\ell \ell}   = \frac{1}{4(4\pi)^2 m_{\eta}^2}\left\lbrace \frac{(g_{2}^2 -g_{1}^2)}{120} \left[ 3g_{1}^2 +3g_{2}^2 +20 y^2 \mathcal{F}_{2}\left(\frac{m_{\chi}}{m_{\eta}}\right) \right] \right. \nonumber
   \\
  & \left. ~~~~~~~~~~~~~~~~~~~~~~~~~~~~~~~~~~~~~~~~~~~~~~~~~~~+ y^4 
 \left[ \mathcal{F}_{1}\left(\frac{m_{\chi}}{m_{\eta}}\right) -\mathcal{F}_{3}\left( \frac{m_{\chi}}{m_{\eta}}\right) \right] \right\rbrace \,,
 \label{App_A_Cll_S2M}
 \\
   &   C^{\rmii{S2D}}_{\ell \ell}   = \frac{1}{4(4\pi)^2 m_{\eta}^2}\left\lbrace \frac{(g_{2}^2 -g_{1}^2)}{120} \left[ 3g_{1}^2 +3g_{2}^2 +20 y^2 \mathcal{F}_{2}\left(\frac{m_{\chi}}{m_{\eta}}\right) \right] + y^4 
 \mathcal{F}_{1}\left(\frac{m_{\chi}}{m_{\eta}}\right) \right\rbrace \, .
 \label{App_A_Cll_S2D}
 \end{align}
Similar to the case of $C_{ed}$ in the S1M and S1D models, the exclusion region takes the shape of a rectangle in the $(m_\chi, y)$ plane. This geometry again originates from a sign change in the matching coefficient near $y \simeq 0.3$ for both S2M and S2D, with only very limited dependence on the mass splitting and $m_\chi$. However, in contrast to the $C_{ed}$ exclusions, the global fit excludes values $y \gtrsim 0.3$ for which $C_{\ell \ell}$ is negative, thereby covering a significantly larger portion of parameter space.
\begin{figure}[t!]
     \begin{minipage}{0.45\linewidth}
      \includegraphics[width=\linewidth]{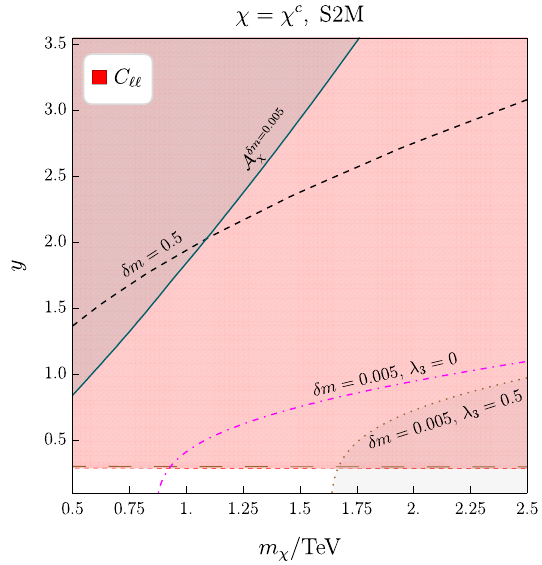}
\label{fig::summary_plots_deltam_doublet_majorana}
    \end{minipage}
    \begin{minipage}{0.45\linewidth}
     \includegraphics[width=\linewidth]{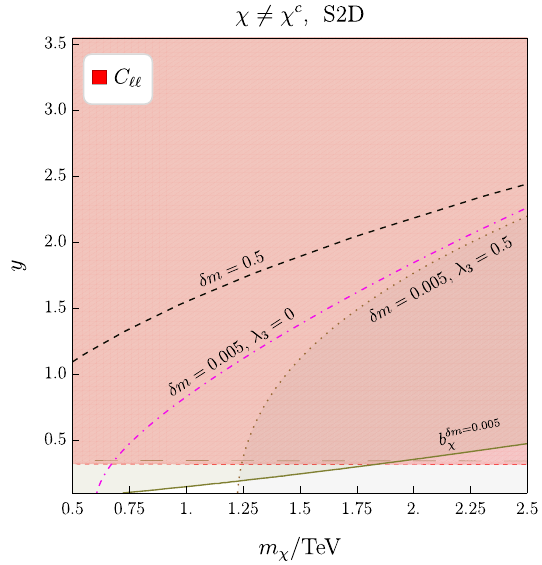}
\label{fig::summary_plots_deltam_doublet_dirac}
    \end{minipage}    
    \caption{Model parameter space in the $(m_\chi, y)$ plane. Same color code and shading conventions as in the previous figure~\ref{fig:summary_s1M_y_vs_mchi_largeDM}. Red-dashed and longer brown-dashed lines correspond to the edge of the bound from $C_{\ell \ell}$ for $\delta m =0.5$ and $\delta m =0.005$ respectively.}
    \label{summary_plots_deltam_doublet}
\end{figure}

In figure~\ref{summary_plots_deltam_doublet}, we display the results for both Majorana and Dirac scenarios in the left and right panels, respectively. The exclusion regions from $C_{\ell \ell}$ for different mass splittings are shown in the same plot: red dashed lines correspond to $\delta m = 0.5$, while the long-dashed brown line represents the minimal splitting $\delta m = 0.005$. The relic density constraints are shown as black dashed, magenta dot-dashed and brown dotted lines for various choices of $\delta m$ and $\lambda_3$. In the case of large splittings, the constraint from $C_{\ell \ell}$ exclude the cosmologically viable parameter space. For smaller splittings, the observed relic density is consistent with the available region $y\lesssim 0.3$ for the Majorana and Dirac dark matter, even for $\lambda_3=0$. As in the previous figures, we shade in gray only the region to the right of the largest value of $\lambda_3$. For the direct detection constraints, we include the anapole moment $\mathcal{A}_\chi$ for the Majorana DM scenario, which becomes relevant for the smallest splitting $\delta m = 0.005$. In the Dirac case, to avoid cluttering the figure, we display only the charge radius constraint for the smallest splitting. Indeed, the interplay between the Wilson coefficient $C_{\ell \ell}$ and the relic density already excludes the case of large splitting, $\delta m = 0.5$.

One can observe that the SMEFT global fit becomes more relevant for models with an SU(2)$_{\rmii{L}}$-charged scalar mediator. For the Majorana case, the constraint from $C_{\ell \ell}$ excludes a larger portion of the parameter space than the bound from the anapole moment. In the Dirac scenario, direct detection remains quite competitive; however, for the smallest splitting, the global fit on $C_{\ell \ell}$ becomes more significant for $m_\chi \gtrsim 2$~TeV.

As a last important comment, the RGE+NLO improved global fit presented in ref.~\cite{Bartocci:2024fmm} is crucial to obtain the exclusions from $C_{ed}$ and $C_{\ell \ell}$. Indeed, the LO results provide ranges for $C_{\ell \ell}$ and $C_{ed}$ that include negative values. This in turn would make the corresponding regions in figures~\ref{fig:high_mass_S1M_S1D} and \ref{summary_plots_deltam_doublet} disappear for the same values of $(m_\chi,\delta m, y)$.

\section{Conclusions}
\label{sec:conclusions}
In this work, we explored the connection between simplified $t$-channel dark matter models and the SMEFT, focusing on scenarios where a fermionic DM particle interacts with Standard Model leptons via a scalar mediator. We assumed a Minimal Flavor Violation framework to constrain flavor structures in the ultraviolet theory. We performed a one-loop matching of these models onto dimension-six SMEFT operators using the matching package \texttt{Matchete}, cross-checking the output analytically at several stages. The analysis included additional steps to organize the results in the Warsaw basis, where manipulations involving SM equations of motion and spinor identities were required to rewrite certain operator structures in standard form. A complete list of the resulting Wilson coefficients was provided for both SU(2)$_{\rmii{L}}$-singlet and doublet scalar mediators.

As a simplifying assumption for constructing the corresponding low-energy SMEFT, we adopted the working hypothesis that the DM and mediator masses are comparable, defining a single high-energy matching scale $\Lambda \simeq m_\chi \simeq m_\eta$. This compressed spectrum regime implies that coannihilation effects must be included when computing the dark matter relic abundance. This is an aspect we accounted for and improved upon with respect to previous literature, e.g.~\cite{Freitas:2014jla,Cepedello:2023yao}, and coannihilations have a sizable impact on the regions excluded from dark matter overproduction, especially for small splittings $\delta m \lesssim 0.1$. Moreover, we included thermal corrections to the scalar mass, Sommerfeld enhancement, and bound-state formation in scalar pair annihilation. For electroweak-charged scalars, these effects are in the ballpark of $\mathcal{O}$(10\%) for our mass range and the involved SM couplings.

We then investigated the interplay between different experimental constraints: global fits of SMEFT Wilson coefficients, bounds from direct detection, and the observed dark matter energy density.
Our results show that, even though the SMEFT matching coefficients are one-loop suppressed, they can still meaningfully constrain the parameter space for $m_\chi \gtrsim 0.5$~TeV and $\mathcal{O}(1)$ portal Yukawa and scalar couplings. In particular, for the models with a SU(2)$_{\rmi{L}}$-singlet mediator, we found that Majorana DM is less constrained than Dirac DM by the global fit on $C_{ee}$, while the bounds from $C_{ed}$ are the same for both Majorana and Dirac cases. Despite the fact that the global fit on $C_{ee}$ excludes large portions of the $(m_\chi, y)$ plane for Dirac DM, the direct detection limits are always more competitive. For the Majorana DM model, the leptonic operator and direct detection exclude smaller regions of the parameter space, and they do so in a complementary way: for large splittings, $C_{ee}$ provides constraints, while for small splittings it is the direct detection bounds that become relevant. For dark matter masses larger than 3~TeV, the lepton-down-quark operator $C_{ed}$ provides an exclusion region for $y \lesssim 0.5$, independent of the DM mass and largely independent on the relative splitting. To the best of our knowledge, this is the first time that lepton-quark four-fermion operators are used to constrain leptophilic dark matter models.

The models with an SU(2)$_{\rmi{L}}$-doublet mediator generate at low energy a significantly constraining $C_{\ell \ell}$. In the $(m_\chi, y)$ parameter space, values $y \lesssim 0.3$ are excluded for both large and small mass splittings. The global fit on the SMEFT matching coefficient $C_{\ell \ell}$ thus provides quite strong constraints, also for the Majorana dark matter option, which are competitive with those from direct detection. Viable regions, after imposing the dark matter energy density constraint, remain available for small mass splittings and are largely independent of the specific value of $\lambda_3$.

There are, of course, some limitations to our analysis, and we identify several possible extensions and improvements that could be addressed in future work. The flavor structure is perhaps the most important simplifying assumption, as it allows for a manageable set of operators and the use of reliable global fit results (including NLO and RGE effects). Although MFV is the most commonly studied option for $t$-channel models, alternative implementations could be considered, for instance, scenarios where the scalar mediator couples differently to each lepton flavor~\cite{Kopp:2014tsa, Garny:2015wea}, or where the dark matter itself carries a flavor index rather than the mediator~\cite{Chen:2015jkt, Acaroglu:2022hrm, Acaroglu:2023cza}. Such modifications would require a different treatment of the global SMEFT fits, necessitating a careful assessment of the flavor structure of the relevant operators~\cite{Faroughy:2020ina, Greljo:2022cah, Greljo:2023adz}. Depending on the flavor construction, SMEFT bounds may constrain one lepton family more strongly than others. 
Moreover, allowing for complex Yukawa couplings would induce additional electromagnetic moments at low energies, potentially strengthening direct detection constraints in the Dirac dark matter scenario. A careful assessment of dark matter scattering off atomic electrons, which occurs at tree level in these models, is not included here, as the direct detection analysis is not the focus of this study. We leave a detailed investigation of these effects for future work.

Another simplification in our work is the assumption of a compressed mass spectrum. Relaxing this assumption would eliminate the relevance of coannihilations in the thermal freeze-out computation. However, in that case, the EFT construction would require a more involved matching procedure before reaching the SMEFT, beginning with integrating out the mediator alone.
Another point to note is that in our analysis we have considered relatively large portal couplings. This motivates the inclusion of higher-order corrections, which may be sizable both in the dark matter annihilation processes and in the SMEFT matching coefficients. In the latter case, this would require performing a two-loop matching at dimension six. To our knowledge, automated tools capable of handling such a task are not yet available.

Finally, one can explore the impact of global fits on other $t$-channel models. By keeping the leptophilic options, there are several possible realizations, each with a potentially different interplay among SMEFT global fits and direct, indirect and collider searches. Examples are scalar or vector dark matter \cite{Arina:2020udz},  where the mediator is a fermionic vector-like particle. 

\section*{Acknowledgements}
We thank Anders E.~Thomsen for support and useful discussions about \texttt{Matchete}. We are grateful to Admir Greljo for insightful comments on the manuscript. We thank Riccardo Bartocci, Matteo Presilla, Stefano Di Noi and Merlin Reichard for discussions relevant to our work. 

\appendix
\renewcommand{\thesection}{\Alph{section}}
\renewcommand{\thesubsection}{\Alph{section}.\arabic{subsection}}
\renewcommand{\theequation}{\Alph{section}.\arabic{equation}}

\section{WCs lists}
\label{appendiix_A}
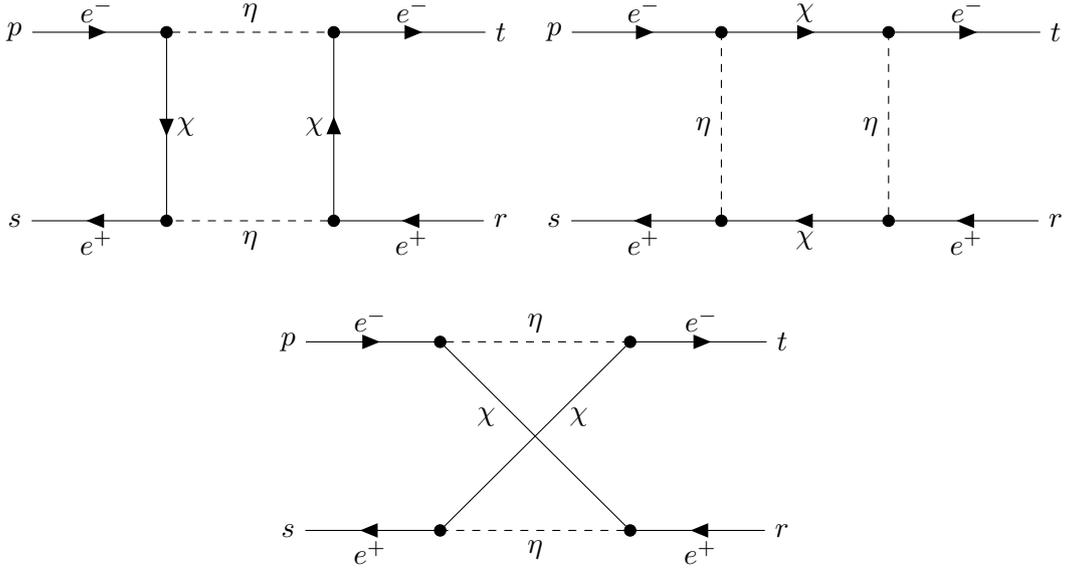
\begin{figure}
   \begin{center}
\begin{tikzpicture}
\begin{feynman}
\vertex[dot] (a1) at (0, 0) {};
\vertex[dot] (b1) at (0, 2.5) {};  
\vertex[dot] (c1) at (2.2, 2.5) {}; 
\vertex[dot] (d1) at (2.2, 0) {};
\vertex (f1a) at (4.4, 2.5) {$t$};
\vertex (f2a) at (4.4, 0) {$r$};
\vertex (i1a) at (-2, 2.5) {$p$}; 
\vertex (i2a) at (-2, 0) {$s$};  

\diagram* {
  (b1) -- [fermion, edge label=\( \chi\)] (a1),
  (a1) -- [fermion, edge label=\( e^{+}\)] (i2a),
  (i1a) -- [fermion, edge label=\( e^{-}\)] (b1),
  (d1) -- [fermion, edge label=\( \chi \)] (c1),
  (b1) -- [scalar, edge label=\( \eta \)] (c1),
  (d1) -- [scalar, edge label=\( \eta \)] (a1),
  (c1) -- [fermion, edge label=\( e^{-}\)] (f1a),
  (f2a) -- [fermion, edge label=\( e^{+}\)] (d1),
};

\vertex[dot] (a2) at (7.3, 0) {};
\vertex[dot] (b2) at (7.3, 2.5) {};
\vertex[dot] (c2) at (9.5, 2.5) {};
\vertex[dot] (d2) at (9.5, 0) {};
\vertex (f1b) at (11.7, 2.5) {$t$};
\vertex (f2b) at (11.7, 0) {$r$};
\vertex (i1b) at (5.1, 2.5) {$p$}; 
\vertex (i2b) at (5.1, 0) {$s$};  

\diagram* {
  (b2) -- [fermion, edge label=\( \chi\)] (c2),
  (a2) -- [fermion, edge label=\( e^{+}\)] (i2b),
  (i1b) -- [fermion, edge label=\( e^{-}\)] (b2),
  (d2) -- [fermion, edge label=\( \chi \)] (a2),
  (a2) -- [scalar, edge label=\( \eta \)] (b2),
  (d2) -- [scalar, edge label=\( \eta \)] (c2),
  (c2) -- [fermion, edge label=\( e^{-}\)] (f1b),
  (f2b) -- [fermion, edge label=\( e^{+}\)] (d2),
};

\end{feynman}
\end{tikzpicture}
\end{center}
\begin{center}
\begin{tikzpicture}
\begin{feynman}
     \vertex[dot]  (a)  at (0, 0) {}{};
     \vertex[dot] (b)  at (0,2.5) {};
     \vertex [dot] (c) at (2.5,2.5) {};
      \vertex [dot] (d) at (2.5,0) {};
      \vertex (f1) at (4.5,2.5) {$t$};
      \vertex (f2) at (4.5,0) {$r$};
      \vertex (i1) at (-2.,2.5) {$p$}; 
       \vertex (i2) at (-2.,0) {$s$};  

\diagram*{
      (a)-- [ edge label=\(\chi \, \, \, \, \,\, \)] (c),
       (a) -- [fermion, edge label=\( e^{+}\)] (i2),
        (i1) -- [fermion, edge label=\( e^{-}\)] (b),
      (b) -- [edge label=\( \,\,\,\, \,\chi \)] (d),
      (d) -- [scalar,edge label=\( \eta \)] (a),
       (b) -- [scalar ,edge label=\( \eta \)] (c),
       (c) -- [fermion, edge label=\( e^{-}\)](f1),
       (f2) -- [fermion, edge label= \( e^{+}\)] (d),
       };
\end{feynman}
\end{tikzpicture}
\end{center}
    \caption{Feynman diagrams that appear in the S1D model are shown in the first row, whereas the diagram below pertains exclusively to the S1M model (Majorana-like field contractions).}
    \label{fig:box_Maj_Dir}
\end{figure}

In this section we collect the complete list of the matching coefficients of the 41 CP-even flavor-conserving operators. We divide the listing in two subsections that correspond to the two scenarios: $\eta$ being a SU(2)$_{\rmii{L}}$ singlet or a SU(2)$_{\rmii{L}}$ doublet in section \ref{Appendix_A_S1M} and \ref{Appendix_A_S2M} respectively. Only the matching coefficient $C_{ee}$ in eqs.~\eqref{App_A_Cee_S1M} and \eqref{App_A_Cee_S1D}, and $C_{\ell \ell}$ in eqs.~\eqref{App_A_Cll_S2M} and \eqref{App_A_Cll_S2D},  are affected when selecting the dark matter to be a Majorana fermion or a Dirac fermion. More specifically, for the Dirac case only the loop function $\mathcal{F}_1$ enters the $y^4$ term. This goes back to the absence of the box diagram with four external right-handed (left-handed) leptons with Majorana contractions $\langle \chi \chi \rangle$ and $\langle \bar{\chi} \bar{\chi} \rangle$ in the model with $\eta$ being a SU(2)$_{\rmii{L}}$ singlet (SU(2)$_{\rmii{L}}$ doublet). The one-loop box diagrams are shown in   figure~\ref{fig:box_Maj_Dir}.

\subsection{Matching coefficients for the model S1M}
\label{Appendix_A_S1M}
\begin{flushleft}
\begin{align}
  &  C_{\ell \ell} = - \frac{g_1^4}{80(4 \pi)^2 m_\eta^2}\, , \quad
     C'_{\ell \ell} = 0 \, ,
     \\
   &  C_{qq}^{(1)}= - \frac{g_1^4}{720(4 \pi)^2 m_\eta^2} \, ,
      \quad  C_{qq}^{(3)}= 
     C_{qq}^{(1)'}= 
     C_{qq}^{(3)'}= 0 \, ,
     \\
 &    C_{\ell q}^{(1)}= \frac{g_1^4}{120(4 \pi)^2 m_\eta^2} \, , \quad
     C_{\ell q}^{(3)}= 0 \, ,
     \\
  &   C^{\rmii{S1M}}_{ee} = \frac{1}{(4 \pi)^2 m_\eta^2} \left\lbrace -\frac{g_1^4}{20} + \frac{y^4}{4} \left[ \mathcal{F}_1 \left( \frac{m_\chi}{m_\eta} \right) + \mathcal{F}_3 \left( \frac{m_\chi}{m_\eta} \right) \right] - \frac{g_1^2 y^2}{6} \mathcal{F}_2 \left( \frac{m_\chi}{m_\eta} \right) \right\rbrace
   \label{App_A_Cee_S1M}
  \\
  &   C^{\rmii{S1D}}_{ee} = \frac{1}{(4 \pi)^2 m_\eta^2} \left\lbrace -\frac{g_1^4}{20} + \frac{y^4}{4}  \mathcal{F}_1 \left( \frac{m_\chi}{m_\eta} \right)  - \frac{g_1^2 y^2}{6} \mathcal{F}_2 \left( \frac{m_\chi}{m_\eta} \right) \right\rbrace
  \label{App_A_Cee_S1D}
     \\
&     C_{uu} = - \frac{g_1^4}{45 (4 \pi)^2 m_\eta^2} \, , \quad
     C_{dd} = - \frac{g_1^4}{180 (4 \pi)^2 m_\eta^2} \, , \quad
     C'_{uu} = C'_{dd} = 0 \, , 
     \\
&     C_{eu} = -2  C_{ed} = \frac{g_1^2}{45 (4 \pi)^2 m_\eta^2} \left[ 3 g_1^2 + 5 y^2 \mathcal{F}_2 \left( \frac{m_\chi}{m_\eta} \right) \right] \, , \quad
     C_{ud}^{(1)}= 
     C_{ud}^{(8)} = 0 \, , 
     \label{C_ED_S1M}
     \\
&     C_{\ell e}=  -\frac{g_1^2}{60 (4 \pi)^2 m_\eta^2} \left[ 3 g_1^2 + 5 y^2 \mathcal{F}_2 \left( \frac{m_\chi}{m_\eta} \right) \right]   \, , \quad 
     C_{\ell u} = -2 C_{\ell d}= \frac{g_1^4}{30 (4 \pi)^2 m_\eta^2}  \, , 
      \\
&     C_{q e}  =   \frac{g_1^2}{180 (4 \pi)^2 m_\eta^2} \left[ 3 g_1^2 + 5 y^2 \mathcal{F}_2 \left( \frac{m_\chi}{m_\eta} \right) \right]  \, , 
     \\
&     C_{q u}^{(1)}  = -2   C_{q d}^{(1)} = -\frac{g_{1}^4}{90(4\pi)^2 m_{\eta}^2} \, ,  \quad 
     C_{q u}^{(8)}   = C_{q d}^{(8)}   = 0 \, , 
     \\
&     C_{H \ell}^{(1)}   =  \frac{1}{40 (4\pi)^2 m_\eta^2} \left[ 10 g_1^4 + |Y_e|^2 y^2 \mathcal{F}_4 \left( \frac{m_\chi}{m_\eta} \right) \right]  \, , \quad
     C_{H \ell}^{(3)}   = \frac{ |Y_e|^2 y^2}{4 (4\pi)^2 m_\eta^2} \mathcal{F}_4 \left( \frac{m_\chi}{m_\eta} \right)  \, ,
     \\
&     C_{H e}  = \frac{g_1^2}{60 (4\pi)^2 m_\eta^2}\left[ 3 g_1^2 + 5 y^2 \mathcal{F}_2 \left( \frac{m_\chi}{m_\eta} \right) \right]   \, ,
     \\
&     C_{H q}^{(1)}  = -\frac{g_1^4}{120 (4\pi)^2 m_\eta^2}  \, , \quad
     C_{H q}^{(3)}  = 0 \, , \quad
     C_{H u}  = -2 C_{H d} =   -\frac{g_1^4}{30 (4\pi)^2 m_\eta^2}\, ,
     \\
&     C_{H G}  =   C_{H W} = C_{H W B} = C_{G}  = C_{W} = 0  \, , \quad
     C_{H B}  = \frac{g_1^2 \lambda_3}{4 (4\pi)^2 m_\eta^2} \, ,
     \label{C_HB_S1M}
     \\
&     C_{H}  = \frac{1}{40 (4\pi)^2 m_\eta^2} \left( \lambda_1 g_1^4 +20 \lambda_3^2 (\lambda-\lambda_3) \right)\, , \quad
     C_{H \Box}  =- \frac{1}{80 (4\pi)^2 m_{\eta}^2} (g_{1}^4 +20 \lambda_{3}^2) \, ,
      \label{C_H_S1M}
     \\
&     C_{H D}  = - \frac{g_{1}^4}{20(4\pi)^2 m_{\eta}^2} \, .
\end{align}
\end{flushleft}
Here, an additional loop function, $\mathcal{F}_4(x)$, has to be defined, and reads 
\begin{equation}
    \mathcal{F}_4(x) = -\frac{1-6x^2+3 x^4 +2 x^6 - 6 x^4 \ln x^2 }{6(1-x^2)^4} \, . 
\end{equation}

\subsection{Matching coefficients S2M}
\label{Appendix_A_S2M}
\begin{align}
   &   C^{\rmii{S2M}}_{\ell \ell}   = \frac{1}{4(4\pi)^2 m_{\eta}^2}\left\lbrace \frac{(g_{2}^2 -g_{1}^2)}{120} \left[ 3g_{1}^2 +3g_{2}^2 +20 y^2 \mathcal{F}_{2}\left(\frac{m_{\chi}}{m_{\eta}}\right) \right] \right. \nonumber
   \\
  & \left. ~~~~~~~~~~~~~~~~~~~~~~~~~~~~~~~~~~~~~~~~~~~~~~~~~~~+ y^4 
 \left[ \mathcal{F}_{1}\left(\frac{m_{\chi}}{m_{\eta}}\right) -\mathcal{F}_{3}\left( \frac{m_{\chi}}{m_{\eta}}\right) \right] \right\rbrace \,,
 \label{App_A_Cll_S2M}
 \\
   &   C^{\rmii{S2D}}_{\ell \ell}   = \frac{1}{4(4\pi)^2 m_{\eta}^2}\left\lbrace \frac{(g_{2}^2 -g_{1}^2)}{120} \left[ 3g_{1}^2 +3g_{2}^2 +20 y^2 \mathcal{F}_{2}\left(\frac{m_{\chi}}{m_{\eta}}\right) \right] + y^4 
 \mathcal{F}_{1}\left(\frac{m_{\chi}}{m_{\eta}}\right) \right\rbrace \,,
 \label{App_A_Cll_S2D}
\\
  &   C'_{\ell \ell} = \frac{1}{4(4\pi)^2 m_\eta^2} \left[\frac{g_2^4}{20} + \frac{y^2 g_2^2}{3} \mathcal{F}_{2}\left(\frac{m_{\chi}}{m_{\eta}}\right)\right] \, ,
     \\
 &    C_{qq}^{(1)}= 
     - \frac{g_{1}^4}{1440(4 \pi)^2 m_\eta^2} \, , \quad 
     C_{qq}^{(3)}= - \frac{g_{2}^4}{160(4 \pi)^2 m_\eta^2}  \, , \quad
     C_{qq}^{(1)'}= 0 \, , \quad
     C_{qq}^{(3)'}= 0 \, ,
     \\
  &  C_{\ell q}^{(1)}= \frac{g_1^2}{720(4 \pi)^2 m_\eta^2} \left[ 3g_{1}^2 +10 y^2 \mathcal{F}_{2}\left(\frac{m_{\chi}}{m_{\eta}}\right) \right] \, ,
  \\
  &    C_{\ell q}^{(3)}= -\frac{g_2^2}{240(4 \pi)^2 m_\eta^2} \left[ 3g_{2}^2 + 10 y^2 \mathcal{F}_{2
     }\left(\frac{m_{\chi}}{m_{\eta}}\right)\right] \, ,
     \\
 &  C_{ee} = -\frac{g_1^4}{40(4 \pi)^2 m_\eta^2} \, , \quad C_{uu} = - \frac{g_1^4}{90 (4 \pi)^2 m_\eta^2}  \, , \quad C_{dd} = - \frac{g_1^4}{360 (4 \pi)^2 m_\eta^2} \, , \quad C'_{uu} =C'_{dd} = 0
     \\
  &   C_{eu} = \frac{g_1^4}{30 (4 \pi)^2 m_\eta^2} \, , \quad
     C_{ed} =  -\frac{g_1^4}{60 (4 \pi)^2 m_\eta^2} \, , \quad 
     C_{ud}^{(1)}= \frac{g_1^4}{90 (4 \pi)^2 m_\eta^2} \, , \quad
     C_{ud}^{(8)} = 0 \, , 
     \\
  &  C_{\ell e}=  -\frac{g_1^2}{120 (4 \pi)^2 m_\eta^2} \left[ 3 g_1^2 + 10 y^2 \mathcal{F}_2 \left( \frac{m_\chi}{m_\eta} \right) \right]   \, , 
     \\
 &    C_{\ell u} = -2 C_{\ell d}= \frac{g_1^2}{180 (4 \pi)^2 m_\eta^2}  \left[ 3 g_1^2 +10 y^2 \mathcal{F}_2 \left( \frac{m_\chi}{m_\eta} \right) \right]  \, , 
      \\
 &   C_{q e}  = 
 \frac{g_1^4}{120 (4 \pi)^2 m_\eta^2}  \, , \quad 
    C_{q u}^{(1)}  = - \frac{g_1^4}{180 (4 \pi)^2 m_\eta^2}  \, , \quad C_{q d}^{(1)}  =  \frac{g_1^4}{360 (4 \pi)^2 m_\eta^2}  \, ,  \quad
     C_{q u}^{(8)}   = C_{q d}^{(8)} =0 \, ,
     \\
 &    C_{H \ell}^{(1)}   = 
 \frac{g_1^2}{240 (4\pi)^2 m_\eta^2} \left[ 3 g_1^2 + 10  y^2 \mathcal{F}_2 \left( \frac{m_\chi}{m_\eta} \right) \right] \, , 
 \\
 &
 C_{H \ell}^{(3)}   = -\frac{g_2^2}{240 (4\pi)^2 m_\eta^2}  \left[3g_{2}^2 +10  y^2 \mathcal{F}_2 \left( \frac{m_\chi}{m_\eta} \right) \right] \, ,
     \\
 &    C_{H e}  = \frac{g_1^2}{40 (4\pi)^2 m_\eta^2}\left[  g_1^2 + 20 y^2 |Y_{e}|^2 \mathcal{F}_4 \left( \frac{m_\chi}{m_\eta} \right) \right]   \, ,
     \\
 &    C_{H q}^{(1)}  = \frac{C_{H q}^{(3)} }{3}=-\frac{g_1^4}{240 (4\pi)^2 m_\eta^2}  \, , \quad  \,     C_{H u}  =-2 C_{H d} =   -\frac{g_1^4}{60 (4\pi)^2 m_\eta^2} \, , 
     \\
 &    C_{H G}  = C_{H W B} = C_{G}  =0 \, , 
 \\
 & C_{H W}  = \frac{\lambda_{3}g_{2}^2}{8(4\pi)^2 m_{\eta}^2} \, , \quad    C_{H B}  = \frac{\lambda_{3}g_{1}^2}{8(4\pi)^2 m_{\eta}^2} \, , \quad    C_{W}  = \frac{g_{2}^3}{120(4\pi)^2 m_{\eta}^{2}} \, ,
     \\
&     C_{H}  = \frac{1}{80 (4\pi)^2 m_\eta^2} \left( \lambda g_{2}^4 + \lambda g_1^4 +80 \lambda_3^2 (\lambda-\lambda_3)\right)\, ,
     \\
  &   C_{H \Box}  = - \frac{1}{160(4\pi)^2 m_{\eta}^2} (3 g_{2}^4 +g_{1}^4 +80 \lambda_{3}^2 ) \, ,
     \\
&     C_{H D}  = -\frac{(g_{1}^4)}{40(4\pi)^2 m_{\eta}^2} \, .
\end{align}

\section{Fierz identities}
\label{sec::Fierz_4f}
The matching procedure with \texttt{Matchete}, as well as the direct computation of the box diagram, gives a spinor structure that reads  
\begin{eqnarray}
     \bar{e}^p_a (C P_{\rmii{L}})_{ab} \bar{e}_{b}^{s T} \; e^{s T}_c \, (C P_{\rmii{R}})_{cd} \, e^p_d  \,,
\end{eqnarray}
where  $a,b,c,d$ are spinor indices.
In the following steps, in order to avoid clutter, we drop the flavor indices and reinstate them at the end.
The decomposition in terms of the basis and its dual for any $4 \times 4$ matrix provides us with \cite{Nishi:2004st}
\begin{eqnarray}
    (C P_{\rmii{L}})_{ab} (C P_{\rmii{R}})_{cd} = \frac{1}{4} \textrm{Tr} \left\lbrace C P_{\rmii{L}} \, \tilde{\Gamma}_n  \, C P_{\rmii{R}} \, \tilde{\Gamma}_m  \right\rbrace  \, (\Gamma^m)_{ad} (\Gamma^n)_{cb}  \, .
\end{eqnarray}
For the algebra in $D=4$, we obtain that all the combinations of the Dirac matrices gives a vanishing trace but for the case
$\tilde{\Gamma}_3=\gamma_\mu P_{\rmii{R}}$ and $\tilde{\Gamma}_4=\gamma_\nu P_{\rmii{L}}$. All the others are zero, upon using the orthogonality of the chiral projectors, the cyclicity of the trace and the relation \[ [C, P_{\rmii{R(L)}}]=0\,. \]
The quantity that we have to evaluate is 
\begin{eqnarray}
    \textrm{Tr} \left\lbrace C P_{\rmii{L}} \, \gamma_\mu P_{\rmii{R}}  \, C P_{\rmii{R}} \, \gamma_\nu P_{\rmii{L}}  \right\rbrace =  \textrm{Tr} \left\lbrace C \, \gamma_\mu  \, C \, \gamma_\nu P_{\rmii{L}}  \right\rbrace = -2 (-2 g_{0 \mu} g_{0 \nu} + 2 g_{2 \mu} g_{2 \nu} + g_{\mu \nu}) \, ,
\end{eqnarray}
where $g^{\mu \nu}$ is the metric tensor.\footnote{We have dropped the vanishing entries of the metric tensor. We use the signature $(+1,-1,-1,-1)$.} The corresponding decomposition reads (the basis elements are $\Gamma_3=\gamma^\mu P_L$ and $\Gamma_4=\gamma^\nu P_R$ )
\begin{eqnarray}
   &&\Bar{e}_a (C P_{\rmii{L}})_{ab} \bar{e}_{b}^T \; e^T_c \, (C P_{\rmii{R}})_{cd} \, e_d    = \frac{1}{2}  \, \Bar{e}_a \, \bar{e}_{b}^T \, e^T_c \, e_d  \; (2 g_{0 \mu} g_{0 \nu} - 2 g_{2 \mu} g_{2 \nu} - g_{\mu \nu}) \, {(\gamma^\nu P_{\rmii{R}})_{ad} (\gamma^\mu P_{\rmii{L}})_{cb}} \, 
   \nonumber
   \\
   &&=  \frac{1}{2}  \, \Bar{e}_a \, \bar{e}_{b}^T \, e^T_c \, e_d  \; \left[ 2 (\gamma^0 {P_{\rmii{R}}})_{ad} (\gamma^0 {P_{\rmii{L}}})_{cb} - 2 (\gamma^2 {P_{\rmii{R}}})_{ad} (\gamma^2 {P_{\rmii{L}}})_{cb} - (\gamma^\mu {P_{\rmii{R}}})_{ad} (\gamma_\mu {P_{\rmii{L}}})_{cb} \right]
    \nonumber
   \\
   &&=  \frac{1}{2}  \, \Bar{e}_a \, \bar{e}_{b}^T \, e^T_c \, e_d  \; \left[ (\gamma^0 P_{\rmii{R}})_{ad} (\gamma^0 P_{\rmii{L}})_{cb} + (\gamma^1 P_{\rmii{R}})_{ad} (\gamma^1 P_{\rmii{L}})_{cb} \right.
   \nonumber
   \\
   && \left. ~~~~~~~~~~~~~~~~~~~~~~~~~~~~~~~~~~~~~-  (\gamma^2 P_{\rmii{R}})_{ad} (\gamma^2 P_{\rmii{L}})_{cb} + (\gamma^3 P_{\rmii{R}})_{ad} (\gamma^3 P_{\rmii{L}})_{cb}\right] \, .
\end{eqnarray}
By using the properties of the gamma matrices $\gamma^0=(\gamma^0)^T$, $\gamma^2=(\gamma^2)^T$, $\gamma^1=-(\gamma^1)^T$, $\gamma^3=-(\gamma^3)^T$, and $(e^T A \bar{e}^T )= - \bar{e} A^T e$, where the minus is due to the Grassmann nature of the fermion fields. Inserting back the flavor structure we obtain: 
\begin{eqnarray}
      \, (\bar{e}^p \, C P_{\rmii{L}} \, \bar{e}^{s T} ) \,  ( e^{s T} \, C P_{\rmii{R}} \, e^p ) = \frac{1}{2}   \, ( \bar{e}^p  \, \gamma^\mu P_{\rmii{R}} e^p) \, (\bar{e}^s \, \gamma_\mu P_{\rmii{R}} \, e^s) \, .
\end{eqnarray}

\bibliographystyle{JHEP.bst}
\bibliography{paper.bib}

@article{Kopp:2014tsa,
    author = "Kopp, Joachim and Michaels, Lisa and Smirnov, Juri",
    title = "{Loopy Constraints on Leptophilic Dark Matter and Internal Bremsstrahlung}",
    eprint = "1401.6457",
    archivePrefix = "arXiv",
    primaryClass = "hep-ph",
    doi = "10.1088/1475-7516/2014/04/022",
    journal = "JCAP",
    volume = "04",
    pages = "022",
    year = "2014"
}

@article{Arina:2020udz,
    author = "Arina, Chiara and Fuks, Benjamin and Mantani, Luca",
    title = "{A universal framework for t-channel dark matter models}",
    eprint = "2001.05024",
    archivePrefix = "arXiv",
    primaryClass = "hep-ph",
    reportNumber = "CP3-20-01, MCNET-20-01",
    doi = "10.1140/epjc/s10052-020-7933-7",
    journal = "Eur. Phys. J. C",
    volume = "80",
    number = "5",
    pages = "409",
    year = "2020"
}

@article{Arina:2020tuw,
    author = "Arina, Chiara and Fuks, Benjamin and Mantani, Luca and Mies, Hanna and Panizzi, Luca and Salko, Jakub",
    title = "{Closing in on $t$-channel simplified dark matter models}",
    eprint = "2010.07559",
    archivePrefix = "arXiv",
    primaryClass = "hep-ph",
    reportNumber = "P3H-20-058, TTK-20-35, CP3-20-47",
    doi = "10.1016/j.physletb.2020.136038",
    journal = "Phys. Lett. B",
    volume = "813",
    pages = "136038",
    year = "2021"
}

@article{Arina:2025zpi,
    author = "Arina, Chiara and others",
    title = "{t-channel dark matter at the LHC}",
    eprint = "2504.10597",
    archivePrefix = "arXiv",
    primaryClass = "hep-ph",
    reportNumber = "CERN-LPCC-2025-001, IRMP-CP3-25-07, TTK-25-07",
    month = "4",
    year = "2025"
}

@article{LopezHonorez:2006gr,
    author = "Lopez Honorez, Laura and Nezri, Emmanuel and Oliver, Josep F. and Tytgat, Michel H. G.",
    title = "{The Inert Doublet Model: An Archetype for Dark Matter}",
    eprint = "hep-ph/0612275",
    archivePrefix = "arXiv",
    reportNumber = "ULB-TH-06-27",
    doi = "10.1088/1475-7516/2007/02/028",
    journal = "JCAP",
    volume = "02",
    pages = "028",
    year = "2007"
}

@article{Barbieri:2006dq,
    author = "Barbieri, Riccardo and Hall, Lawrence J. and Rychkov, Vyacheslav S.",
    title = "{Improved naturalness with a heavy Higgs: An Alternative road to LHC physics}",
    eprint = "hep-ph/0603188",
    archivePrefix = "arXiv",
    reportNumber = "UCB-PTH-06-04, LBNL-59894",
    doi = "10.1103/PhysRevD.74.015007",
    journal = "Phys. Rev. D",
    volume = "74",
    pages = "015007",
    year = "2006"
}

@article{Deshpande:1977rw,
    author = "Deshpande, Nilendra G. and Ma, Ernest",
    title = "{Pattern of Symmetry Breaking with Two Higgs Doublets}",
    reportNumber = "OITS-81",
    doi = "10.1103/PhysRevD.18.2574",
    journal = "Phys. Rev. D",
    volume = "18",
    pages = "2574",
    year = "1978"
}

@article{Biondini:2022ggt,
    author = "Biondini, Simone and Schicho, Philipp and Tenkanen, Tuomas V. I.",
    title = "{Strong electroweak phase transition in t-channel simplified dark matter models}",
    eprint = "2207.12207",
    archivePrefix = "arXiv",
    primaryClass = "hep-ph",
    reportNumber = "HIP-2022-19/TH, NORDITA 2022-050",
    doi = "10.1088/1475-7516/2022/10/044",
    journal = "JCAP",
    volume = "10",
    pages = "044",
    year = "2022"
}

@article{Nishi:2004st,
    author = "Nishi, C. C.",
    title = "{Simple derivation of general Fierz-like identities}",
    eprint = "hep-ph/0412245",
    archivePrefix = "arXiv",
    doi = "10.1119/1.2074087",
    journal = "Am. J. Phys.",
    volume = "73",
    pages = "1160--1163",
    year = "2005"
}

@article{Kim:2016kxt,
    author = "Kim, Seyong and Laine, M.",
    title = "{On thermal corrections to near-threshold annihilation}",
    eprint = "1609.00474",
    archivePrefix = "arXiv",
    primaryClass = "hep-ph",
    doi = "10.1088/1475-7516/2017/01/013",
    journal = "JCAP",
    volume = "01",
    pages = "013",
    year = "2017"
}

@article{Guedes:2024vuf,
    author = "Guedes, Guilherme and Olgoso, Pablo",
    title = "{From the EFT to the UV: the complete SMEFT one-loop dictionary}",
    eprint = "2412.14253",
    archivePrefix = "arXiv",
    primaryClass = "hep-ph",
    reportNumber = "DESY-24-2",
    month = "12",
    year = "2024"
}

@article{Liu:2021mhn,
    author = "Liu, Jia and Wang, Xiao-Ping and Xie, Ke-Pan",
    title = "{Searching for lepton portal dark matter with colliders and gravitational waves}",
    eprint = "2104.06421",
    archivePrefix = "arXiv",
    primaryClass = "hep-ph",
    doi = "10.1007/JHEP06(2021)149",
    journal = "JHEP",
    volume = "06",
    pages = "149",
    year = "2021"
}

@article{Becker:2023vwd,
    author = "Becker, Mathias and Copello, Emanuele and Harz, Julia and Tamarit, Carlos",
    title = "{Dark matter freeze-in from non-equilibrium QFT: towards a consistent treatment of thermal effects}",
    eprint = "2312.17246",
    archivePrefix = "arXiv",
    primaryClass = "hep-ph",
    reportNumber = "MITP-23-085",
    doi = "10.1088/1475-7516/2025/03/071",
    journal = "JCAP",
    volume = "03",
    pages = "071",
    year = "2025"
}

@article{Biondini:2020ric,
    author = "Biondini, S. and Ghiglieri, J.",
    title = "{Freeze-in produced dark matter in the ultra-relativistic regime}",
    eprint = "2012.09083",
    archivePrefix = "arXiv",
    primaryClass = "hep-ph",
    doi = "10.1088/1475-7516/2021/03/075",
    journal = "JCAP",
    volume = "03",
    pages = "075",
    year = "2021"
}

@article{Mohan:2019zrk,
    author = "Mohan, Kirtimaan A. and Sengupta, Dipan and Tait, Tim M. P. and Yan, Bin and Yuan, C. P.",
    title = "{Direct detection and LHC constraints on a $t$-channel simplified model of Majorana dark matter at one loop}",
    eprint = "1903.05650",
    archivePrefix = "arXiv",
    primaryClass = "hep-ph",
    reportNumber = "UCI-HEP-TR-2019-02, MSUHEP-19-00",
    doi = "10.1007/JHEP05(2019)115",
    journal = "JHEP",
    volume = "05",
    pages = "115",
    year = "2019",
    note = "[Erratum: JHEP 05, 232 (2023)]"
}

@article{Barman:2021hhg,
    author = "Barman, Basabendu and Bhattacharya, Subhaditya and Girmohanta, Sudhakantha and Jahedi, Sahabub",
    title = "{Effective Leptophilic WIMPs at the e$^{+}$e$^{-}$ collider}",
    eprint = "2109.10936",
    archivePrefix = "arXiv",
    primaryClass = "hep-ph",
    reportNumber = "Stony Brook preprint YITP-SB-2021-16, PI/UAN-2021-700FT",
    doi = "10.1007/JHEP04(2022)146",
    journal = "JHEP",
    volume = "04",
    pages = "146",
    year = "2022"
}

@article{Parker_2018,
   title={Measurement of the fine-structure constant as a test of the Standard Model},
   volume={360},
   ISSN={1095-9203},
   url={http://dx.doi.org/10.1126/science.aap7706},
   DOI={10.1126/science.aap7706},
   number={6385},
   journal={Science},
   publisher={American Association for the Advancement of Science (AAAS)},
   author={Parker, Richard H. and Yu, Chenghui and Zhong, Weicheng and Estey, Brian and Müller, Holger},
   year={2018},
   month=apr, pages={191–195} }

@article{LZ:2022lsv,
    author = "Aalbers, J. and others",
    collaboration = "LZ",
    title = "{First Dark Matter Search Results from the LUX-ZEPLIN (LZ) Experiment}",
    eprint = "2207.03764",
    archivePrefix = "arXiv",
    primaryClass = "hep-ex",
    doi = "10.1103/PhysRevLett.131.041002",
    journal = "Phys. Rev. Lett.",
    volume = "131",
    number = "4",
    pages = "041002",
    year = "2023"
}

@article{Muong-2:2025xyk,
    author = "Aguillard, D. P. and others",
    collaboration = "Muon g-2",
    title = "{Measurement of the Positive Muon Anomalous Magnetic Moment to 127 ppb}",
    eprint = "2506.03069",
    archivePrefix = "arXiv",
    primaryClass = "hep-ex",
    reportNumber = "FERMILAB-PUB-25-0364-PPD",
    month = "6",
    year = "2025"
}

@article{DOnofrio:2015gop,
    author = "D'Onofrio, Michela and Rummukainen, Kari",
    title = "{Standard model cross-over on the lattice}",
    eprint = "1508.07161",
    archivePrefix = "arXiv",
    primaryClass = "hep-ph",
    reportNumber = "HIP-2015-30-TH",
    doi = "10.1103/PhysRevD.93.025003",
    journal = "Phys. Rev. D",
    volume = "93",
    number = "2",
    pages = "025003",
    year = "2016"
}

@article{Biondini:2023hek,
    author = "Biondini, Simone",
    title = "{Interplay between improved interaction rates and modified cosmological histories for dark matter}",
    eprint = "2309.00323",
    archivePrefix = "arXiv",
    primaryClass = "hep-ph",
    doi = "10.3389/fphy.2023.1285986",
    journal = "Front. in Phys.",
    volume = "11",
    pages = "1285986",
    year = "2023"
}

@article{Garny:2014waa,
    author = "Garny, Mathias and Ibarra, Alejandro and Rydbeck, Sara and Vogl, Stefan",
    title = "{Majorana Dark Matter with a Coloured Mediator: Collider vs Direct and Indirect Searches}",
    eprint = "1403.4634",
    archivePrefix = "arXiv",
    primaryClass = "hep-ph",
    reportNumber = "DESY-14-029, TUM-HEP-935-14, CERN-PH-TH-2014-041",
    doi = "10.1007/JHEP06(2014)169",
    journal = "JHEP",
    volume = "06",
    pages = "169",
    year = "2014"
}

@article{Planck:2018vyg,
    author = "Aghanim, N. and others",
    collaboration = "Planck",
    title = "{Planck 2018 results. VI. Cosmological parameters}",
    eprint = "1807.06209",
    archivePrefix = "arXiv",
    primaryClass = "astro-ph.CO",
    doi = "10.1051/0004-6361/201833910",
    journal = "Astron. Astrophys.",
    volume = "641",
    pages = "A6",
    year = "2020",
    note = "[Erratum: Astron.Astrophys. 652, C4 (2021)]"
}

@article{Edsjo:1997bg,
    author = "Edsjo, Joakim and Gondolo, Paolo",
    title = "{Neutralino relic density including coannihilations}",
    eprint = "hep-ph/9704361",
    archivePrefix = "arXiv",
    reportNumber = "UUITP-11-97, MPI-PHT-97-27",
    doi = "10.1103/PhysRevD.56.1879",
    journal = "Phys. Rev. D",
    volume = "56",
    pages = "1879--1894",
    year = "1997"
}

@article{Biondini:2025ihi,
    author = "Biondini, S. and Eriksson, M. and Laine, M.",
    title = "{Computing singlet scalar freeze-out with plasmon and plasmino states}",
    eprint = "2505.05206",
    archivePrefix = "arXiv",
    primaryClass = "hep-ph",
    month = "5",
    year = "2025"
}

@article{Hambye:2009pw,
    author = "Hambye, T. and Ling, F. -S. and Lopez Honorez, L. and Rocher, J.",
    title = "{Scalar Multiplet Dark Matter}",
    eprint = "0903.4010",
    archivePrefix = "arXiv",
    primaryClass = "hep-ph",
    reportNumber = "ULB-TH-09-03, FTUAM-09-04",
    doi = "10.1007/JHEP05(2010)066",
    journal = "JHEP",
    volume = "07",
    pages = "090",
    year = "2009",
    note = "[Erratum: JHEP 05, 066 (2010)]"
}

@article{Gondolo:1990dk,
    author = "Gondolo, Paolo and Gelmini, Graciela",
    title = "{Cosmic abundances of stable particles: Improved analysis}",
    reportNumber = "UCLA-90-TEP-68",
    doi = "10.1016/0550-3213(91)90438-4",
    journal = "Nucl. Phys. B",
    volume = "360",
    pages = "145--179",
    year = "1991"
}

@article{Ellis:2014ipa,
    author = "Ellis, John and Olive, Keith A. and Zheng, Jiaming",
    title = "{The Extent of the Stop Coannihilation Strip}",
    eprint = "1404.5571",
    archivePrefix = "arXiv",
    primaryClass = "hep-ph",
    reportNumber = "KCL-PH-TH-2014-17, LCTS-2014-16, CERN-PH-TH-2014-067, FTPI-MINN-14-11, UMN-TH-3333-14",
    doi = "10.1140/epjc/s10052-014-2947-7",
    journal = "Eur. Phys. J. C",
    volume = "74",
    pages = "2947",
    year = "2014"
}

@article{Harz:2014gaa,
    author = "Harz, J. and Herrmann, B. and Klasen, M. and Kova\v{r}\'\i{}k, K. and Meinecke, M.",
    title = "{SUSY-QCD corrections to stop annihilation into electroweak final states including Coulomb enhancement effects}",
    eprint = "1410.8063",
    archivePrefix = "arXiv",
    primaryClass = "hep-ph",
    doi = "10.1103/PhysRevD.91.034012",
    journal = "Phys. Rev. D",
    volume = "91",
    number = "3",
    pages = "034012",
    year = "2015"
}

@article{Isidori:2023pyp,
    author = "Isidori, Gino and Wilsch, Felix and Wyler, Daniel",
    title = "{The standard model effective field theory at work}",
    eprint = "2303.16922",
    archivePrefix = "arXiv",
    primaryClass = "hep-ph",
    reportNumber = "ZU-TH 14/23",
    doi = "10.1103/RevModPhys.96.015006",
    journal = "Rev. Mod. Phys.",
    volume = "96",
    number = "1",
    pages = "015006",
    year = "2024"
}

@article{Garny:2015wea,
    author = "Garny, Mathias and Ibarra, Alejandro and Vogl, Stefan",
    title = "{Signatures of Majorana dark matter with t-channel mediators}",
    eprint = "1503.01500",
    archivePrefix = "arXiv",
    primaryClass = "hep-ph",
    reportNumber = "CERN-PH-TH-2015-036, TUM-HEP-985-15",
    doi = "10.1142/S0218271815300190",
    journal = "Int. J. Mod. Phys. D",
    volume = "24",
    number = "07",
    pages = "1530019",
    year = "2015"
}

@article{Bringmann:2012vr,
    author = "Bringmann, Torsten and Huang, Xiaoyuan and Ibarra, Alejandro and Vogl, Stefan and Weniger, Christoph",
    title = "{Fermi LAT Search for Internal Bremsstrahlung Signatures from Dark Matter Annihilation}",
    eprint = "1203.1312",
    archivePrefix = "arXiv",
    primaryClass = "hep-ph",
    reportNumber = "TUM-HEP-828-12, MPP-2012-54",
    doi = "10.1088/1475-7516/2012/07/054",
    journal = "JCAP",
    volume = "07",
    pages = "054",
    year = "2012"
}

@article{Kayser:1983wm,
    author = "Kayser, Boris and Goldhaber, Alfred S.",
    title = "{{CPT} and {CP} Properties of Majorana Particles, and the Consequences}",
    reportNumber = "Print-83-0630 (NSF)",
    doi = "10.1103/PhysRevD.28.2341",
    journal = "Phys. Rev. D",
    volume = "28",
    pages = "2341",
    year = "1983"
}

@article{Radescu:1985wf,
    author = "Radescu, E. E.",
    title = "{Comments on the Electromagnetic Properties of Majorana Fermions}",
    reportNumber = "JINR-E2-85-341",
    doi = "10.1103/PhysRevD.32.1266",
    journal = "Phys. Rev. D",
    volume = "32",
    pages = "1266",
    year = "1985"
}

@article{Fukushima:2013efa,
    author = "Fukushima, Keita and Kumar, Jason",
    title = "{Dipole Moment Bounds on Dark Matter Annihilation}",
    eprint = "1307.7120",
    archivePrefix = "arXiv",
    primaryClass = "hep-ph",
    doi = "10.1103/PhysRevD.88.056017",
    journal = "Phys. Rev. D",
    volume = "88",
    number = "5",
    pages = "056017",
    year = "2013"
}

@article{Freitas:2014jla,
    author = "Freitas, Ayres and Westhoff, Susanne",
    title = "{Leptophilic Dark Matter in Lepton Interactions at LEP and ILC}",
    eprint = "1408.1959",
    archivePrefix = "arXiv",
    primaryClass = "hep-ph",
    reportNumber = "PITT-PACC-1404",
    doi = "10.1007/JHEP10(2014)116",
    journal = "JHEP",
    volume = "10",
    pages = "116",
    year = "2014"
}

@article{Jenkins:2013zja,
    author = "Jenkins, Elizabeth E. and Manohar, Aneesh V. and Trott, Michael",
    title = "{Renormalization Group Evolution of the Standard Model Dimension Six Operators I: Formalism and lambda Dependence}",
    eprint = "1308.2627",
    archivePrefix = "arXiv",
    primaryClass = "hep-ph",
    doi = "10.1007/JHEP10(2013)087",
    journal = "JHEP",
    volume = "10",
    pages = "087",
    year = "2013"
}

@article{Biondini:2017ufr,
    author = "Biondini, S. and Laine, M.",
    title = "{Re-derived overclosure bound for the inert doublet model}",
    eprint = "1706.01894",
    archivePrefix = "arXiv",
    primaryClass = "hep-ph",
    doi = "10.1007/JHEP08(2017)047",
    journal = "JHEP",
    volume = "08",
    pages = "047",
    year = "2017"
}

@article{Biondini:2018pwp,
      author         = "Biondini, S. and Laine, M.",
      title          = "{Thermal dark matter co-annihilating with a strongly
                        interacting scalar}",
      journal        = "JHEP",
      volume         = "04",
      year           = "2018",
      pages          = "072",
      eprint         = "1801.05821",
      archivePrefix  = "arXiv",
      primaryClass   = "hep-ph",
      SLACcitation   = "%%CITATION = ARXIV:1801.05821;%%"
}

@article{Garny:2018ali,
    author = "Garny, Mathias and Heisig, Jan",
    title = "{Interplay of super-WIMP and freeze-in production of dark matter}",
    eprint = "1809.10135",
    archivePrefix = "arXiv",
    primaryClass = "hep-ph",
    reportNumber = "TUM-HEP 1166/18, TTK-18-39",
    doi = "10.1103/PhysRevD.98.095031",
    journal = "Phys. Rev. D",
    volume = "98",
    number = "9",
    pages = "095031",
    year = "2018"
}

@article{Becker:2022iso,
    author = "Becker, Mathias and Copello, Emanuele and Harz, Julia and Mohan, Kirtimaan A. and Sengupta, Dipan",
    title = "{Impact of Sommerfeld effect and bound state formation in simplified t-channel dark matter models}",
    eprint = "2203.04326",
    archivePrefix = "arXiv",
    primaryClass = "hep-ph",
    reportNumber = "ADP-22-6/T1177, MSUHEP-22-002, TUM-HEP-1387-22",
    doi = "10.1007/JHEP08(2022)145",
    journal = "JHEP",
    volume = "08",
    pages = "145",
    year = "2022"
}

@article{Arina:2023msd,
    author = {Arina, Chiara and Fuks, Benjamin and Heisig, Jan and Kr\"amer, Michael and Mantani, Luca and Panizzi, Luca},
    title = "{Comprehensive exploration of t-channel simplified models of dark matter}",
    eprint = "2307.10367",
    archivePrefix = "arXiv",
    primaryClass = "hep-ph",
    reportNumber = "TTK-23-19",
    doi = "10.1103/PhysRevD.108.115007",
    journal = "Phys. Rev. D",
    volume = "108",
    number = "11",
    pages = "115007",
    year = "2023"
}

@article{Garny:2021qsr,
    author = "Garny, Mathias and Heisig, Jan",
    title = "{Bound-state effects on dark matter coannihilation: Pushing the boundaries of conversion-driven freeze-out}",
    eprint = "2112.01499",
    archivePrefix = "arXiv",
    primaryClass = "hep-ph",
    reportNumber = "TUM-HEP 1379/21, TTK-21-52",
    doi = "10.1103/PhysRevD.105.055004",
    journal = "Phys. Rev. D",
    volume = "105",
    number = "5",
    pages = "055004",
    year = "2022"
}

@article{Biondini:2018ovz,
      author         = "Biondini, S. and Vogl, Stefan",
      title          = "{Coloured coannihilations: Dark matter phenomenology
                        meets non-relativistic EFTs}",
      journal        = "JHEP",
      volume         = "02",
      year           = "2019",
      pages          = "016",
      doi            = "10.1007/JHEP02(2019)016",
      eprint         = "1811.02581",
      archivePrefix  = "arXiv",
      primaryClass   = "hep-ph",
      SLACcitation   = "%%CITATION = ARXIV:1811.02581;%%"
}

@article{Biondini:2019int,
      author         = "Biondini, Simone and Vogl, Stefan",
      title          = "{Scalar dark matter coannihilating with a coloured
                        fermion}",
      journal        = "JHEP",
      volume         = "11",
      year           = "2019",
      pages          = "147",
      doi            = "10.1007/JHEP11(2019)147",
      eprint         = "1907.05766",
      archivePrefix  = "arXiv",
      primaryClass   = "hep-ph",
      SLACcitation   = "%%CITATION = ARXIV:1907.05766;%%"
}

@article{Harz:2018csl,
      author         = "Harz, Julia and Petraki, Kalliopi",
      title          = "{Radiative bound-state formation in unbroken perturbative
                        non-Abelian theories and implications for dark matter}",
      journal        = "JHEP",
      volume         = "07",
      year           = "2018",
      pages          = "096",
      doi            = "10.1007/JHEP07(2018)096",
      eprint         = "1805.01200",
      archivePrefix  = "arXiv",
      primaryClass   = "hep-ph",
      reportNumber   = "Nikhef-2018-023",
      SLACcitation   = "%%CITATION = ARXIV:1805.01200;%%"
}

@article{Feng:2009mn,
    author = "Feng, Jonathan L. and Kaplinghat, Manoj and Tu, Huitzu and Yu, Hai-Bo",
    title = "{Hidden Charged Dark Matter}",
    eprint = "0905.3039",
    archivePrefix = "arXiv",
    primaryClass = "hep-ph",
    reportNumber = "UCI-TR-2009-06",
    doi = "10.1088/1475-7516/2009/07/004",
    journal = "JCAP",
    volume = "07",
    pages = "004",
    year = "2009"
}

@article{Hisano:2004ds,
    author = "Hisano, Junji and Matsumoto, Shigeki. and Nojiri, Mihoko M. and Saito, Osamu",
    title = "{Non-perturbative effect on dark matter annihilation and gamma ray signature from galactic center}",
    eprint = "hep-ph/0412403",
    archivePrefix = "arXiv",
    reportNumber = "ICRR-REPORT-513-2004-11, YITP-04-73",
    doi = "10.1103/PhysRevD.71.063528",
    journal = "Phys. Rev. D",
    volume = "71",
    pages = "063528",
    year = "2005"
}

@article{Iengo:2009ni,
    author = "Iengo, Roberto",
    title = "{Sommerfeld enhancement: General results from field theory diagrams}",
    eprint = "0902.0688",
    archivePrefix = "arXiv",
    primaryClass = "hep-ph",
    doi = "10.1088/1126-6708/2009/05/024",
    journal = "JHEP",
    volume = "05",
    pages = "024",
    year = "2009"
}

@article{Cassel:2009wt,
    author = "Cassel, S.",
    title = "{Sommerfeld factor for arbitrary partial wave processes}",
    eprint = "0903.5307",
    archivePrefix = "arXiv",
    primaryClass = "hep-ph",
    reportNumber = "OUTP-0910P",
    doi = "10.1088/0954-3899/37/10/105009",
    journal = "J. Phys. G",
    volume = "37",
    pages = "105009",
    year = "2010"
}

@article{Feng:2010zp,
    author = "Feng, Jonathan L. and Kaplinghat, Manoj and Yu, Hai-Bo",
    title = "{Sommerfeld Enhancements for Thermal Relic Dark Matter}",
    eprint = "1005.4678",
    archivePrefix = "arXiv",
    primaryClass = "hep-ph",
    reportNumber = "UCI-TR-2010-06",
    doi = "10.1103/PhysRevD.82.083525",
    journal = "Phys. Rev. D",
    volume = "82",
    pages = "083525",
    year = "2010"
}

@article{vonHarling:2014kha,
    author = "von Harling, Benedict and Petraki, Kalliopi",
    title = "{Bound-state formation for thermal relic dark matter and unitarity}",
    eprint = "1407.7874",
    archivePrefix = "arXiv",
    primaryClass = "hep-ph",
    reportNumber = "NIKHEF-2014-018",
    doi = "10.1088/1475-7516/2014/12/033",
    journal = "JCAP",
    volume = "12",
    pages = "033",
    year = "2014"
}

@article{Detmold:2014qqa,
    author = "Detmold, William and McCullough, Matthew and Pochinsky, Andrew",
    title = "{Dark Nuclei I: Cosmology and Indirect Detection}",
    eprint = "1406.2276",
    archivePrefix = "arXiv",
    primaryClass = "hep-ph",
    reportNumber = "MIT-CTP-4554",
    doi = "10.1103/PhysRevD.90.115013",
    journal = "Phys. Rev. D",
    volume = "90",
    number = "11",
    pages = "115013",
    year = "2014"
}

@article{Baker:2015qna,
    author = "Baker, Michael J. and others",
    title = "{The Coannihilation Codex}",
    eprint = "1510.03434",
    archivePrefix = "arXiv",
    primaryClass = "hep-ph",
    reportNumber = "MITP-15-078",
    doi = "10.1007/JHEP12(2015)120",
    journal = "JHEP",
    volume = "12",
    pages = "120",
    year = "2015"
}

@article{Laine:2015kra,
    author = "Laine, M. and Meyer, M.",
    title = "{Standard Model thermodynamics across the electroweak crossover}",
    eprint = "1503.04935",
    archivePrefix = "arXiv",
    primaryClass = "hep-ph",
    doi = "10.1088/1475-7516/2015/07/035",
    journal = "JCAP",
    volume = "07",
    pages = "035",
    year = "2015"
}

@article{Griest:1990kh,
    author = "Griest, Kim and Seckel, David",
    title = "{Three exceptions in the calculation of relic abundances}",
    reportNumber = "CFPA-TH-90-001A, BA-90-79",
    doi = "10.1103/PhysRevD.43.3191",
    journal = "Phys. Rev. D",
    volume = "43",
    pages = "3191--3203",
    year = "1991"
}

@article{Falkowski:2020pma,
    author = "Falkowski, Adam and Gonz\'alez-Alonso, Mart\'\i{}n and Naviliat-Cuncic, Oscar",
    title = "{Comprehensive analysis of beta decays within and beyond the Standard Model}",
    eprint = "2010.13797",
    archivePrefix = "arXiv",
    primaryClass = "hep-ph",
    reportNumber = "IFIC/20-49, FTUV/20-1027",
    doi = "10.1007/JHEP04(2021)126",
    journal = "JHEP",
    volume = "04",
    pages = "126",
    year = "2021"
}

@article{Crivellin:2017rmk,
    author = "Crivellin, Andreas and Davidson, Sacha and Pruna, Giovanni Marco and Signer, Adrian",
    title = "{Renormalisation-group improved analysis of $\mu\to e$ processes in a systematic effective-field-theory approach}",
    eprint = "1702.03020",
    archivePrefix = "arXiv",
    primaryClass = "hep-ph",
    reportNumber = "PSI-PR-17-01, ZU-TH-01-17",
    doi = "10.1007/JHEP05(2017)117",
    journal = "JHEP",
    volume = "05",
    pages = "117",
    year = "2017"
}

@article{Falkowski:2017pss,
    author = "Falkowski, Adam and Gonz\'alez-Alonso, Mart\'\i{}n and Mimouni, Kin",
    title = "{Compilation of low-energy constraints on 4-fermion operators in the SMEFT}",
    eprint = "1706.03783",
    archivePrefix = "arXiv",
    primaryClass = "hep-ph",
    doi = "10.1007/JHEP08(2017)123",
    journal = "JHEP",
    volume = "08",
    pages = "123",
    year = "2017"
}

@article{Falkowski:2015krw,
    author = "Falkowski, Adam and Mimouni, Kin",
    title = "{Model independent constraints on four-lepton operators}",
    eprint = "1511.07434",
    archivePrefix = "arXiv",
    primaryClass = "hep-ph",
    doi = "10.1007/JHEP02(2016)086",
    journal = "JHEP",
    volume = "02",
    pages = "086",
    year = "2016"
}

@article{Cepedello:2023yao,
    author = "Cepedello, Ricardo and Esser, Fabian and Hirsch, Martin and Sanz, Veronica",
    title = "{SMEFT goes dark: Dark Matter models for four-fermion operators}",
    eprint = "2302.03485",
    archivePrefix = "arXiv",
    primaryClass = "hep-ph",
    doi = "10.1007/JHEP09(2023)081",
    journal = "JHEP",
    volume = "09",
    pages = "081",
    year = "2023"
}

@article{Fuentes-Martin:2022jrf,
    author = {Fuentes-Mart\'\i{}n, Javier and K\"onig, Matthias and Pag\`es, Julie and Thomsen, Anders Eller and Wilsch, Felix},
    title = "{A proof of concept for matchete: an automated tool for matching effective theories}",
    eprint = "2212.04510",
    archivePrefix = "arXiv",
    primaryClass = "hep-ph",
    reportNumber = "MITP-22-105, TUM-HEP-1443/22, ZU-TH-58/22",
    doi = "10.1140/epjc/s10052-023-11726-1",
    journal = "Eur. Phys. J. C",
    volume = "83",
    number = "7",
    pages = "662",
    year = "2023"
}

@article{Buchmuller:1985jz,
    author = "Buchmuller, W. and Wyler, D.",
    title = "{Effective Lagrangian Analysis of New Interactions and Flavor Conservation}",
    reportNumber = "CERN-TH-4254/85",
    doi = "10.1016/0550-3213(86)90262-2",
    journal = "Nucl. Phys. B",
    volume = "268",
    pages = "621--653",
    year = "1986"
}

@article{Cirelli:2024ssz,
    author = "Cirelli, Marco and Strumia, Alessandro and Zupan, Jure",
    title = "{Dark Matter}",
    eprint = "2406.01705",
    archivePrefix = "arXiv",
    primaryClass = "hep-ph",
    month = "6",
    year = "2024"
}

@article{Schumann:2019eaa,
    author = "Schumann, Marc",
    title = "{Direct Detection of WIMP Dark Matter: Concepts and Status}",
    eprint = "1903.03026",
    archivePrefix = "arXiv",
    primaryClass = "astro-ph.CO",
    doi = "10.1088/1361-6471/ab2ea5",
    journal = "J. Phys. G",
    volume = "46",
    number = "10",
    pages = "103003",
    year = "2019"
}

@article{Grzadkowski:2010es,
    author = "Grzadkowski, B. and Iskrzynski, M. and Misiak, M. and Rosiek, J.",
    title = "{Dimension-Six Terms in the Standard Model Lagrangian}",
    eprint = "1008.4884",
    archivePrefix = "arXiv",
    primaryClass = "hep-ph",
    reportNumber = "IFT-9-2010, TTP10-35",
    doi = "10.1007/JHEP10(2010)085",
    journal = "JHEP",
    volume = "10",
    pages = "085",
    year = "2010"
}

@article{Morgante:2018tiq,
    author = "Morgante, Enrico",
    title = "{Simplified Dark Matter Models}",
    eprint = "1804.01245",
    archivePrefix = "arXiv",
    primaryClass = "hep-ph",
    reportNumber = "DESY-18-047, DESY 18-047",
    doi = "10.1155/2018/5012043",
    journal = "Adv. High Energy Phys.",
    volume = "2018",
    pages = "5012043",
    year = "2018"
}

@article{LHCNewPhysicsWorkingGroup:2011mji,
    author = "Alves, Daniele",
    editor = "Arkani-Hamed, Nima and others",
    collaboration = "LHC New Physics Working Group",
    title = "{Simplified Models for LHC New Physics Searches}",
    eprint = "1105.2838",
    archivePrefix = "arXiv",
    primaryClass = "hep-ph",
    reportNumber = "SLAC-PUB-15045, FERMILAB-PUB-11-842-A-PPD",
    doi = "10.1088/0954-3899/39/10/105005",
    journal = "J. Phys. G",
    volume = "39",
    pages = "105005",
    year = "2012"
}

@article{Alwall:2008ag,
    author = "Alwall, Johan and Schuster, Philip and Toro, Natalia",
    title = "{Simplified Models for a First Characterization of New Physics at the LHC}",
    eprint = "0810.3921",
    archivePrefix = "arXiv",
    primaryClass = "hep-ph",
    reportNumber = "SLAC-PUB-13425, SU-ITP-08-24",
    doi = "10.1103/PhysRevD.79.075020",
    journal = "Phys. Rev. D",
    volume = "79",
    pages = "075020",
    year = "2009"
}

@article{Brivio:2017vri,
    author = "Brivio, Ilaria and Trott, Michael",
    title = "{The Standard Model as an Effective Field Theory}",
    eprint = "1706.08945",
    archivePrefix = "arXiv",
    primaryClass = "hep-ph",
    doi = "10.1016/j.physrep.2018.11.002",
    journal = "Phys. Rept.",
    volume = "793",
    pages = "1--98",
    year = "2019"
}

@article{Arcadi:2021glq,
    author = "Arcadi, Giorgio and Calibbi, Lorenzo and Fedele, Marco and Mescia, Federico",
    title = "{Systematic approach to B-physics anomalies and t-channel dark matter}",
    eprint = "2103.09835",
    archivePrefix = "arXiv",
    primaryClass = "hep-ph",
    reportNumber = "TTP21-007, P3H-21-017",
    doi = "10.1103/PhysRevD.104.115012",
    journal = "Phys. Rev. D",
    volume = "104",
    number = "11",
    pages = "115012",
    year = "2021"
}

@article{Herrero-Garcia:2018koq,
    author = "Herrero-Garcia, Juan and Molinaro, Emiliano and Schmidt, Michael A.",
    title = "{Dark matter direct detection of a fermionic singlet at one loop}",
    eprint = "1803.05660",
    archivePrefix = "arXiv",
    primaryClass = "hep-ph",
    reportNumber = "ADP-18-8/T1056, CP3-Origins-2018-009, ADP-18-8-T1056, CP3-ORIGINS-2018-009, CP3-Origins-2018-009-DNRF90",
    doi = "10.1140/epjc/s10052-018-5935-5",
    journal = "Eur. Phys. J. C",
    volume = "78",
    number = "6",
    pages = "471",
    year = "2018",
    note = "[Erratum: None 82, 53 (2022)]"
}

@article{Ma:2006km,
    author = "Ma, Ernest",
    title = "{Verifiable radiative seesaw mechanism of neutrino mass and dark matter}",
    eprint = "hep-ph/0601225",
    archivePrefix = "arXiv",
    reportNumber = "UCRHEP-T403",
    doi = "10.1103/PhysRevD.73.077301",
    journal = "Phys. Rev. D",
    volume = "73",
    pages = "077301",
    year = "2006"
}

@article{Cepedello:2022pyx,
    author = "Cepedello, Ricardo and Esser, Fabian and Hirsch, Martin and Sanz, Veronica",
    title = "{Mapping the SMEFT to discoverable models}",
    eprint = "2207.13714",
    archivePrefix = "arXiv",
    primaryClass = "hep-ph",
    doi = "10.1007/JHEP09(2022)229",
    journal = "JHEP",
    volume = "09",
    pages = "229",
    year = "2022"
}

@article{Arcadi:2017kky,
    author = "Arcadi, Giorgio and Dutra, Ma\'\i{}ra and Ghosh, Pradipta and Lindner, Manfred and Mambrini, Yann and Pierre, Mathias and Profumo, Stefano and Queiroz, Farinaldo S.",
    title = "{The waning of the WIMP? A review of models, searches, and constraints}",
    eprint = "1703.07364",
    archivePrefix = "arXiv",
    primaryClass = "hep-ph",
    doi = "10.1140/epjc/s10052-018-5662-y",
    journal = "Eur. Phys. J. C",
    volume = "78",
    number = "3",
    pages = "203",
    year = "2018"
}

@article{Bell:2010ei,
    author = "Bell, Nicole F. and Dent, James B. and Jacques, Thomas D. and Weiler, Thomas J.",
    title = "{W/Z Bremsstrahlung as the Dominant Annihilation Channel for Dark Matter}",
    eprint = "1009.2584",
    archivePrefix = "arXiv",
    primaryClass = "hep-ph",
    doi = "10.1103/PhysRevD.83.013001",
    journal = "Phys. Rev. D",
    volume = "83",
    pages = "013001",
    year = "2011"
}

@article{Bringmann:2007nk,
    author = "Bringmann, Torsten and Bergstrom, Lars and Edsjo, Joakim",
    title = "{New Gamma-Ray Contributions to Supersymmetric Dark Matter Annihilation}",
    eprint = "0710.3169",
    archivePrefix = "arXiv",
    primaryClass = "hep-ph",
    doi = "10.1088/1126-6708/2008/01/049",
    journal = "JHEP",
    volume = "01",
    pages = "049",
    year = "2008"
}

@article{Gargalionis:2024jaw,
    author = "Gargalionis, John and Quevillon, Jeremie and Vuong, Pham Ngoc Hoa and You, Tevong",
    title = "{Linear Standard Model extensions in the SMEFT at one loop and Tera-Z}",
    eprint = "2412.01759",
    archivePrefix = "arXiv",
    primaryClass = "hep-ph",
    reportNumber = "DESY-24-184, ADP-24-19/T1258, KCL-PH-TH-2024-72",
    month = "12",
    year = "2024"
}

@article{Becker:2025lkc,
    author = "Becker, Mathias and Lozano, Maria Jose Fernandez and Harz, Julia and Tamarit, Carlos",
    title = "{Multiple Soft Scatterings in Scalar Dark Matter Freeze-In}",
    eprint = "2506.11185",
    archivePrefix = "arXiv",
    primaryClass = "hep-ph",
    reportNumber = "MITP-25-046",
    month = "6",
    year = "2025"
}

@article{ATLAS:2019lng,
    author = "Aad, Georges and others",
    collaboration = "ATLAS",
    title = "{Searches for electroweak production of supersymmetric particles with compressed mass spectra in $\sqrt{s}=$ 13 TeV $pp$ collisions with the ATLAS detector}",
    eprint = "1911.12606",
    archivePrefix = "arXiv",
    primaryClass = "hep-ex",
    reportNumber = "CERN-EP-2019-242",
    doi = "10.1103/PhysRevD.101.052005",
    journal = "Phys. Rev. D",
    volume = "101",
    number = "5",
    pages = "052005",
    year = "2020"
}

@article{Ibarra:2022nzm,
    author = "Ibarra, Alejandro and Reichard, Merlin and Nagai, Ryo",
    title = "{Anapole moment of Majorana fermions and implications for direct detection of neutralino dark matter}",
    eprint = "2207.01014",
    archivePrefix = "arXiv",
    primaryClass = "hep-ph",
    doi = "10.1007/JHEP01(2023)086",
    journal = "JHEP",
    volume = "01",
    pages = "086",
    year = "2023"
}

@article{Brown:1996qs,
    author = "Brown, Lowell S. and Sawyer, R. F.",
    title = "{Nuclear reaction rates in a plasma}",
    eprint = "astro-ph/9610256",
    archivePrefix = "arXiv",
    reportNumber = "UW-PT-96-30",
    doi = "10.1103/RevModPhys.69.411",
    journal = "Rev. Mod. Phys.",
    volume = "69",
    pages = "411--436",
    year = "1997"
}

@article{Mitridate:2017izz,
    author = "Mitridate, Andrea and Redi, Michele and Smirnov, Juri and Strumia, Alessandro",
    title = "{Cosmological Implications of Dark Matter Bound States}",
    eprint = "1702.01141",
    archivePrefix = "arXiv",
    primaryClass = "hep-ph",
    reportNumber = "CERN-TH-2017-030, IFUP-TH-2017",
    doi = "10.1088/1475-7516/2017/05/006",
    journal = "JCAP",
    volume = "05",
    pages = "006",
    year = "2017"
}

@article{Fuentes_Mart_n_2021,
   title={DsixTools 2.0: the effective field theory toolkit},
   volume={81},
   ISSN={1434-6052},
   url={http://dx.doi.org/10.1140/epjc/s10052-020-08778-y},
   DOI={10.1140/epjc/s10052-020-08778-y},
   number={2},
   journal={The European Physical Journal C},
   publisher={Springer Science and Business Media LLC},
   author={Fuentes-Martín, Javier and Ruiz-Femenía, Pedro and Vicente, Avelino and Virto, Javier},
   year={2021},
   month=feb }

@article{Di_Noi_2023,
   title={RGESolver: a C++ library to perform renormalization group evolution in the Standard Model Effective Theory},
   volume={83},
   ISSN={1434-6052},
   url={http://dx.doi.org/10.1140/epjc/s10052-023-11189-4},
   DOI={10.1140/epjc/s10052-023-11189-4},
   number={3},
   journal={The European Physical Journal C},
   publisher={Springer Science and Business Media LLC},
   author={Di Noi, Stefano and Silvestrini, Luca},
   year={2023},
   month=mar }

@article{Hall:1990ac,
    author = "Hall, L. J. and Randall, Lisa",
    title = "{Weak scale effective supersymmetry}",
    reportNumber = "UCB-PTH-90/13, LBL-28879",
    doi = "10.1103/PhysRevLett.65.2939",
    journal = "Phys. Rev. Lett.",
    volume = "65",
    pages = "2939--2942",
    year = "1990"
}

@article{Horigome:2021qof,
    author = "Horigome, Shun-Ichi and Katayose, Taisuke and Matsumoto, Shigeki and Saha, Ipsita",
    title = "{Leptophilic fermion WIMP: Role of future lepton colliders}",
    eprint = "2102.08645",
    archivePrefix = "arXiv",
    primaryClass = "hep-ph",
    reportNumber = "IPMU21-0013",
    doi = "10.1103/PhysRevD.104.055001",
    journal = "Phys. Rev. D",
    volume = "104",
    number = "5",
    pages = "055001",
    year = "2021"
}

@article{DAmbrosio:2002vsn,
    author = "D'Ambrosio, G. and Giudice, G. F. and Isidori, G. and Strumia, A.",
    title = "{Minimal flavor violation: An Effective field theory approach}",
    eprint = "hep-ph/0207036",
    archivePrefix = "arXiv",
    reportNumber = "CERN-TH-2002-147, IFUP-TH-2002-17",
    doi = "10.1016/S0550-3213(02)00836-2",
    journal = "Nucl. Phys. B",
    volume = "645",
    pages = "155--187",
    year = "2002"
}

@article{Chivukula:1987py,
    author = "Chivukula, R. Sekhar and Georgi, Howard",
    title = "{Composite Technicolor Standard Model}",
    reportNumber = "BUHEP-87-2, HUTP-87/A003",
    doi = "10.1016/0370-2693(87)90713-1",
    journal = "Phys. Lett. B",
    volume = "188",
    pages = "99--104",
    year = "1987"
}

@article{Gerard:1982mm,
    author = "Gerard, J. M.",
    title = "{FERMION MASS SPECTRUM IN SU(2)-L x U(1)}",
    reportNumber = "CERN-TH-3454",
    doi = "10.1007/BF01572477",
    journal = "Z. Phys. C",
    volume = "18",
    pages = "145",
    year = "1983"
}

@article{Ethier:2021ydt,
    author = "Ethier, Jacob J. and Gomez-Ambrosio, Raquel and Magni, Giacomo and Rojo, Juan",
    title = "{SMEFT analysis of vector boson scattering and diboson data from the LHC Run II}",
    eprint = "2101.03180",
    archivePrefix = "arXiv",
    primaryClass = "hep-ph",
    reportNumber = "Nikhef-2020-039, Nikhef-2020-039 , IPPP/20/71, VBSCAN-PUB-01-21",
    doi = "10.1140/epjc/s10052-021-09347-7",
    journal = "Eur. Phys. J. C",
    volume = "81",
    number = "6",
    pages = "560",
    year = "2021"
}

@article{ATLAS:2019lff,
    author = "Aad, Georges and others",
    collaboration = "ATLAS",
    title = "{Search for electroweak production of charginos and sleptons decaying into final states with two leptons and missing transverse momentum in $\sqrt{s}=13$ TeV $pp$ collisions using the ATLAS detector}",
    eprint = "1908.08215",
    archivePrefix = "arXiv",
    primaryClass = "hep-ex",
    reportNumber = "CERN-EP-2019-106",
    doi = "10.1140/epjc/s10052-019-7594-6",
    journal = "Eur. Phys. J. C",
    volume = "80",
    number = "2",
    pages = "123",
    year = "2020"
}

@article{Acaroglu:2023cza,
    author = "Acaro\u{g}lu, Harun and Blanke, Monika and Tabet, Mustafa",
    title = "{Opening the Higgs portal to lepton-flavoured dark matter}",
    eprint = "2309.10700",
    archivePrefix = "arXiv",
    primaryClass = "hep-ph",
    reportNumber = "TTP23-028, P3H-23-050, DO-TH 23/11",
    doi = "10.1007/JHEP11(2023)079",
    journal = "JHEP",
    volume = "11",
    pages = "079",
    year = "2023"
}

@article{Acaroglu:2022hrm,
    author = "Acaro\u{g}lu, Harun and Agrawal, Prateek and Blanke, Monika",
    title = "{Lepton-flavoured scalar dark matter in Dark Minimal Flavour Violation}",
    eprint = "2211.03809",
    archivePrefix = "arXiv",
    primaryClass = "hep-ph",
    reportNumber = "TTP22-063; P3H-22-104",
    doi = "10.1007/JHEP05(2023)106",
    journal = "JHEP",
    volume = "05",
    pages = "106",
    year = "2023"
}

@article{Chen:2015jkt,
    author = "Chen, Mu-Chun and Huang, Jinrui and Takhistov, Volodymyr",
    title = "{Beyond Minimal Lepton Flavored Dark Matter}",
    eprint = "1510.04694",
    archivePrefix = "arXiv",
    primaryClass = "hep-ph",
    reportNumber = "UCI-TR-2015-17, LA-UR-15-27938",
    doi = "10.1007/JHEP02(2016)060",
    journal = "JHEP",
    volume = "02",
    pages = "060",
    year = "2016"
}

@article{Faroughy:2020ina,
    author = "Faroughy, Darius A. and Isidori, Gino and Wilsch, Felix and Yamamoto, Kei",
    title = "{Flavour symmetries in the SMEFT}",
    eprint = "2005.05366",
    archivePrefix = "arXiv",
    primaryClass = "hep-ph",
    doi = "10.1007/JHEP08(2020)166",
    journal = "JHEP",
    volume = "08",
    pages = "166",
    year = "2020"
}

@article{Greljo:2023adz,
    author = "Greljo, Admir and Palavri\'c, Ajdin",
    title = "{Leading directions in the SMEFT}",
    eprint = "2305.08898",
    archivePrefix = "arXiv",
    primaryClass = "hep-ph",
    doi = "10.1007/JHEP09(2023)009",
    journal = "JHEP",
    volume = "09",
    pages = "009",
    year = "2023"
}

@article{Greljo:2022cah,
    author = "Greljo, Admir and Palavri\'c, Ajdin and Thomsen, Anders Eller",
    title = "{Adding Flavor to the SMEFT}",
    eprint = "2203.09561",
    archivePrefix = "arXiv",
    primaryClass = "hep-ph",
    doi = "10.1007/JHEP10(2022)005",
    journal = "JHEP",
    volume = "10",
    pages = "010",
    year = "2022"
}

@article{Bartocci:2024fmm,
    author = {Bartocci, Riccardo and Biek\"otter, Anke and Hurth, Tobias},
    title = "{Renormalisation group evolution effects on global SMEFT analyses}",
    eprint = "2412.09674",
    archivePrefix = "arXiv",
    primaryClass = "hep-ph",
    reportNumber = "MITP-24-087",
    doi = "10.1007/JHEP05(2025)203",
    journal = "JHEP",
    volume = "05",
    pages = "203",
    year = "2025"
}

@article{Ellis:2020ivx,
    author = "Ellis, Sebastian A. R. and Quevillon, J\'er\'emie and Vuong, Pham Ngoc Hoa and You, Tevong and Zhang, Zhengkang",
    title = "{The Fermionic Universal One-Loop Effective Action}",
    eprint = "2006.16260",
    archivePrefix = "arXiv",
    primaryClass = "hep-ph",
    reportNumber = "CERN-TH-2020-104, CALT-TH-2020-027",
    doi = "10.1007/JHEP11(2020)078",
    journal = "JHEP",
    volume = "11",
    pages = "078",
    year = "2020"
}

@article{Brivio:2021alv,
    author = {Brivio, Ilaria and Bruggisser, Sebastian and Geoffray, Emma and Killian, Wolfgang and Kr\"amer, Michael and Luchmann, Michel and Plehn, Tilman and Summ, Benjamin},
    title = "{From models to SMEFT and back?}",
    eprint = "2108.01094",
    archivePrefix = "arXiv",
    primaryClass = "hep-ph",
    doi = "10.21468/SciPostPhys.12.1.036",
    journal = "SciPost Phys.",
    volume = "12",
    number = "1",
    pages = "036",
    year = "2022"
}

@article{Haisch:2020ahr,
    author = "Haisch, Ulrich and Ruhdorfer, Maximilian and Salvioni, Ennio and Venturini, Elena and Weiler, Andreas",
    title = "{Singlet night in Feynman-ville: one-loop matching of a real scalar}",
    eprint = "2003.05936",
    archivePrefix = "arXiv",
    primaryClass = "hep-ph",
    reportNumber = "CERN-TH-2020-038, TUM-HEP-1254-20",
    doi = "10.1007/JHEP04(2020)164",
    journal = "JHEP",
    volume = "04",
    pages = "164",
    year = "2020",
    note = "[Erratum: JHEP 07, 066 (2020)]"
}

@article{Jiang:2018pbd,
    author = "Jiang, Minyuan and Craig, Nathaniel and Li, Ying-Ying and Sutherland, Dave",
    title = "{Complete one-loop matching for a singlet scalar in the Standard Model EFT}",
    eprint = "1811.08878",
    archivePrefix = "arXiv",
    primaryClass = "hep-ph",
    doi = "10.1007/JHEP02(2019)031",
    journal = "JHEP",
    volume = "02",
    pages = "031",
    year = "2019",
    note = "[Erratum: JHEP 01, 135 (2021)]"
}

@article{Gherardi:2020det,
    author = "Gherardi, Valerio and Marzocca, David and Venturini, Elena",
    title = "{Matching scalar leptoquarks to the SMEFT at one loop}",
    eprint = "2003.12525",
    archivePrefix = "arXiv",
    primaryClass = "hep-ph",
    doi = "10.1007/JHEP07(2020)225",
    journal = "JHEP",
    volume = "07",
    pages = "225",
    year = "2020",
    note = "[Erratum: JHEP 01, 006 (2021)]"
}

@article{Wells:2017vla,
    author = "Wells, James D. and Zhang, Zhengkang",
    title = "{Effective field theory approach to trans-TeV supersymmetry: covariant matching, Yukawa unification and Higgs couplings}",
    eprint = "1711.04774",
    archivePrefix = "arXiv",
    primaryClass = "hep-ph",
    reportNumber = "LCTP-17-06",
    doi = "10.1007/JHEP05(2018)182",
    journal = "JHEP",
    volume = "05",
    pages = "182",
    year = "2018"
}

@article{Huo:2015nka,
    author = "Huo, Ran",
    title = "{Effective Field Theory of Integrating out Sfermions in the MSSM: Complete One-Loop Analysis}",
    eprint = "1509.05942",
    archivePrefix = "arXiv",
    primaryClass = "hep-ph",
    reportNumber = "IPMU15-0172",
    doi = "10.1103/PhysRevD.97.075013",
    journal = "Phys. Rev. D",
    volume = "97",
    number = "7",
    pages = "075013",
    year = "2018"
}

@article{Fuentes-Martin:2016uol,
    author = "Fuentes-Martin, Javier and Portoles, Jorge and Ruiz-Femenia, Pedro",
    title = "{Integrating out heavy particles with functional methods: a simplified framework}",
    eprint = "1607.02142",
    archivePrefix = "arXiv",
    primaryClass = "hep-ph",
    reportNumber = "IFIC-16-28, TUM-HEP-1047-16",
    doi = "10.1007/JHEP09(2016)156",
    journal = "JHEP",
    volume = "09",
    pages = "156",
    year = "2016"
}

@article{Kraml:2025fpv,
    author = "Kraml, Sabine and Lessa, Andre and Prakash, Suraj and Wilsch, Felix",
    title = "{SUSY meets SMEFT: Complete one-loop matching of the general MSSM}",
    eprint = "2506.05201",
    archivePrefix = "arXiv",
    primaryClass = "hep-ph",
    reportNumber = "TTK-25-14, P3H-25-036",
    month = "6",
    year = "2025"
}

@article{Rogers:1970xx,
  title = "{Bound Eigenstates of the Static Screened Coulomb Potential}",
  author = "{Rogers, F. J. and Graboske, H. C. and Harwood, D. J.}",
  journal = {Phys. Rev. A},
  volume = {1},
  issue = {6},
  pages = {1577--1586},
  numpages = {0},
  year = {1970},
  month = {Jun},
  publisher = {American Physical Society},
  doi = {10.1103/PhysRevA.1.1577},
  url = {https://link.aps.org/doi/10.1103/PhysRevA.1.1577}
}

@article{Bollig:2021psb,
    author = "Bollig, Julian and Vogl, Stefan",
    title = "{Impact of bound states on non-thermal dark matter production}",
    eprint = "2112.01491",
    archivePrefix = "arXiv",
    primaryClass = "hep-ph",
    doi = "10.1088/1475-7516/2022/10/031",
    journal = "JCAP",
    volume = "10",
    pages = "031",
    year = "2022"
}

@article{ParticleDataGroup:2016lqr,
    author = "Patrignani, C. and others",
    collaboration = "Particle Data Group",
    title = "{Review of Particle Physics}",
    doi = "10.1088/1674-1137/40/10/100001",
    journal = "Chin. Phys. C",
    volume = "40",
    number = "10",
    pages = "100001",
    year = "2016"
}

@article{Electroweak:2003ram,
    author = "{The LEP Collaborations: ALEPH, DELPHI,  L3, OPAL and the  LEP Electroweak Working Group}",
    collaboration = "LEP, ALEPH, DELPHI, L3, OPAL, LEP Electroweak Working Group, SLD Electroweak Group, SLD Heavy Flavor Group",
    title = "{A Combination of preliminary electroweak measurements and constraints on the standard model}",
    eprint = "hep-ex/0312023",
    archivePrefix = "arXiv",
    reportNumber = "SLAC-R-701, LEPEWWG-2003-02, ALEPH-2003-017-PHYSICS-2003-005, DELPHI-2003-072-PHYS-937, L3-NOTE-2825, OPAL-PR-392, SLD-PHYSICS-NOTE-78, CERN-EP-2003-091, LEP-EWWG-2003-02",
    month = "12",
    year = "2003"
}

@article{ALEPH:2013dgf,
    author = "Schael, S. and others",
    collaboration = "ALEPH, DELPHI, L3, OPAL, LEP Electroweak",
    title = "{Electroweak Measurements in Electron-Positron Collisions at W-Boson-Pair Energies at LEP}",
    eprint = "1302.3415",
    archivePrefix = "arXiv",
    primaryClass = "hep-ex",
    reportNumber = "CERN-PH-EP-2013-022",
    doi = "10.1016/j.physrep.2013.07.004",
    journal = "Phys. Rept.",
    volume = "532",
    pages = "119--244",
    year = "2013"
}

@article{VENUS:1997cjg,
    author = "Hanai, H. and others",
    collaboration = "VENUS",
    title = "{Measurement of tau polarization in e+ e- annihilation at s**(1/2) = 58-GeV}",
    eprint = "hep-ex/9703003",
    archivePrefix = "arXiv",
    reportNumber = "KEK-PREPRINT-96-171, KOBE-HEP-97-01, NGTHEP-97-01, OULNS-96-05, TMU-HEP-97-20",
    doi = "10.1016/S0370-2693(97)00506-6",
    journal = "Phys. Lett. B",
    volume = "403",
    pages = "155--162",
    year = "1997"
}

@article{Petraki:2015hla,
    author = "Petraki, Kalliopi and Postma, Marieke and Wiechers, Michael",
    title = "{Dark-matter bound states from Feynman diagrams}",
    eprint = "1505.00109",
    archivePrefix = "arXiv",
    primaryClass = "hep-ph",
    reportNumber = "NIKHEF-2015-013",
    doi = "10.1007/JHEP06(2015)128",
    journal = "JHEP",
    volume = "06",
    pages = "128",
    year = "2015"
}

@article{Dreiner:2012xm,
    author = {Dreiner, Herbert and Huck, Moritz and Kr\"amer, Michael and Schmeier, Daniel and Tattersall, Jamie},
    title = "{Illuminating Dark Matter at the ILC}",
    eprint = "1211.2254",
    archivePrefix = "arXiv",
    primaryClass = "hep-ph",
    doi = "10.1103/PhysRevD.87.075015",
    journal = "Phys. Rev. D",
    volume = "87",
    number = "7",
    pages = "075015",
    year = "2013"
}

@article{Bartocci:2023nvp,
    author = {Bartocci, Riccardo and Biek\"otter, Anke and Hurth, Tobias},
    title = "{A global analysis of the SMEFT under the minimal MFV assumption}",
    eprint = "2311.04963",
    archivePrefix = "arXiv",
    primaryClass = "hep-ph",
    reportNumber = "MITP/23-063",
    doi = "10.1007/JHEP05(2024)074",
    journal = "JHEP",
    volume = "05",
    pages = "074",
    year = "2024"
}

@article{Manohar:1997qy,
    author = "Manohar, Aneesh V.",
    title = "{The HQET / NRQCD Lagrangian to order alpha / m-3}",
    eprint = "hep-ph/9701294",
    archivePrefix = "arXiv",
    reportNumber = "UCSD-PTH-97-01",
    doi = "10.1103/PhysRevD.56.230",
    journal = "Phys. Rev. D",
    volume = "56",
    pages = "230--237",
    year = "1997"
}

@article{Grzadkowski_2010,
   title={Dimension-six terms in the Standard Model Lagrangian},
   volume={2010},
   ISSN={1029-8479},
   url={http://dx.doi.org/10.1007/JHEP10(2010)085},
   DOI={10.1007/jhep10(2010)085},
   number={10},
   journal={Journal of High Energy Physics},
   publisher={Springer Science and Business Media LLC},
   author={Grzadkowski, B. and Iskrzyński, M. and Misiak, M. and Rosiek, J.},
   year={2010},
   month=oct }

@article{Fuentes-Martin:2022vvu,
    author = {Fuentes-Mart\'\i{}n, Javier and K\"onig, Matthias and Pag\`es, Julie and Thomsen, Anders Eller and Wilsch, Felix},
    title = "{Evanescent operators in one-loop matching computations}",
    eprint = "2211.09144",
    archivePrefix = "arXiv",
    primaryClass = "hep-ph",
    reportNumber = "MITP-22-091, TUM-HEP-1428/22, ZU-TH-48/22",
    doi = "10.1007/JHEP02(2023)031",
    journal = "JHEP",
    volume = "02",
    pages = "031",
    year = "2023"
}

@article{Banta_2022,
   title={Non-decoupling new particles},
   volume={2022},
   ISSN={1029-8479},
   url={http://dx.doi.org/10.1007/JHEP02(2022)029},
   DOI={10.1007/jhep02(2022)029},
   number={2},
   journal={Journal of High Energy Physics},
   publisher={Springer Science and Business Media LLC},
   author={Banta, Ian and Cohen, Timothy and Craig, Nathaniel and Lu, Xiaochuan and Sutherland, Dave},
   year={2022},
   month=feb }

@article{Brivio_2016,
   title={The complete HEFT Lagrangian after the LHC Run I},
   volume={76},
   ISSN={1434-6052},
   url={http://dx.doi.org/10.1140/epjc/s10052-016-4211-9},
   DOI={10.1140/epjc/s10052-016-4211-9},
   number={7},
   journal={The European Physical Journal C},
   publisher={Springer Science and Business Media LLC},
   author={Brivio, I. and Gonzalez-Fraile, J. and Gonzalez-Garcia, M. C. and Merlo, L.},
   year={2016},
   month=jul }

@article{FERUGLIO_1993,
   title={THE CHIRAL APPROACH TO THE ELECTROWEAK INTERACTIONS},
   volume={08},
   ISSN={1793-656X},
   url={http://dx.doi.org/10.1142/S0217751X93001946},
   DOI={10.1142/s0217751x93001946},
   number={28},
   journal={International Journal of Modern Physics A},
   publisher={World Scientific Pub Co Pte Lt},
   author={FERUGLIO, F.},
   year={1993},
   month=nov, pages={4937–4972} }

@misc{deblas2024globalsmeftfitsfuture,
      title={Global SMEFT Fits at Future Colliders}, 
      author={Jorge de Blas and Yong Du and Christophe Grojean and Jiayin Gu and Victor Miralles and Michael E. Peskin and Junping Tian and Marcel Vos and Eleni Vryonidou},
      year={2024},
      eprint={2206.08326},
      archivePrefix={arXiv},
      primaryClass={hep-ph},
      url={https://arxiv.org/abs/2206.08326}, 
}

@article{DiFranzo:2013vra,
    author = "DiFranzo, Anthony and Nagao, Keiko I. and Rajaraman, Arvind and Tait, Tim M. P.",
    title = "{Simplified Models for Dark Matter Interacting with Quarks}",
    eprint = "1308.2679",
    archivePrefix = "arXiv",
    primaryClass = "hep-ph",
    reportNumber = "UCI-HEP-TR-2013-17, KEK-TH-1659",
    doi = "10.1007/JHEP11(2013)014",
    journal = "JHEP",
    volume = "11",
    pages = "014",
    year = "2013",
    note = "[Erratum: JHEP 01, 162 (2014)]"
}

@article{Bai:2013iqa,
    author = "Bai, Yang and Berger, Joshua",
    title = "{Fermion Portal Dark Matter}",
    eprint = "1308.0612",
    archivePrefix = "arXiv",
    primaryClass = "hep-ph",
    reportNumber = "SLAC-PUB-15704",
    doi = "10.1007/JHEP11(2013)171",
    journal = "JHEP",
    volume = "11",
    pages = "171",
    year = "2013"
}

@article{Garny:2013ama,
    author = "Garny, Mathias and Ibarra, Alejandro and Pato, Miguel and Vogl, Stefan",
    title = "{Internal bremsstrahlung signatures in light of direct dark matter searches}",
    eprint = "1306.6342",
    archivePrefix = "arXiv",
    primaryClass = "hep-ph",
    reportNumber = "DESY-13-115, TUM-HEP-896-13",
    doi = "10.1088/1475-7516/2013/12/046",
    journal = "JCAP",
    volume = "12",
    pages = "046",
    year = "2013"
}

@article{Garny:2011ii,
    author = "Garny, Mathias and Ibarra, Alejandro and Vogl, Stefan",
    title = "{Dark matter annihilations into two light fermions and one gauge boson: General analysis and antiproton constraints}",
    eprint = "1112.5155",
    archivePrefix = "arXiv",
    primaryClass = "hep-ph",
    reportNumber = "DESY-11-257, TUM-HEP-823-11",
    doi = "10.1088/1475-7516/2012/04/033",
    journal = "JCAP",
    volume = "04",
    pages = "033",
    year = "2012"
}

@article{Fuentes-Martin:2020udw,
    author = {Fuentes-Martin, Javier and K\"onig, Matthias and Pag\`es, Julie and Thomsen, Anders Eller and Wilsch, Felix},
    title = "{SuperTracer: A Calculator of Functional Supertraces for One-Loop EFT Matching}",
    eprint = "2012.08506",
    archivePrefix = "arXiv",
    primaryClass = "hep-ph",
    reportNumber = "MITP-20-076, TUM-HEP-1302/20, ZU-TH-54/20",
    doi = "10.1007/JHEP04(2021)281",
    journal = "JHEP",
    volume = "04",
    pages = "281",
    year = "2021"
}

@article{Cohen:2020fcu,
    author = "Cohen, Timothy and Lu, Xiaochuan and Zhang, Zhengkang",
    title = "{Functional Prescription for EFT Matching}",
    eprint = "2011.02484",
    archivePrefix = "arXiv",
    primaryClass = "hep-ph",
    reportNumber = "CALT-TH-2020-047",
    doi = "10.1007/JHEP02(2021)228",
    journal = "JHEP",
    volume = "02",
    pages = "228",
    year = "2021"
}

@article{CMS:2021ctt,
    author = "Sirunyan, Albert M and others",
    collaboration = "CMS",
    title = "{Search for resonant and nonresonant new phenomena in high-mass dilepton final states at $ \sqrt{s} $ = 13 TeV}",
    eprint = "2103.02708",
    archivePrefix = "arXiv",
    primaryClass = "hep-ex",
    reportNumber = "CMS-EXO-19-019, CERN-EP-2021-026",
    doi = "10.1007/JHEP07(2021)208",
    journal = "JHEP",
    volume = "07",
    pages = "208",
    year = "2021"
}

@article{ATLAS:2020zms,
    author = "Aad, Georges and others",
    collaboration = "ATLAS",
    title = "{Search for heavy Higgs bosons decaying into two tau leptons with the ATLAS detector using $pp$ collisions at $\sqrt{s}=13$ TeV}",
    eprint = "2002.12223",
    archivePrefix = "arXiv",
    primaryClass = "hep-ex",
    reportNumber = "CERN-EP-2020-014",
    doi = "10.1103/PhysRevLett.125.051801",
    journal = "Phys. Rev. Lett.",
    volume = "125",
    number = "5",
    pages = "051801",
    year = "2020"
}

@article{ATLAS:2018dpp,
    author = "Aaboud, M. and others",
    collaboration = "ATLAS",
    title = "{Search for Higgs boson pair production in the $\gamma\gamma b\bar{b}$ final state with 13 TeV $pp$ collision data collected by the ATLAS experiment}",
    eprint = "1807.04873",
    archivePrefix = "arXiv",
    primaryClass = "hep-ex",
    reportNumber = "CERN-EP-2018-130",
    doi = "10.1007/JHEP11(2018)040",
    journal = "JHEP",
    volume = "11",
    pages = "040",
    year = "2018"
}

@article{ATLAS:2018rnh,
    author = "Aaboud, Morad and others",
    collaboration = "ATLAS",
    title = "{Search for pair production of Higgs bosons in the $b\bar{b}b\bar{b}$ final state using proton-proton collisions at $\sqrt{s} = 13$ TeV with the ATLAS detector}",
    eprint = "1804.06174",
    archivePrefix = "arXiv",
    primaryClass = "hep-ex",
    reportNumber = "CERN-EP-2018-029",
    doi = "10.1007/JHEP01(2019)030",
    journal = "JHEP",
    volume = "01",
    pages = "030",
    year = "2019"
}

@article{ATLAS:2018uni,
    author = "Aaboud, Morad and others",
    collaboration = "ATLAS",
    title = "{Search for resonant and non-resonant Higgs boson pair production in the ${b\bar{b}\tau^+\tau^-}$ decay channel in $pp$ collisions at $\sqrt{s}=13$ TeV with the ATLAS detector}",
    eprint = "1808.00336",
    archivePrefix = "arXiv",
    primaryClass = "hep-ex",
    reportNumber = "CERN-EP-2018-164",
    doi = "10.1103/PhysRevLett.121.191801",
    journal = "Phys. Rev. Lett.",
    volume = "121",
    number = "19",
    pages = "191801",
    year = "2018",
    note = "[Erratum: Phys.Rev.Lett. 122, 089901 (2019)]"
}

@article{CMS:2017hea,
    author = "Sirunyan, Albert M and others",
    collaboration = "CMS",
    title = "{Search for Higgs boson pair production in events with two bottom quarks and two tau leptons in proton{\textendash}proton collisions at $\sqrt s$ =13TeV}",
    eprint = "1707.02909",
    archivePrefix = "arXiv",
    primaryClass = "hep-ex",
    reportNumber = "CMS-HIG-17-002, CERN-EP-2017-126",
    doi = "10.1016/j.physletb.2018.01.001",
    journal = "Phys. Lett. B",
    volume = "778",
    pages = "101--127",
    year = "2018"
}

@article{CMS:2022cpr,
    author = "Tumasyan, Armen and others",
    collaboration = "CMS",
    title = "{Search for Higgs Boson Pair Production in the Four b Quark Final State in Proton-Proton Collisions at s=13{\,}{\,}TeV}",
    eprint = "2202.09617",
    archivePrefix = "arXiv",
    primaryClass = "hep-ex",
    reportNumber = "CMS-HIG-20-005, CERN-EP-2022-004",
    doi = "10.1103/PhysRevLett.129.081802",
    journal = "Phys. Rev. Lett.",
    volume = "129",
    number = "8",
    pages = "081802",
    year = "2022"
}

@article{Cohen:2020xca,
    author = "Cohen, Timothy and Craig, Nathaniel and Lu, Xiaochuan and Sutherland, Dave",
    title = "{Is SMEFT Enough?}",
    eprint = "2008.08597",
    archivePrefix = "arXiv",
    primaryClass = "hep-ph",
    doi = "10.1007/JHEP03(2021)237",
    journal = "JHEP",
    volume = "03",
    pages = "237",
    year = "2021"
}

@article{Ibarra:2024mpq,
    author = "Ibarra, Alejandro and Reichard, Merlin and Tomar, Gaurav",
    title = "{Probing dark matter electromagnetic properties in direct detection experiments}",
    eprint = "2408.15760",
    archivePrefix = "arXiv",
    primaryClass = "hep-ph",
    doi = "10.1088/1475-7516/2025/02/072",
    journal = "JCAP",
    volume = "02",
    pages = "072",
    year = "2025"
}

\end{document}